\documentclass[twocolumn,astrosymb]{aastex62}

\usepackage{morefloats}
\usepackage{amsmath}
\usepackage{amssymb}
\usepackage{latexsym}
\usepackage{multirow}
\usepackage{enumitem}
\hypersetup{linkcolor=red,citecolor=blue,filecolor=cyan,urlcolor=teal}
\received{February 5, 2019}
\revised{\today}
\accepted{N/A}
\submitjournal{ApJS}
\shorttitle{ELECTRON PARTITION METHODOLOGY}
\shortauthors{Wilson III et al.}
\makeatletter
\renewcommand\@makefnmark{\hbox{\@textsuperscript{\normalfont\color{violet}\@thefnmark}}}
\makeatother

\begin{document}

\title{Electron energy partition across interplanetary shocks: I.  Methodology and Data Product}
\correspondingauthor{L.B. Wilson III}
\email{lynn.b.wilsoniii@gmail.com}

\author[0000-0002-4313-1970]{Lynn B. Wilson III}
\affiliation{NASA Goddard Space Flight Center, Heliophysics Science Division, Greenbelt, MD, USA.}

\author[0000-0002-4768-189X]{Li-Jen Chen}
\affiliation{NASA Goddard Space Flight Center, Heliophysics Science Division, Greenbelt, MD, USA.}

\author[0000-0002-6783-7759]{Shan Wang}
\affiliation{Astronomy Department, University of Maryland, College Park, Maryland, USA.}
\affiliation{NASA Goddard Space Flight Center, Heliophysics Science Division, Greenbelt, MD, USA.}

\author[0000-0003-0682-2753]{Steven J. Schwartz}
\affiliation{Laboratory for Atmospheric and Space Physics, University of Colorado, Boulder, Boulder, CO, USA.}

\author[0000-0002-2425-7818]{Drew L. Turner}
\affiliation{Space Sciences Department, The Aerospace Corporation, El Segundo, CA, USA.}

\author[0000-0002-7728-0085]{Michael L. Stevens}
\affiliation{Harvard-Smithsonian Center for Astrophysics, Harvard University, Cambridge, MA, USA.}

\author[0000-0002-7077-930X]{Justin C. Kasper}
\affiliation{University of Michigan, Ann Arbor, School of Climate and Space Sciences and Engineering, Ann Arbor, MI, USA.}

\author[0000-0003-2555-5953]{Adnane Osmane}
\affiliation{Department of Physics, University of Helsinki, Helsinki, Finland.}

\author[0000-0003-0939-8775]{Damiano Caprioli}
\affiliation{Department of Astronomy and Astrophysics, University of Chicago, Chicago, IL, USA.}

\author[0000-0002-1989-3596]{Stuart D. Bale}
\affiliation{University of California Berkeley, Space Sciences Laboratory, Berkeley, CA, USA.}

\author[0000-0002-1573-7457]{Marc P. Pulupa}
\affiliation{University of California Berkeley, Space Sciences Laboratory, Berkeley, CA, USA.}

\author[0000-0002-6536-1531]{Chadi S. Salem}
\affiliation{University of California Berkeley, Space Sciences Laboratory, Berkeley, CA, USA.}

\author[0000-0002-4288-5084]{Katherine A. Goodrich}
\affiliation{University of California Berkeley, Space Sciences Laboratory, Berkeley, CA, USA.}

\begin{abstract}
  Analysis of 15,314 electron velocity distribution functions (VDFs) within $\pm$2 hours of 52 interplanetary (IP) shocks observed by the \emph{Wind} spacecraft near 1 AU are introduced.  The electron VDFs are fit to the sum of three model functions for the cold dense core, hot tenuous halo, and field-aligned beam/strahl component.  The best results were found by modeling the core as either a bi-kappa or a symmetric (or asymmetric) bi-self-similar velocity distribution function, while both the halo and beam/strahl components were best fit to bi-kappa velocity distribution function.  This is the first statistical study to show that the core electron distribution is better fit to a self-similar velocity distribution function than a bi-Maxwellian under all conditions.  The self-similar distribution deviation from a Maxwellian is a measure of inelasticity in particle scattering from waves and/or turbulence.  The range of values defined by the lower and upper quartiles for the kappa exponents are $\kappa{\scriptstyle_{ec}}$ $\sim$ 5.40--10.2 for the core, $\kappa{\scriptstyle_{eh}}$ $\sim$ 3.58--5.34 for the halo, and $\kappa{\scriptstyle_{eb}}$ $\sim$ 3.40--5.16 for the beam/strahl.  The lower-to-upper quartile range of symmetric bi-self-similar core exponents are $s{\scriptstyle_{ec}}$ $\sim$ 2.00--2.04, and asymmetric bi-self-similar core exponents are $p{\scriptstyle_{ec}}$ $\sim$ 2.20--4.00 for the parallel exponent, and $q{\scriptstyle_{ec}}$ $\sim$ 2.00--2.46 for the perpendicular exponent.  The nuanced details of the fit procedure and description of resulting data product are also presented.  The statistics and detailed analysis of the results are presented in Paper II and Paper III of this three-part study.
\end{abstract}


\phantomsection   
\section{Background and Motivation}  \label{sec:introduction}

\begin{figure*}
  \centering
    {\includegraphics[trim = 0mm 0mm 0mm 0mm, clip, width=170mm]{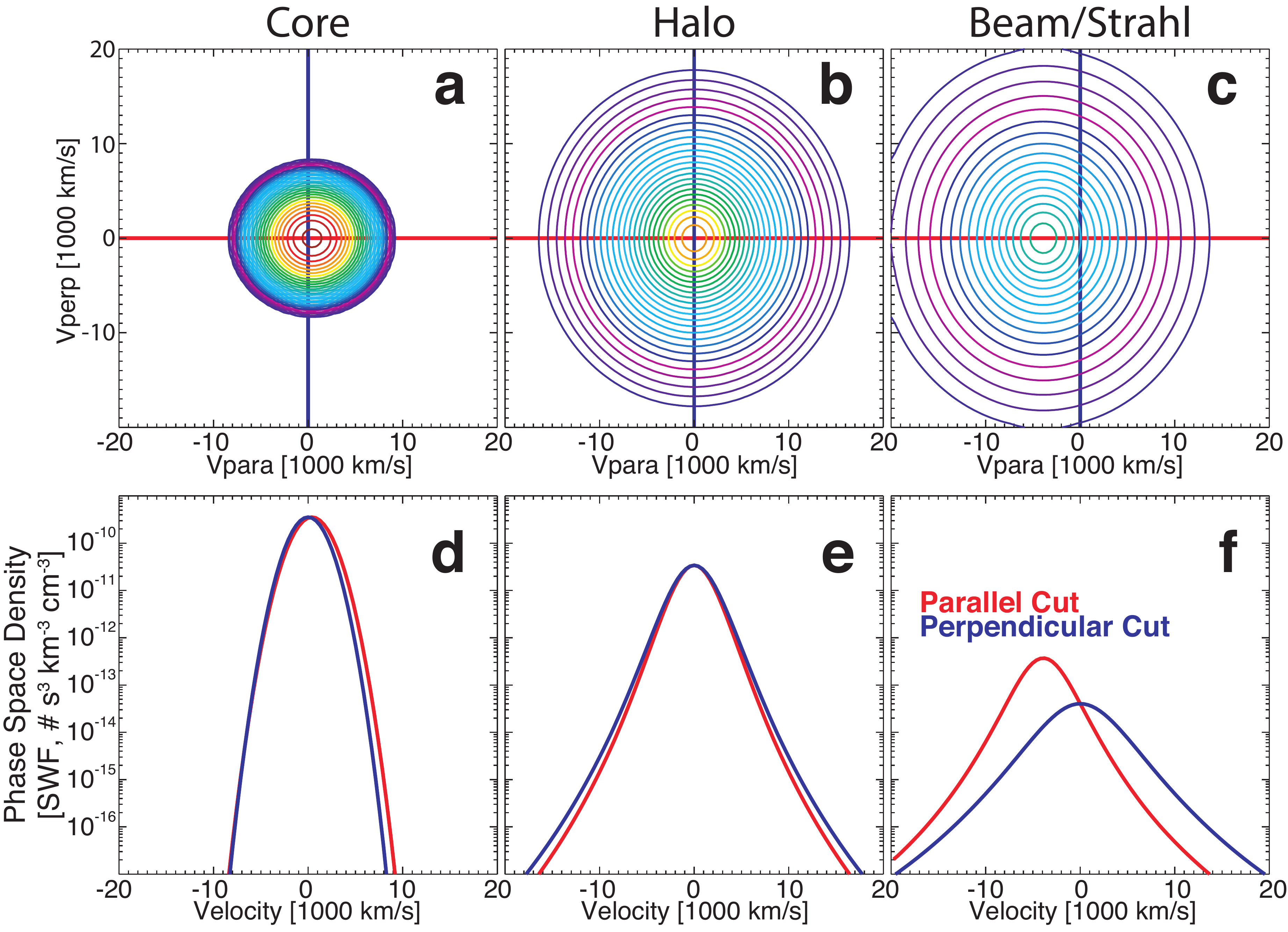}}
    \caption{Illustrative example VDFs of the core, halo, and beam/strahl components of the electron VDFs in the solar wind near 1 AU.  The top row (panels a through c) shows contours of constant phase space density [$cm^{-3} \ km^{-3} \ s^{+3}$] of a two-dimensional cut through a three-dimensional VDF.  The plane and coordinate basis are defined by the quasi-static magnetic field, $\mathbf{B}{\scriptstyle_{o}}$, and the ion bulk flow velocity, $\mathbf{V}{\scriptstyle_{i}}$.  The vertical axis is defined by the unit vector $\left( \mathbf{B}{\scriptstyle_{o}} \times \mathbf{V}{\scriptstyle_{i}} \right) \times \mathbf{B}{\scriptstyle_{o}}$ and the horizontal by $\mathbf{B}{\scriptstyle_{o}}$.  The bottom row (panels d through f) shows one-dimensional cuts of the VDF along the horizontal (solid red line) and along the vertical (solid blue line).  The location of these cuts are defined by the color-coded crosshairs in the top row panels.  The VDF is shown in the ion bulk flow rest frame.}
    \label{fig:ExampleCartoonVDFComponents}
\end{figure*}

\indent  The solar wind is an ionized gas experiencing collective effects where Coulomb collisions occur, but the rates are often so low that, for instance, two constituent particle species, $s'$ and $s$, are not in thermodynamic or thermal equilibrium, i.e., $\left(T{\scriptstyle_{s'}}/T{\scriptstyle_{s}}\right){\scriptstyle_{tot}}$ $\neq$ 1 for $s'$ $\neq$ $s$, and the relevant scale lengths are orders of magnitude smaller than the collisional mean free path \citep[e.g.,][]{wilsoniii18b}.  Therefore, for any process dependent upon scales like the thermal gyroradii, $\rho{\scriptstyle_{cs}}$, or inertial lengths, $\lambda{\scriptstyle_{s}}$, the media is considered collisionless (see Appendix \ref{app:Definitions} for definitions).  That the solar wind is a non-equilibrium, weakly collisional, kinetic gas results in multi-component velocity distribution functions (VDFs) for both ions \citep[e.g.,][]{kasper06a, kasper12a, kasper13a, maruca11a, maruca13a, wicks16a} and electrons \citep[e.g.,][]{lin98a, maksimovic97a, maksimovic98a, pierrard99a, pierrard01a, pulupa14a, schwartz83b, stverak08a, stverak09a}.

\indent  The electron VDFs in the solar wind below $\sim$1 keV are comprised of a cold core with energies $E{\scriptstyle_{ec}}$ $\lesssim$ 15 eV \citep[e.g.,][]{bale13a, maksimovic97a, maksimovic98a, pilipp87a, pilipp87b, pilipp87c, pilipp90a, pulupa14a}, a hot, tenuous halo with $E{\scriptstyle_{eh}}$ $\gtrsim$ 20 eV \citep[e.g.,][]{maksimovic97a, maksimovic98a, pulupa14a, stverak08a, stverak09a}, and an anti-sunward, field-aligned beam called the strahl with $E{\scriptstyle_{eb}}$ $\sim$few 10s of eV \citep[e.g.,][]{bale13a, crooker03a, graham17a, graham18a, horaites18a, stverak09a} (e.g., see Figure \ref{fig:ExampleCartoonVDFComponents} for illustrative example).  The electrons also dominate the solar wind heat flux \citep[e.g.,][]{bale13a, crooker03a, pagel05a, pagel07}, arising from the consistent skewness in the VDFs, specifically the halo and/or strahl components.  Note that there also exists a suprathermal super halo with $E{\scriptstyle_{esh}}$ $\gtrsim$ 1 keV \citep[e.g.,][]{lin98a, wang12a, wang15a}, but these higher energy electrons are not examined herein.

\indent  The three electron components below $\sim$1 keV are predicted and observed to be coupled through multiple processes from wave-particle interactions \citep[e.g.,][]{pierrard11a, pierrard16a, phillips89a, phillips89b, saito07, saito08, vocks03a, vocks05a, yoon14a, yoon12a, yoon15b, yoon16b} to adiabatic transport effects \citep[e.g.,][]{schwartz83b} to collisional effects \citep[e.g.,][]{pilipp87a, pilipp87b, pilipp87c, schwartz83b}.  They have also been shown to behave differently across collisionless shocks depending upon shock strength \citep[e.g.,][]{wilsoniii09a, wilsoniii10a}.

\indent  An illustrative example, showing the three electron components typically observed in the solar wind near 1 AU below $\sim$1.2 keV, is shown in Figure \ref{fig:ExampleCartoonVDFComponents}.  The components parameters are exaggerated\footnote{The following were enhanced to increase contrast and for ease of viewing differences:  parallel core temperature, perpendicular halo temperature, and parallel core drift speed.} for illustrative purposes but based upon the fit results of the VDF shown in Figure \ref{fig:ExampleGoodVDF}.  The core is modeled by a symmetric bi-self-similar VDF and the halo and beam/strahl by a bi-kappa VDF (see Section \ref{subsec:VelocityDistributionFunctions}).  In this case, the self-similar exponent reduced to 2 so the VDF reduced to a bi-Maxwellian (see Section \ref{subsec:VelocityDistributionFunctions}).  This example is phenomenologically consistent with the majority of solar wind electron VDFs \citep[e.g.,][]{phillips89a, phillips89b, pilipp87a, pilipp87b, pilipp87c, stverak08a, stverak09a}.

\indent  Despite its collisionless, non-equilibrium nature the solar wind can support the existence of shock waves.  That the particles are in neither thermal or thermodynamic equilibrium leads to a non-homogeneous partition of energy among not only electrons and ions but also among the components of each species, e.g., the core electrons do not have the same response as the halo to collisionless shock waves.  The reason for the non-homogeneous partition of energy lies in the energy-dependent mechanisms that transfer the bulk flow kinetic energy lost across the shock ramp to other forms like heat or particle acceleration \citep[e.g., see][and references therein]{coroniti70b, kennel85a, sagdeev66, tidman71a, treumann09a, wilsoniii16a, wilsoniii17c}.  The mechanisms can also be dependent upon pitch-angle and species \citep[e.g.,][]{artemyev13j, artemyev14e, artemyev15d, artemyev16b, artemyev17a, artemyev17b, artemyev18a, sagdeev66}.  Most collisionless shocks are subsonic to electrons, yet electrons still respond to the shock showing even Mach number dependent effects \citep[e.g.,][]{feldman82a, feldman83a, feldman83b, masters11a, thomsen85a, thomsen87b, thomsen93a, wilsoniii10a}.  This is all further complicated by recent observations showing that the evolution of the electron VDF through a collisionless shock is not a trivial, uniform inflation of the entire distribution, but a multi-stage process that deforms and redistributes/exchanges energy for different energies and pitch-angles at different stages \citep[e.g.,][]{chen18a, goodrich18c, goodrich19a}.  There is no currently known way to quantify these non-homogenous changes to capture the energy- and pitch-angle-dependent effects, therefore the next best systematic approach for a statistical study is to parameterize the electron components by their velocity moments.  This is further supported by the fact that nearly all theories describing the evolution of electron VDFs rely upon either the velocity moments or a model velocity distribution function \citep[e.g.,][]{livadiotis15a, livadiotis17b, nicolaou18a, schunk75a, schunk77a, schwartz83b, schwartz88a, shizgal18a}.

\indent  In this first part of a multi-part study we describe the methodology and numerical analysis techniques used to model the solar wind eVDFs below $\sim$1.2 keV observed by the \emph{Wind} spacecraft near 1 AU around 52 interplanetary (IP) shocks.  This is the first statistical study to show that the core electron distribution is better fit to a self-similar velocity distribution function than a bi-Maxwellian under all conditions.  The analysis differs from numerous previous studies in its approach and the model functions used, each of which are justified herein using physically significant arguments.  A benefit of the analysis is an improved, semi-analytic relationship between the spacecraft potential and ion number density.  The paper also includes procedural documentation to disclose the nuances and issues associated with applying a nonlinear least squares fitting algorithm to in situ VDF data in the solar wind.  This serves as a reference for use of the resulting data product described herein.  In Paper II \citep[][]{wilsoniii19b} the statistical results of the model fits are presented with comparison to previous studies and associated discussions.  In Paper III \citep[][]{wilsoniii19c} the analysis and interpretation of the model fit results are presented.

\indent  This paper is outlined as follows:  Section \ref{sec:DefinitionsDataSets} introduces the data sets and event selection; Section \ref{sec:FitMethodology} introduces the methodology of the fit analysis, model functions, parameter constraints, quality control, and summary of fit results; Section \ref{sec:ExponentsandDrifts} discusses the statistics of the fit exponents and drift velocities; and Section \ref{sec:Discussion} discusses the results and interpretations with reference to further analysis in the following Papers II and III.  Appendices are also included to provide additional details of the parameter definitions (Appendix \ref{app:Definitions}), spacecraft potential and detector calibration (Appendix \ref{app:DetectorCalibration}), numerical analysis procedure (Appendix \ref{app:NumericalAnalysis}), numerical instabilities (Appendix \ref{app:NumericalInstability}), direct fit method comparisons (Appendix \ref{app:NumericalMethodComparisons}), and the data product produced by this effort (Appendix \ref{app:DataProduct}).

\section{Data Sets and Event Selection}  \label{sec:DefinitionsDataSets}

\indent  In this section we introduce the instrument data sets and shock database used to examine the data observed by the \emph{Wind} spacecraft \citep{harten95a} near 1 AU.  The data described herein spanned from 00:55:40 UTC on 1995-02-26 to 23:04:00 UTC on 2000-02-20 (see Supplemental Material for list of dates).  The symbol/parameter definitions are found in Appendix \ref{app:Definitions}.

\begin{deluxetable*}{| l | c | c | c | c | c | c | c |}
  \tabletypesize{\small}    
  \tablecaption{Shock Parameters \label{tab:ShockParameters}}
  \tablehead{\colhead{Parameter} & \colhead{$X{\scriptstyle_{min}}$}\tablenotemark{a} & \colhead{$X{\scriptstyle_{max}}$}\tablenotemark{b} & \colhead{$\bar{X}$}\tablenotemark{c} & \colhead{$\tilde{X}$}\tablenotemark{d} & \colhead{$X{\scriptstyle_{25\%}}$}\tablenotemark{e} & \colhead{$X{\scriptstyle_{75\%}}$}\tablenotemark{f} & \colhead{$\sigma{\scriptstyle_{x}}$}\tablenotemark{g}}
  \startdata
  \hline
  $\langle \lvert \mathbf{B}{\scriptstyle_{o}} \rvert \rangle{\scriptstyle_{up}}$ [nT] & 1.04 & 17.4 & 5.96 & 5.59 & 3.99 & 7.10 & 3.01 \\
  $\langle n{\scriptstyle_{i}} \rangle{\scriptstyle_{up}}$ [$cm^{-3}$] & 0.60 & 21.3 & 8.34 & 8.00 & 3.70 & 12.1 & 5.32 \\
  $\langle \beta{\scriptstyle_{tot}} \rangle{\scriptstyle_{up}}$ [N/A] & 0.03 & 3.86 & 0.50 & 0.38 & 0.19 & 0.60 & 0.60 \\
  $\langle \lvert V{\scriptstyle_{shn}} \rvert \rangle{\scriptstyle_{up}}$ [km/s] & 155 & 699 & 460 & 456 & 383 & 535 & 123 \\
  $\langle \lvert U{\scriptstyle_{shn}} \rvert \rangle{\scriptstyle_{up}}$ [km/s] & 36.9 & 401 & 126 & 110 & 83.3 & 145 & 70.2 \\
  $\theta{\scriptstyle_{Bn}}$ [deg] & 17.1 & 88.6 & 56.8 & 54.6 & 42.7 & 73.3 & 19.5 \\
  $\langle M{\scriptstyle_{A}} \rangle{\scriptstyle_{up}}$ [N/A] & 1.06 & 15.6 & 2.79 & 2.41 & 1.86 & 3.06 & 2.10 \\
  $\langle M{\scriptstyle_{f}} \rangle{\scriptstyle_{up}}$ [N/A] & 1.01 & 6.39 & 2.12 & 1.86 & 1.58 & 2.35 & 0.94 \\
  $\langle M{\scriptstyle_{f}} \rangle{\scriptstyle_{up}} / M{\scriptstyle_{cr}}$ [N/A] & 0.41 & 5.14 & 1.08 & 0.91 & 0.77 & 1.19 & 0.70 \\
  $\langle M{\scriptstyle_{f}} \rangle{\scriptstyle_{up}} / M{\scriptstyle_{ww}}$ [N/A] & 0.06 & 2.49 & 0.36 & 0.18 & 0.11 & 0.32 & 0.51 \\
  $\langle M{\scriptstyle_{f}} \rangle{\scriptstyle_{up}} / M{\scriptstyle_{gr}}$ [N/A] & 0.04 & 1.91 & 0.28 & 0.14 & 0.09 & 0.25 & 0.39 \\
  $\langle M{\scriptstyle_{f}} \rangle{\scriptstyle_{up}} / M{\scriptstyle_{nw}}$ [N/A] & 0.04 & 1.76 & 0.26 & 0.13 & 0.08 & 0.23 & 0.36 \\
  \hline
  \enddata
  \tablenotetext{a}{minimum}
  \tablenotetext{b}{maximum}
  \tablenotetext{c}{mean}
  \tablenotetext{d}{median}
  \tablenotetext{e}{lower quartile}
  \tablenotetext{f}{upper quartile}
  \tablenotetext{g}{standard deviation}
  \tablecomments{For symbol definitions, see Appendix \ref{app:Definitions}.}
\end{deluxetable*}

\indent  Quasi-static magnetic field vectors ($\mathbf{B}{\scriptstyle_{o}}$) were measured by the \emph{Wind}/MFI dual, triaxial fluxgate magnetometers \citep[][]{lepping95} using the three second cadence data for each particle distribution.  The components/directions of some parameters are defined with respect to $\mathbf{B}{\scriptstyle_{o}}$ using the subscript $j$.  That is, the parallel ($j$ $=$ $\parallel$) and the perpendicular components ($j$ $=$ $\perp$) of any vector or pseudo-tensor (e.g., temperature) are defined with respect to $\mathbf{B}{\scriptstyle_{o}}$.

\indent  The electron velocity distribution functions (VDFs) were measured by the \emph{Wind}/3DP low energy (i.e., few eV to $\sim$1.2 keV) electron electrostatic analyzer \citep[][]{lin95a} or EESA Low.  The instrument operated in both burst and survey modes for the data presented herein, which has cadences of $\sim$3 seconds and $\sim$24--78 seconds, respectively.  The energy and angular resolutions are commandable but the instrument typically operates with $\Delta \ E$/$E$ $\sim$ 20\% and $\Delta \ \phi$ $\sim$ 5$^{\circ}$--22.5$^{\circ}$ depending on the poloidal anode\footnote{The ecliptic plane bins have higher angular resolution than the zenith.} \citep[e.g., see][for instrument details]{wilsoniii09a, wilsoniii10a}.

\indent  The EESA Low measurements are contaminated with photoelectrons from the spacecraft, something for which must be accounted to obtain accurate velocity moments or any other results.  The details of how the spacecraft potential, $\phi{\scriptstyle_{sc}}$, was numerically determined for each VDF is described in Appendix \ref{app:DetectorCalibration}.  The VDFs are transformed into the ion frame prior to any fit using relativistically correct Lorentz transformations, where the steps are as follows:  (1) convert the units of the VDFs to phase space density [\# cm$^{-3}$ s$^{+3}$ km$^{-3}$]; (2) correct the energies by $\phi{\scriptstyle_{sc}}$; (3) convert the energy-angle bins to velocity coordinates; and (4) transform the velocities into the ion rest frame using proper Lorentz transformations.  Nothing need be done to VDFs once in units of phase space density as phase space density is a Lorentz invariant \citep[][]{vankampen69a} (see Appendices \ref{app:DetectorCalibration} and \ref{app:NumericalAnalysis} for details).

\indent  We also examined solar wind proton and alpha-particle velocity moments determined by a nonlinear least squares fitting algorithm \citep[e.g.,][]{kasper06a, maruca13a} observed by the \emph{Wind}/SWE Faraday Cups \citep[][]{ogilvie95}.  Similar quality requirements for the SWE results to that discussed in \citet[][]{wilsoniii18b} were used herein.

\indent  The VDFs examined are found within $\pm$2 hours of 52 IP shocks found in the \emph{Wind} shock database from the Harvard Smithsonian Center for Astrophysics\footnote{\url{https://www.cfa.harvard.edu/shocks/wi\_data/}}.  Of those 52 IP shocks, there were 16 quasi-parallel ($\theta{\scriptstyle_{Bn}}$ $\leq$ 45$^{\circ}$), 36 quasi-perpendicular ($\theta{\scriptstyle_{Bn}}$ $>$ 45$^{\circ}$), 45 low Mach number ($\langle M{\scriptstyle_{f}} \rangle{\scriptstyle_{up}}$ $<$ 3), and 7 high Mach number ($\langle M{\scriptstyle_{f}} \rangle{\scriptstyle_{up}}$ $\geq$ 3) shocks.  The shock parameters for the 52 IP shocks examined in this three-part set of papers are shown in Table \ref{tab:ShockParameters} (see Supplemental Material for full list of values for each shock).  The IP shocks examined were selected because of burst mode 3DP availability.  See Appendix \ref{app:Definitions} for definitions of symbols and/or parameters.

\phantomsection   
\section{Fit Methodology}  \label{sec:FitMethodology}

\indent  This section (and Appendix \ref{app:NumericalAnalysis}) introduces and discusses the nuances of the approach and software used to numerically compute the model fit parameters for every electron VDF examined.  The nuances and details are provided for reproducibility and documentation for the data product discussed in Appendix \ref{app:DataProduct}.

\indent  The data are fit to a user defined model function using a nonlinear least squares fitting algorithm called the Levenberg-Marquardt Algorithm (LMA) \citep[][]{more78a}.  The generalized LMA software used for the present study is called MPFIT \citep[][]{markwardt09a}.  The specific details for its use are outlined in Appendix \ref{app:NumericalAnalysis}.

\indent  The components of the electron VDFs are fit to bi-Maxwellian, bi-kappa, or bi-self-similar model functions (see Section \ref{subsec:VelocityDistributionFunctions}).  The components can be fit separately because the solar wind is a non-equilibrium, weakly collisional, kinetic gas.  That is, in the absence of a magnetic field, each electron component could, in principle, stream past the other components for nearly an astronomical unit without significant interaction.  Thus, there is physical justification to fit to the sum of three model functions (see Appendix \ref{app:NumericalAnalysis} for details).

\indent  Given that the bi-self-similar reduces to the bi-Maxwellian in the limit as the exponential argument goes to 2 and that it consistently yielded lower reduced chi-squared values, $\tilde{\chi}{\scriptstyle_{s}}^{2}$, the symmetric bi-self-similar function was used as the default core model function.  In the downstream of strong (i.e., $\langle M{\scriptstyle_{f}} \rangle{\scriptstyle_{up}}$ $\gtrsim$ 2.5) IP shocks it was found that the asymmetric bi-self-similar function produced the best results and so was the default core model function\footnote{The parallel and perpendicular profiles at low energies differ greatly in these regions and required the use of the asymmetric function to accommodate the differences.  Using a symmetric function resulted in very poor fit qualities, as defined in Section \ref{subsec:FitQualityAnalysis}.}.  Note that of all the core VDFs fit to a symmetric bi-self-similar function, $\sim$80.5\% that satisfied 2.0 $\leq$ $s{\scriptstyle_{ec}}$ $\leq$ 2.05.  That is, the majority of the distributions would be nearly indistinguishable from a bi-Maxwellian on visual inspection.  The halo and beam/strahl were modeled with a bi-kappa model function for all VDFs examined since they always have a power-law tail and previous work found kappa model functions to be the best approximation \citep[e.g.,][]{maksimovic05a, stverak09a}.

\indent  For each IP shock, an iterative process was followed to correct for the spacecraft potential, $\phi{\scriptstyle_{sc}}$ (details found in Appendix \ref{app:DetectorCalibration}), and define fit parameter initial guess values and constraints to yield stable solutions for the most VDFs (detailed steps found in Appendix \ref{app:NumericalAnalysis} and list of initial guess values and constraints found in Supplemental Material ASCII files described in Appendix \ref{app:DataProduct}).  The process of defining the initial guess values and constraints is discussed in Section \ref{subsec:FitPhysicalConstraints} and the quantified estimates of the fit quality is discussed in Section \ref{subsec:FitQualityAnalysis}.

\indent  A total of 15,314 electron VDFs were observed by the \emph{Wind} spacecraft within $\pm$2 hours of 52 IP shocks.  Of those 15,314 VDFs, 15,210 progressed to fit analysis and stable model function parameters were found for 14,847($\sim$98\%) core fits, 13,871($\sim$91\%) halo fits, and 9567($\sim$63\%) beam/strahl fits.  The reason for the large disparity in beam/strahl fits compared to the other two components will be discussed in Section \ref{subsec:FitQualityAnalysis} and Appendix \ref{app:NumericalAnalysis}.

\phantomsection   
\subsection{Velocity Distribution Functions} \label{subsec:VelocityDistributionFunctions}

\indent  This section introduces and defines the model functions used to fit to the particle velocity distribution functions (VDFs) in this study with examples provided to illustrate shape and dependences on parameters.

\begin{figure*}
  \centering
    {\includegraphics[trim = 0mm 0mm 0mm 0mm, clip, width=170mm]{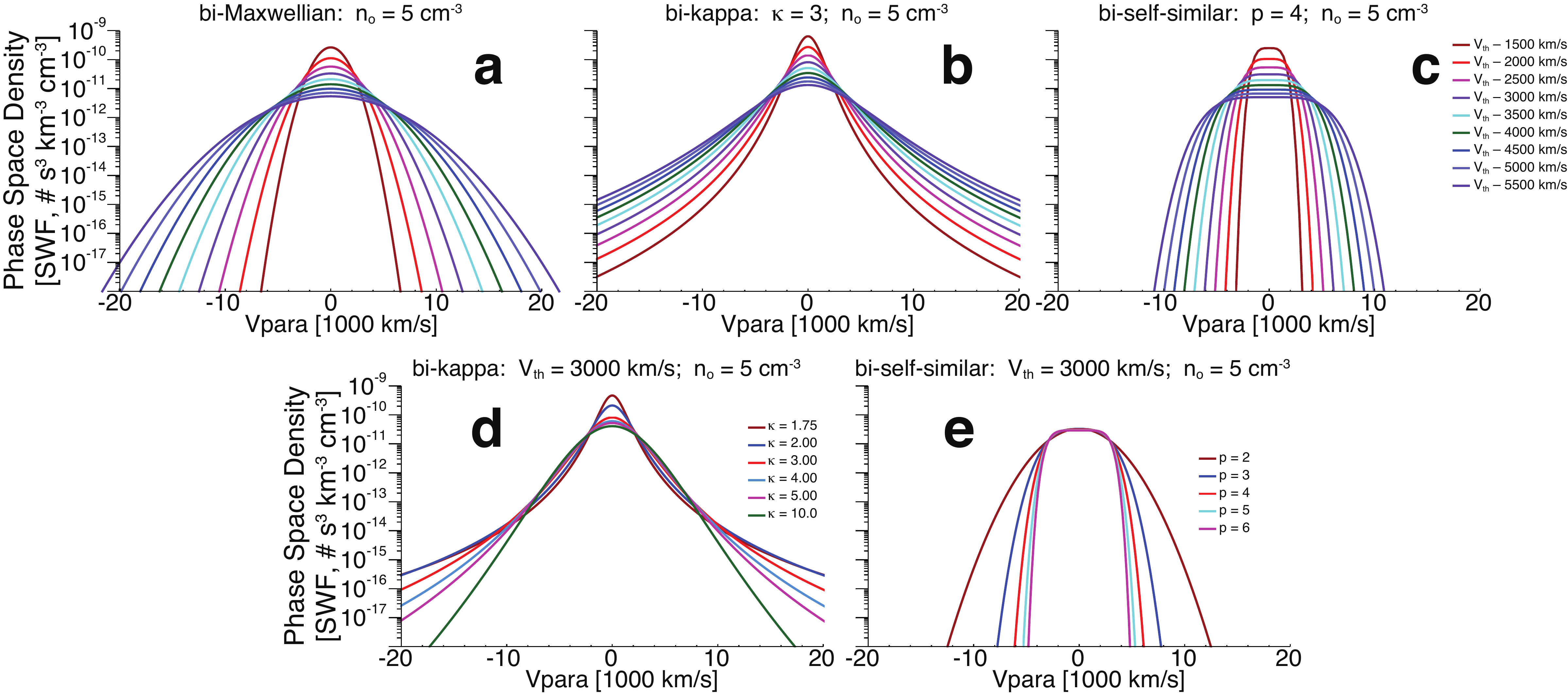}}
    \caption{Examples of one-dimensional cuts through multiple model VDF functions to illustrate the functional dependence on various parameters.  The top row (panels a through c) shows the dependence on the thermal speed, denoted generically as $V{\scriptstyle_{th}}$ here.  The bottom row (panels d and e) show the exponent dependencies.  Panel a shows bi-Maxwellian VDFs (Equation \ref{eq:gauss_7}), panels b and d show bi-kappa VDFs (Equation \ref{eq:bikappa_0a}), and panels c and e show bi-self-similar VDFs (Equations \ref{eq:app4_8} and \ref{eq:genbissvdf_3b}).  All examples shown have the same number density of 5 $cm^{-3}$, denoted generically as $n{\scriptstyle_{o}}$ here.}
    \label{fig:Example1DCutsVDF}
\end{figure*}

\indent  The most common velocity distribution function (VDF) used to model particle VDFs in space plasmas is the bi-Maxwellian \citep[e.g.,][]{feldman79a, feldman79b, feldman83b, kasper06a}, given by:

\begin{subequations}
  \begin{align}
    f\left( V{\scriptstyle_{\parallel}}, V{\scriptstyle_{\perp}} \right) & = A{\scriptstyle_{M}} \ e^{^{\displaystyle - \left[ \left( \frac{ V{\scriptstyle_{\parallel}} - v{\scriptstyle_{o \parallel}} }{ V{\scriptstyle_{T \parallel}} } \right)^{2} + \left( \frac{ V{\scriptstyle_{\perp}} - v{\scriptstyle_{o \perp}} }{ V{\scriptstyle_{T \perp}} } \right)^{2} \right]}} \label{eq:gauss_7} \\
    \intertext{where $A{\scriptstyle_{M}}$ is given by}
    A{\scriptstyle_{M}} & = \frac{ n{\scriptstyle_{o}} }{ \pi^{3/2} V{\scriptstyle_{T \perp}}^{2} V{\scriptstyle_{T \parallel}} } \label{eq:gauss_7b}
  \end{align}
\end{subequations}

\noindent  where $v{\scriptstyle_{o, j}}$ is the drift speed of the peak relative to zero along the j$^{th}$ component, $V{\scriptstyle_{T, j}}^{2}$ is the thermal speed given by Equation \ref{eq:params_2}, $V{\scriptstyle_{j}}$ is the velocity ordinate of the j$^{th}$ component, and $n{\scriptstyle_{o}}$ is the number density.

\indent  The second most popular model VDF is the kappa distribution.  The kappa velocity distribution has gained popularity in recent years owing to improvements in particle detectors and the ubiquitous non-Maxwellian tails observed for both ions and electrons \citep[e.g.,][]{lazar15a, lazar15b, lazar16a, lazar17a, lazar18a, livadiotis15a, livadiotis18a, mace10a, pulupa14b, saeed17a, shaaban18a}, but references to and use of kappa or kappa-like (e.g., modified Lorentzian) distributions have been around for decades \citep[e.g.,][]{feldman83a, maksimovic97a, salem03a, vasyliunas68a}.  It is beyond the scope of this study to explain the physical interpretation/origin of this function but there are several detailed discussions already published on the topic \citep[e.g.,][]{livadiotis15a, livadiotis18a}.  A generalized power-law particle distribution is given by a bi-kappa VDF \citep[e.g.,][]{livadiotis15a, mace10a}, for electrons here as:

\begin{subequations}
  \begin{align}
    f\left( V{\scriptstyle_{\perp}}, V{\scriptstyle_{\parallel}} \right) & = A{\scriptstyle_{\kappa}} \left\{ 1 + \frac{ B{\scriptstyle_{\kappa}} }{ \left( \kappa - \tfrac{3}{2} \right) } \right\}^{-( \kappa + 1 )} \label{eq:bikappa_0a} \\
    \intertext{where $A{\scriptstyle_{\kappa}}$ is given by}
    A{\scriptstyle_{\kappa}} & = \left[ \frac{ 1 }{ \pi \left( \kappa - \tfrac{3}{2} \right) } \right]^{3/2} \frac{ n{\scriptstyle_{o}} \ \Gamma\left( \kappa + 1 \right) }{ V{\scriptstyle_{T \perp}}^{2} \ V{\scriptstyle_{T \parallel}} \ \Gamma\left( \kappa - \tfrac{1}{2} \right) } \label{eq:bikappa_0b} \\
    \intertext{and $B{\scriptstyle_{\kappa}}$ is given by}
    B{\scriptstyle_{\kappa}} & = \left[ \left( \frac{ V{\scriptstyle_{\parallel}} - v{\scriptstyle_{o \parallel}} }{ V{\scriptstyle_{T \parallel}} } \right)^{2} + \left( \frac{ V{\scriptstyle_{\perp}} - v{\scriptstyle_{o \perp}} }{ V{\scriptstyle_{T \perp}} } \right)^{2} \right] \label{eq:bikappa_0c}
  \end{align}
\end{subequations}

\noindent  where $\Gamma\left( z \right)$ is the Riemann gamma function of argument $z$ and $V{\scriptstyle_{T j}}$ is again the most probable speed of a 1D Gaussian for consistency, i.e., it does not depend upon $\kappa$.

\indent  The last model VDF is called a \emph{self-similar distribution} which results when a VDF evolves under the action of inelastic scattering \citep[e.g.,][]{dum74a, dum75a, goldman84a, horton76a, horton79a, jain79a} or flows through disordered porous media \citep[e.g.,][]{matyka16a}.  The symmetric form is given by:

\begin{subequations}
  \begin{align}
    f\left( V{\scriptstyle_{\parallel}}, V{\scriptstyle_{\perp}} \right) & = A{\scriptstyle_{SS}} \ e^{^{\displaystyle - \left[ \left( \frac{ V{\scriptstyle_{\parallel}} - v{\scriptstyle_{o \parallel}} }{ V{\scriptstyle_{T \parallel}} } \right)^{s} + \left( \frac{ V{\scriptstyle_{\perp}} - v{\scriptstyle_{o \perp}} }{ V{\scriptstyle_{T \perp}} } \right)^{s} \right]}} \label{eq:app4_8} \\
    \intertext{where $A{\scriptstyle_{SS}}$ is given by}
    A{\scriptstyle_{SS}} & = \left[ 2 \Gamma\left( \frac{1 + s}{s} \right) \right]^{-3} \frac{ n{\scriptstyle_{o}} }{ V{\scriptstyle_{T \perp}}^{2} V{\scriptstyle_{T \parallel}} } \label{eq:app4_8b}
  \end{align}
\end{subequations}

\noindent  Note that $V{\scriptstyle_{T j}}$ is again the most probable speed of a 1D Gaussian for consistency, i.e., it does not depend upon $s$.  Further, one can see that Equation \ref{eq:app4_8} reduces to Equation \ref{eq:gauss_7} in the limit where $s \rightarrow 2$.  The function in Equation \ref{eq:app4_8} will be referred to as the symmetric self-similar distribution function.

\indent  A slightly more general approach can be taken where the exponents are not uniform, which will be referred to as the asymmetric self-similar distribution function.  The asymmetric functional form is given by:

\begin{subequations}
  \begin{align}
    f\left( V{\scriptstyle_{\parallel}}, V{\scriptstyle_{\perp}} \right) & = A{\scriptstyle_{AS}} \ e^{^{\displaystyle - \left[ \left( \frac{ V{\scriptstyle_{\parallel}} - v{\scriptstyle_{o \parallel}} }{ V{\scriptstyle_{T \parallel}} } \right)^{p} + \left( \frac{ V{\scriptstyle_{\perp}} - v{\scriptstyle_{o \perp}} }{ V{\scriptstyle_{T \perp}} } \right)^{q} \right]}} \label{eq:genbissvdf_3b} \\
    \intertext{where $A{\scriptstyle_{AS}}$ is given by}
    A{\scriptstyle_{AS}} & = \frac{ n{\scriptstyle_{o}}  \ \Gamma^{-1}\left( \frac{1+p}{p} \right) \ \Gamma^{-2}\left( \frac{1+q}{q} \right) }{ 2^{3} \ V{\scriptstyle_{T_{\parallel}}} \ V{\scriptstyle_{T_{\perp}}}^{2} } \label{eq:genbissvdf_3c}
  \end{align}
\end{subequations}

\noindent  Again, this will reduce to a bi-Maxwellian in the limit where $p \rightarrow 2$ and $q \rightarrow 2$.  Note that in the event that the the exponents $s$, $p$, or $q$ are not even integers, the velocity ordinates, $\left( V{\scriptstyle_{\parallel}} - v{\scriptstyle_{o \parallel}} \right)$ and $\left( V{\scriptstyle_{\perp}} - v{\scriptstyle_{o \perp}} \right)$, will become absolute values to avoid complex roots and negative values of $f\left( V{\scriptstyle_{\parallel}}, V{\scriptstyle_{\perp}} \right)$.  Example one-dimensional cuts of these three model VDFs can be found in Figure \ref{fig:Example1DCutsVDF} for comparison.

\indent  The self-similar exponents are mostly a new variable, since most previous work modeled the core electrons as a bi-Maxwellian \citep[e.g.,][]{bale13a, pulupa14b, stverak08a, stverak09a}.  There are a few studies that used one-dimensional self-similar functions to model a select few electron VDFs near collisionless shocks \citep[e.g.,][]{feldman83a, feldman83b} finding values consistent with those presented in Table \ref{tab:Exponents}.  However, these studies did not define the normalization parameter in terms of the number density and thermal speeds (e.g., see Equations \ref{eq:app4_8} and \ref{eq:genbissvdf_3b}) but rather found a numerical value from empirical fits, i.e., the normalization parameter was not coupled to the physical parameters of the fit function.  At least one study in the solar wind did define the normalization constant, but they only considered a one-dimensional, isotropic distribution \citep[e.g.,][]{marsch85a}.  Although several theoretical works predicted ranges of possible self-similar exponent values under various extrema scenarios \citep[e.g.,][]{dum74a, dum75a, goldman84a, horton76a, horton79a, jain79a}, this is the first time the model has been used on a statistically significant set of VDFs.

\begin{figure*}
  \centering
    {\includegraphics[trim = 0mm 0mm 0mm 0mm, clip, height=150mm]{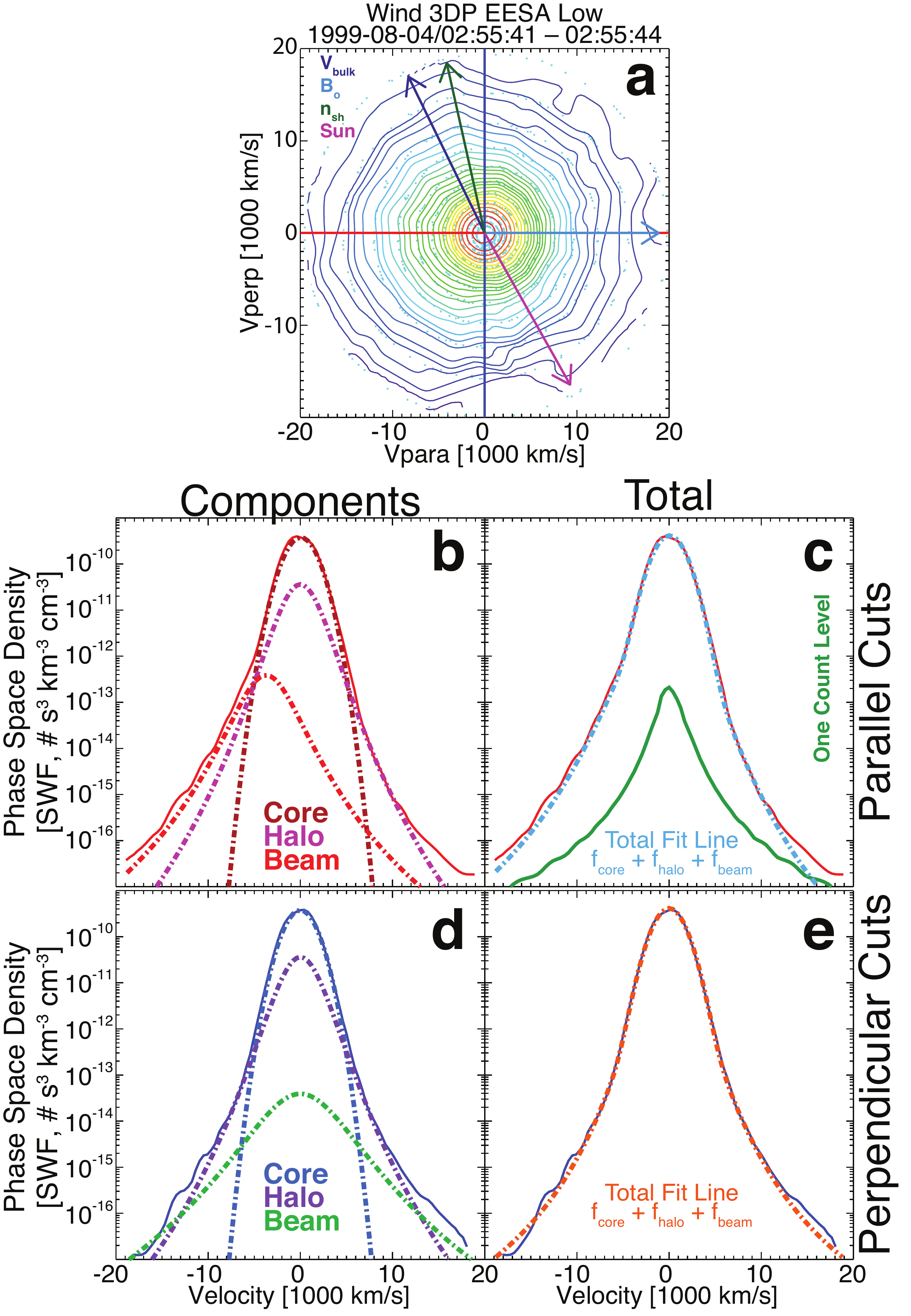}}
    \caption{An example VDF observed at 02:55:41.008 UTC on 1999-08-04 by the \emph{Wind}/3DP EESA Low detector.  Panel a shows a 2D cut through the 3D VDF as contours of constant phase space density, where the cut plane is defined by the unit vectors $\left( \mathbf{B}{\scriptstyle_{o}} \times \mathbf{V}{\scriptstyle_{i}} \right) \times \mathbf{B}{\scriptstyle_{o}}$ on the vertical and $\mathbf{B}{\scriptstyle_{o}}$ on the horizontal, where $\mathbf{B}{\scriptstyle_{o}}$ $=$ $\left( +6.41, \ -7.64, \ -8.48 \right)$ [nT, GSE].  The origin in velocity space is defined by $\mathbf{V}{\scriptstyle_{i}}$ $=$ $\left( -388.38, \ +3.13, \ -32.63 \right)$ [$km \ s^{-1}$, GSE].  The value of $\phi{\scriptstyle_{sc}}$ for this VDF is 6.35 eV.  Projected onto panel a are the following vectors: ion bulk flow velocity $\mathbf{V}{\scriptstyle_{i}}$ or $\mathbf{V}{\scriptstyle_{bulk}}$ (purple arrow), $\mathbf{B}{\scriptstyle_{o}}$ (cyan arrow), shock normal vector $\mathbf{n}{\scriptstyle_{sh}}$ (green arrow), and the sun direction (magenta arrow).  The small cyan dots show the location of actual measurements prior to regularized gridding with Delaunay triangulation.  Panels b and c show the 1D parallel cuts along the horizontal (solid red line is data in both panels) and panels d and e show the 1D perpendicular cuts along the vertical (solid blue line is data in both panels).  Panels b and d show the individual electron component fit results while panels c and e show the sum of the fit results all as dashed lines and with color-coded labels.  Panel c shows the one-count level for reference.}
    \label{fig:ExampleGoodVDF}
\end{figure*}

\indent  The following is an illustrative example that shows how the signal-to-noise ratio of particle detectors strongly depends upon the number density and thermal speed and that hot, tenuous plasmas are much more difficult to measure and accurately model.  Examine the one-dimensional cuts shown in Figures \ref{fig:Example1DCutsVDF} and \ref{fig:ExampleGoodVDF}.  The toy models in Figure \ref{fig:Example1DCutsVDF} are shown to illustrate the effect of thermal speed and exponents on the model fit function peaks and shapes.  Notice that increasing the thermal speed of the Maxwellian from $V{\scriptstyle_{Te}}$ $=$ 1500 km/s to 5500 km/s drops the peak phase space density by nearly two orders of magnitude.  The cut line also passes the $\pm$20,000 km/s velocity boundary (i.e., roughly the upper energy bound of the EESA Low instrument) at a phase space density roughly one order of magnitude higher than the colder examples.  That is, the change in thermal speed reduced the dynamic range of observed phase space densities by three orders of magnitude.  Suppose one examines a more extreme example with $n{\scriptstyle_{e}}$ $=$ 15 $cm^{-3}$ and $V{\scriptstyle_{Te}}$ $=$ 10,000 km/s.  In this case, the difference between the peak and the lowest phase space density within the $\pm$20,000 km/s velocity boundary would only be a factor of $\sim$55, i.e., slightly more than one order of magnitude.

\indent  For reference, the list of potential free parameters are as follows (see Appendix \ref{app:Definitions} for symbol definitions):
\begin{itemize}[itemsep=0pt,parsep=0pt,topsep=0pt]
  \item[]  \textbf{Core}
  \begin{itemize}[itemsep=0pt,parsep=0pt,topsep=0pt]
    \item  $n{\scriptstyle_{ec}}$
    \item  $V{\scriptstyle_{Tec, j}}$ or $T{\scriptstyle_{ec, j}}$
    \item  $v{\scriptstyle_{oec, j}}$
    \item  $s{\scriptstyle_{ec}}$
    \item  $p{\scriptstyle_{ec}}$
    \item  $q{\scriptstyle_{ec}}$
    \item  $\kappa{\scriptstyle_{ec}}$
  \end{itemize}
  \item[]  \textbf{Halo}
  \begin{itemize}[itemsep=0pt,parsep=0pt,topsep=0pt]
    \item  $n{\scriptstyle_{eh}}$
    \item  $V{\scriptstyle_{Teh, j}}$ or $T{\scriptstyle_{eh, j}}$
    \item  $v{\scriptstyle_{oeh, j}}$
    \item  $\kappa{\scriptstyle_{eh}}$
  \end{itemize}
  \item[]  \textbf{Beam/Strahl}
  \begin{itemize}[itemsep=0pt,parsep=0pt,topsep=0pt]
    \item  $n{\scriptstyle_{eb}}$
    \item  $V{\scriptstyle_{Teb, j}}$ or $T{\scriptstyle_{eb, j}}$
    \item  $v{\scriptstyle_{oeb, j}}$
    \item  $\kappa{\scriptstyle_{eb}}$
  \end{itemize}
\end{itemize}

\indent  For more details about derivation and normalization constants, see the Supplemental Material.

\phantomsection   
\subsection{Fit Parameter Constraints}  \label{subsec:FitPhysicalConstraints}

\indent  This section involves the discussion of the constraints/limits placed on fit parameters for each electron component and justifies them based on physically significant assumptions.

\begin{figure}
  \centering
    {\includegraphics[trim = 0mm 0mm 0mm 0mm, clip, width=80mm]{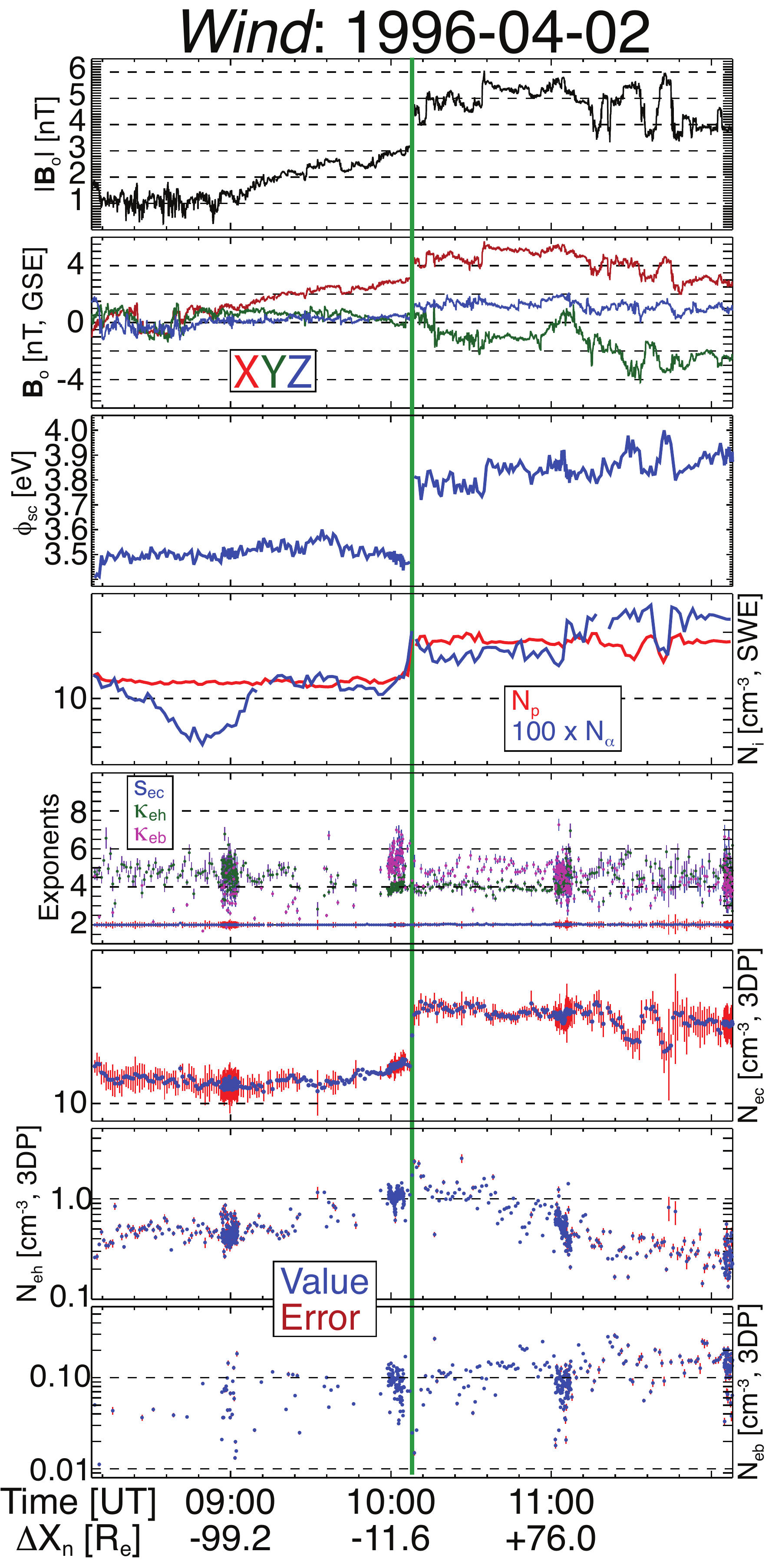}}
    \caption{An example IP shock crossing observed on 1996-04-02 by the \emph{Wind} spacecraft.  The panels are as follows from top-to-bottom: $\lvert \mathbf{B}{\scriptstyle_{o}} \rvert$ [nT], $\mathbf{B}{\scriptstyle_{o}}$ [nT, GSE]; value of spacecraft potential used for fits $\phi{\scriptstyle_{sc}}$ [eV]; $n{\scriptstyle_{p}}$ (red line) and $100 \times n{\scriptstyle_{\alpha}}$ (blue line) [$cm^{-3}$, SWE]; $s{\scriptstyle_{ec}}$ (blue dots), $\kappa{\scriptstyle_{eh}}$ (green dots), and $\kappa{\scriptstyle_{eb}}$ (magenta dots); $n{\scriptstyle_{ec}}$ values (blue dots) and uncertainty (red error bars) [$cm^{-3}$, 3DP Fit]; $n{\scriptstyle_{eh}}$ [$cm^{-3}$, 3DP Fit]; and $n{\scriptstyle_{eb}}$ [$cm^{-3}$, 3DP Fit].  The error bars for the four electron fit parameter panels are defined by the percent deviation discussed in Section \ref{subsec:FitQualityAnalysis}.  The error for this date satisfied 0.2\% $<$ $\delta \mathcal{R}$ $\leq$ 54\% with a median of 10.3\%.}
    \label{fig:ExampleIPShock}
\end{figure}

\indent  As an illustrative example, Figure \ref{fig:ExampleIPShock} shows the densities of the protons, alpha-particles, and three electron components (blue squares) and the associated uncertainties (red error bars) for a subcritical, quasi-perpendicular IP shock (see Supplemental Materials for shock parameters) observed by \emph{Wind} on 1996-04-02 at 10:07:57.525 UTC.  For this event, the plasma parameters are listed below in the following form \textit{Min}--\textit{Max}(\textit{Mean})[\textit{Median}]
\begin{itemize}[itemsep=0pt,parsep=0pt,topsep=0pt]
  \item[]  \textbf{Upstream}
  \begin{itemize}[itemsep=0pt,parsep=0pt,topsep=0pt]
    \item  $\lvert \mathbf{B}{\scriptstyle_{o}} \rvert$ $\sim$ 0.53--3.14(1.96)[1.53] nT;
    \item  $n{\scriptstyle_{p}}$ $\sim$ 11.3--15.8(12.0)[11.9] $cm^{-3}$;
    \item  $n{\scriptstyle_{\alpha}}$ $\sim$ 0.06--0.18(0.10)[0.11] $cm^{-3}$;
    \item  $s{\scriptstyle_{ec}}$ $\sim$ 2.00--2.09(2.00)[2.00] N/A;
    \item  $\kappa{\scriptstyle_{eh}}$ $\sim$ 2.83--12.2(4.46)[4.40] N/A;
    \item  $\kappa{\scriptstyle_{eb}}$ $\sim$ 1.67--12.6(4.85)[5.10] N/A;
    \item  $n{\scriptstyle_{ec}}$ $\sim$ 10.7--13.0(11.7)[11.5] $cm^{-3}$;
    \item  $n{\scriptstyle_{eh}}$ $\sim$ 0.06--1.44(0.69)[0.54] $cm^{-3}$;
    \item  $n{\scriptstyle_{eb}}$ $\sim$ 0.02--0.17(0.09)[0.09] $cm^{-3}$;
  \end{itemize}
  \item[]  \textbf{Downstream}
  \begin{itemize}[itemsep=0pt,parsep=0pt,topsep=0pt]
    \item  $\lvert \mathbf{B}{\scriptstyle_{o}} \rvert$ $\sim$ 3.45--5.99(4.85)[5.19] nT;
    \item  $n{\scriptstyle_{p}}$ $\sim$ 14.9--19.7(18.0)[18.1] $cm^{-3}$;
    \item  $n{\scriptstyle_{\alpha}}$ $\sim$ 0.14--0.27(0.19)[0.19] $cm^{-3}$;
    \item  $s{\scriptstyle_{ec}}$ $\sim$ 2.00--2.07(2.01)[2.01] N/A;
    \item  $\kappa{\scriptstyle_{eh}}$ $\sim$ 2.72--6.96(4.39)[4.29] N/A;
    \item  $\kappa{\scriptstyle_{eb}}$ $\sim$ 2.74--7.27(4.45)[4.50] N/A;
    \item  $n{\scriptstyle_{ec}}$ $\sim$ 13.6--18.4(16.7)[16.8] $cm^{-3}$;
    \item  $n{\scriptstyle_{eh}}$ $\sim$ 0.02--2.53(0.56)[0.44] $cm^{-3}$;
    \item  $n{\scriptstyle_{eb}}$ $\sim$ 0.01--0.29(0.12)[0.11] $cm^{-3}$;
  \end{itemize}
\end{itemize}

\noindent  Note that there are two time periods after 11:00 UTC where a few fit results satisfy $n{\scriptstyle_{eb}} / n{\scriptstyle_{eh}}$ $\geq$ 1.  Figure \ref{fig:ExampleIPShock} is illustrative of some of the error analysis employed in the present study and that the beam/strahl fit more often fails than the core or halo as evidenced by the number of points.  Below the details of how the fit parameters are constrained/limited are outlined with physical arguments.

\indent  First, the present study differs from some previous studies in that the fits are performed on the two-dimensional VDF rather than separate fits on one-dimensional cuts of the two-dimensional VDF \citep[e.g.,][]{maksimovic05a, pulupa14a, pulupa14b}.  One of the limitations of the latter approach is that the distribution function is not necessarily a separable function, which can introduce difficulty for the physical interpretation of the results.  However, the latter approach has numerous advantages including the stability of the solutions and ease with which the solutions are found with nonlinear least squares software, i.e., it is generally easier to fit to a one-dimensional cut than a two-dimensional distribution.

\indent  The present study uses the former approach to avoid the difficulties introduced for non-separable functions.  For instance, when fitting to the parallel one-dimensional cut the amplitude of the VDF is directly tied to the amplitude of the perpendicular cut.  The amplitude of all standard model two-dimensional, gyrotropic VDFs is dependent upon $n{\scriptstyle_{s}}$, $V{\scriptstyle_{Ts, \parallel}}^{-1}$, and $V{\scriptstyle_{Ts, \perp}}^{-2}$.  While it is computationally possible to fix the amplitude to the observed amplitude of the data for each cut and only vary the respective thermal speeds/temperatures and exponents, the inversion to find $n{\scriptstyle_{s}}$ can be problematic if care is not taken.  For instance, the normalization constants differ for one-dimensional cuts from the two-dimensional gyrotropic VDF (e.g., see Equation \ref{eq:gauss_7}).  Although this approach involves fewer free parameters and should thus be easier to fit, it is much more restrictive in parameter space, i.e., $n{\scriptstyle_{s}}$ only varies indirectly through the variation of the thermal speeds/temperatures and exponents.

\indent  Given that fitting to a two-dimensional gyrotropic VDF has more free parameters and orders of magnitude more degrees of freedom, a stable solution requires reasonable constraints/limits on the variable parameters.  There are some obvious boundaries determined by instrumental and physical constraints.  As shown in the previous section, the difference between the highest and lowest phase space densities is important for the signal-to-noise ratio but it is also relevant to fitting model functions to the data.  For instance, if an electron distribution had a population with $V{\scriptstyle_{Te}}$ $\geq$ 10,000 km/s the weights would not provide sufficient contrast between the peak and tails to constrain a stable and reliable fit without multiple imposed constraints.  In contrast, electron VDFs with thermal speeds below $\sim$1000 km/s fall below the lowest energy of the detector and so would be artificially hotter if they were observed \citep[e.g.,][]{paschmann98a}.  A similar effect is often observed by spacecraft with electrostatic analyzers designed for the magnetosphere, not the comparatively cold, fast solar wind beam \citep[e.g.,][]{mcfadden08a, mcfadden08b, pollock16a}.

\indent  Statistical studies of the solar wind have shown that the maximum range of the total electron temperature is $T{\scriptstyle_{e, j}}$ $\sim$ 2.29--77.2 eV or $V{\scriptstyle_{Te, j}}$ $\sim$ 450--2600 km/s \citep[e.g.,][]{wilsoniii18b}.  Previous studies have found that the electron halo temperatures satisfy $T{\scriptstyle_{e, j}}$ $\sim$ 14--560 eV or $V{\scriptstyle_{Teh, j}}$ $\sim$ 1100--7000 km/s \citep[e.g.,][]{feldman75a, feldman78b, feldman79b, lazar17a, maksimovic97a, maksimovic05a, skoug00a, tao16a, tao16b}.  Previous studies have also found that the electron beam/strahl temperatures satisfy $T{\scriptstyle_{eb, j}}$ $\sim$ 20--150 eV or $V{\scriptstyle_{Te, j}}$ $\sim$ 1300--3600 km/s \citep[e.g.,][]{ogilvie00a, tao16a, tao16b, vinas10a}.  Thus, a range of allowed core thermal speeds from $\sim$1000 km/s to $\sim$10,000 km/s can be assumed.

\indent  There are similar instrumental constraints on the drift speed of the three components.  The core, however, is not likely to exhibit drift speeds (in the ion rest frame) in excess of several hundred km/s \citep[e.g.,][]{pulupa14a}.  In the present work, most fit results show less than 50 km/s, i.e., only 1838 of 14847 or $\sim$12\% have drift speeds exceeding 50 km/s, consistent with previous work\footnote{Note that in the present work the dipole correction to $\phi{\scriptstyle_{sc}}$ was not applied, which affects the drift velocity and heat flux velocity moments.  Thus, the core drift velocities in our work suffer the greatest from this correction.}.  In contrast, owing the physical interpretation of the strahl/beam component most (8848 of 9567 or $\sim$92\%) have drift speeds in excess of 1000 km/s.  The range of allowed core, halo, and beam/strahl drift speeds loosely ranged from $\sim$1000 km/s to $\sim$10,000 km/s for most events.  In some events, a lower bound was imposed to prevent unphysical fit results, e.g., beam/strahl component with near zero drift speed (see Supplemental Material ASCII files described in Appendix \ref{app:DataProduct} for ranges for specific events).  Note that $V{\scriptstyle_{oes, \perp}}$ was fixed during the fitting, i.e., it was not allowed to vary.  Originally this parameter was free to vary but resulted in fewer stable fits and rarely varied more than few km/s.  In some events, an explicit $V{\scriptstyle_{oec, \perp}}$ was set as the initial guess values determined from examination of the distributions, but this is for a small minority of events (333 of 14847 or $\sim$2\%).

\indent  It has also been empirically found that the EESA Low detector has issues when $n{\scriptstyle_{ce}}$ $\lesssim$ 0.5 $cm^{-3}$ or $n{\scriptstyle_{ce}}$ $\gtrsim$ 50 $cm^{-3}$ for typical solar wind thermal speeds\footnote{Technically, this is an issue for nearly all electrostatic analyzers designed and flown to date.  This is largely unavoidable without increasing the dynamic range of the detector significantly.}.  This is rarely an issue as only 41 of the 14847 or $\sim$0.3\% VDFs analyzed have fit results falling outside the range $\sim$0.5--50 $cm^{-3}$.  Note that the total electron density, $n{\scriptstyle_{e}}$ $=$ $n{\scriptstyle_{ec}}$ $+$ $n{\scriptstyle_{eh}}$ $+$ $n{\scriptstyle_{eb}}$ $\sim$ $n{\scriptstyle_{e}}$ $=$ $n{\scriptstyle_{p}}$ $+$ 2$n{\scriptstyle_{\alpha}}$, is constrained by the total ion density from SWE and the total electron density from the upper hybrid line observed by the WAVES radio receiver \citep[][]{bougeret95a}, when possible (see Appendix \ref{app:DetectorCalibration} for more details).

\indent  Physically, the halo and beam/strahl components are suprathermal, thus they should not have the dominant contribution to the total phase space density of the VDF.  Therefore, it is physically consistent to assume that the fit results should satisfy $n{\scriptstyle_{eh}} / n{\scriptstyle_{ec}}$ $<$ 1 and $n{\scriptstyle_{eb}} / n{\scriptstyle_{ec}}$ $<$ 1.  The solutions were constrained to satisfy $n{\scriptstyle_{eh}} / n{\scriptstyle_{ec}}$ $<$ 0.5 and $n{\scriptstyle_{eb}} / n{\scriptstyle_{ec}}$ $<$ 1 based upon results found in previous studies  near 1 AU \citep[e.g.,][]{feldman75a, maksimovic97a, maksimovic05a, pierrard16a, skoug00a, stverak09a, tao16a, vinas10a}.

\begin{figure*}
  \centering
    {\includegraphics[trim = 0mm 0mm 0mm 0mm, clip, height=150mm]{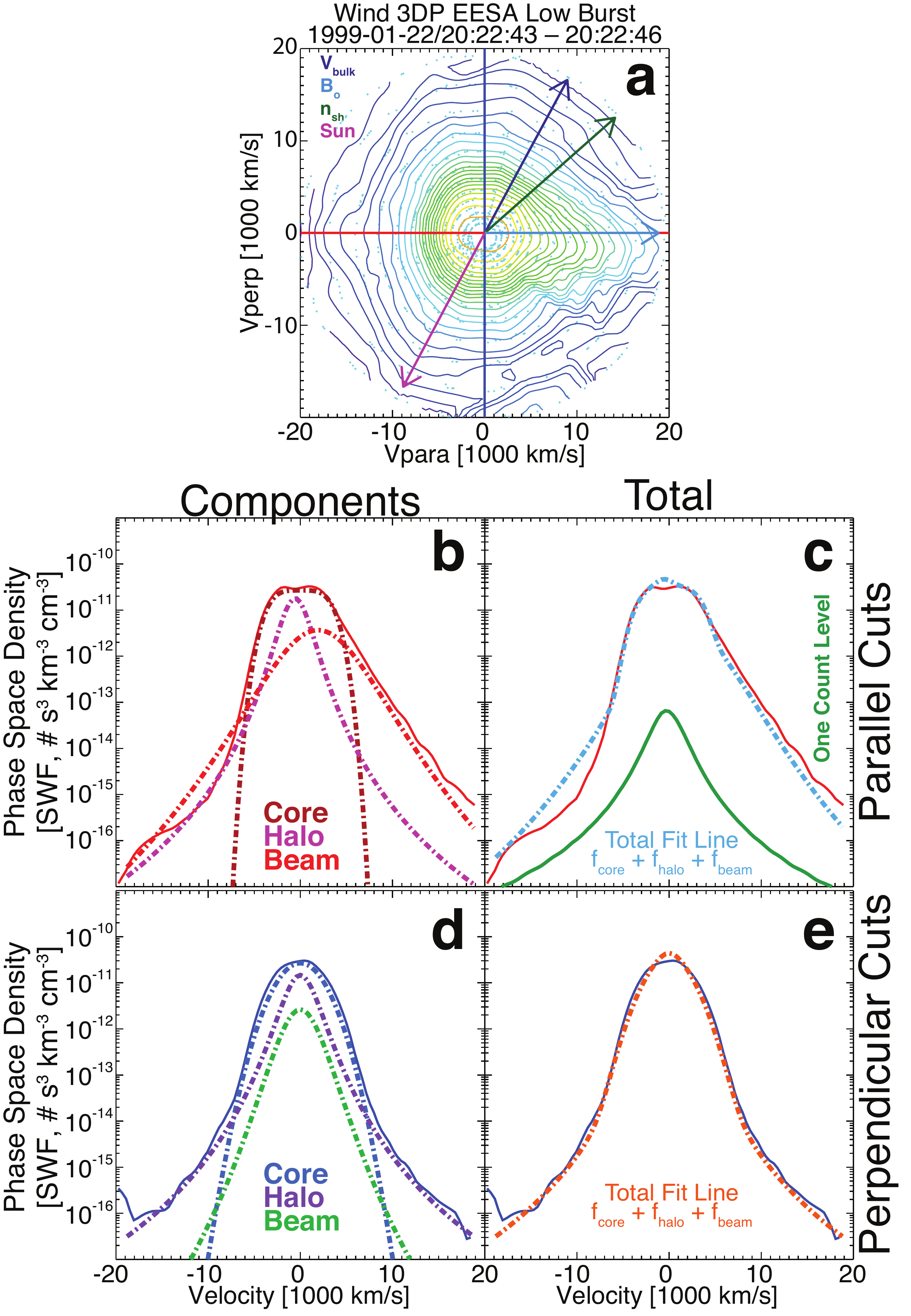}}
    \caption{Another example VDF observed at 20:22:43.490 UTC on 1999-01-22 by the \emph{Wind}/3DP EESA Low detector in burst mode.  The format is the same as Figure \ref{fig:ExampleGoodVDF} where this VDF has $\mathbf{B}{\scriptstyle_{o}}$ $=$ $\left( -6.95, \ +9.78, \ -8.77 \right)$ [nT, GSE], $\mathbf{V}{\scriptstyle_{i}}$ $=$ $\left( -619.12, \ +26.66, \ +21.19 \right)$ [$km \ s^{-1}$, GSE], and $\phi{\scriptstyle_{sc}}$ $=$ 9.45 eV.}
    \label{fig:ExampleBadVDF}
\end{figure*}

\indent  In numerous previous studies that assumed a three component solar wind electron VDF near 1 AU \citep[e.g.,][]{maksimovic05a, pulupa14a, pulupa14b, stverak09a}, constraints were sometimes assumed such as that the fits satisfy $n{\scriptstyle_{eb}} / n{\scriptstyle_{eh}}$ $<$ 1.  There is no restriction on this ratio\footnote{The number of good ratios differs from the number of beam/strahl fits because some VDFs had a stable halo or beam/strahl but not the converse.} imposed during the fit process and 1824 of 9313 or $\sim$20\% of the fits satisfy $n{\scriptstyle_{eb}} / n{\scriptstyle_{eh}}$ $\geq$ 1.  In fact, it was found that imposing the constraint, $n{\scriptstyle_{eb}} / n{\scriptstyle_{eh}}$ $<$ 1, during the fit process actually greatly reduced the number of stable solutions found for the beam/strahl component\footnote{Note that there was a post-fit constraint imposed limiting $n{\scriptstyle_{eb}} / n{\scriptstyle_{eh}}$ $<$ 3 because it was found empirically that most fits exceeding this threshold were bad/unphysical.  However, not all were bad as evidenced by the example in Figure \ref{fig:ExampleIntenseBeamVDF}.}.  Previous work did show that the ratio $n{\scriptstyle_{eb}} / n{\scriptstyle_{eh}}$ decreases with increasing radial distance from the sun dropping below unity before 1 AU, on average, but the ranges overlapped allowing for $n{\scriptstyle_{eb}} / n{\scriptstyle_{eh}}$ $\geq$ 1 \citep[e.g.,][]{stverak09a}.

\indent  Another constraint that is often assumed/used is that the strahl/beam component be only anti-sunward along $\mathbf{B}{\scriptstyle_{o}}$ \citep[e.g.,][]{maksimovic05a, pulupa14a, pulupa14b, stverak09a}, though some magnetic field topologies have sunward directed beam/strahl components \citep[e.g.,][]{owens17a}.  This constraint is imposed in this study but it is important to note that some IP shocks examined have observable electron foreshocks.  A consequence is that the halo component of the fit results effectively absorbs both the halo and the shock-reflected electron component in the events where this is directed sunward along $\mathbf{B}{\scriptstyle_{o}}$ (this is very rare).  If the shock-reflected electron component is directed anti-sunward they will be included in the beam/strahl fit (this is much more common).  The net result for the former is a smaller $\left(T{\scriptstyle_{\perp}}/T{\scriptstyle_{\parallel}}\right){\scriptstyle_{eh}}$ and on the latter a larger $\left(T{\scriptstyle_{\perp}}/T{\scriptstyle_{\parallel}}\right){\scriptstyle_{eb}}$ and $n{\scriptstyle_{eb}}$.

\indent  The lower bound of possible $\kappa{\scriptstyle_{es}}$ values is defined for mathematical/physical reasons as being $\gtrsim$3/2 \citep[e.g.,][]{livadiotis15a, livadiotis18a}.  The upper bound is set to 100 solely because above that value the difference between a bi-Maxwellian and bi-kappa VDF is smaller than the accuracy of the measurements.  Although the upper bound is allowed to extend to 100 the typical upper bound observed near 1 AU is $<$ 20 \citep[e.g.,][]{lazar17a, maksimovic97a, pierrard16a, stverak09a, tao16a, tao16b}.  The range of possible values for $s{\scriptstyle_{ec}}$, $p{\scriptstyle_{ec}}$, or $q{\scriptstyle_{ec}}$ falls between 2 and 10 for physical reasons \citep[e.g.,][]{dum74a, dum75a, goldman84a, horton76a, horton79a, jain79a}.

\indent  Finally, by definition the halo and beam/strahl components represent the lowest energy suprathermal components of the electrons.  Therefore, it is natural to assume that $T{\scriptstyle_{eh}} / T{\scriptstyle_{ec}}$ $>$ 1.  There is no explicit restriction on this ratio imposed and only 384 of 13867 or $\sim$3\% of the fits satisfy $T{\scriptstyle_{eh}} / T{\scriptstyle_{ec}}$ $<$ 1 and these occur downstream of strong shocks where core heating dominates.  However, there are numerous events where limits/constraints were imposed on the component temperatures individually.  So the low percentage is not entirely unexpected.  In contrast, there were no corresponding attempts to limit $T{\scriptstyle_{eh}} / T{\scriptstyle_{eb}}$ in any way other than to fit to the data.

\phantomsection   
\subsection{Quality Analysis}  \label{subsec:FitQualityAnalysis}

\indent  The initial approach was to use the reduced chi-squared value $\tilde{\chi}{\scriptstyle_{s}}^{2}$ of component $s$ (see Appendix \ref{app:NumericalInstability} for definition) as a test of the quality of the fit.  However, it was quickly determined that some fit lines matched well with the data but had $\tilde{\chi}{\scriptstyle_{s}}^{2}$ $>$ 10 while others did not fit well at all despite having $\tilde{\chi}{\scriptstyle_{s}}^{2}$ $\lesssim$ 1.  The issue is partly related to the calibration of the detector and thus the quality of the $\mathcal{W}$ values (see Appendix \ref{app:DetectorCalibration} for more details).  The issue is also related to fitting a gyrotropic model function to data that is not, in general, gyrotropic.  A possible improvement would fold the entire VDF into a forced gyrotropy prior to fitting to improve counting statistics and the comparison between data and model functions, but that is beyond the scope of the current study.  Therefore, a new quantity was defined to provide an additional definition of the quality of any given fit by direct comparison.

\begin{figure*}
  \centering
    {\includegraphics[trim = 0mm 0mm 0mm 0mm, clip, height=150mm]{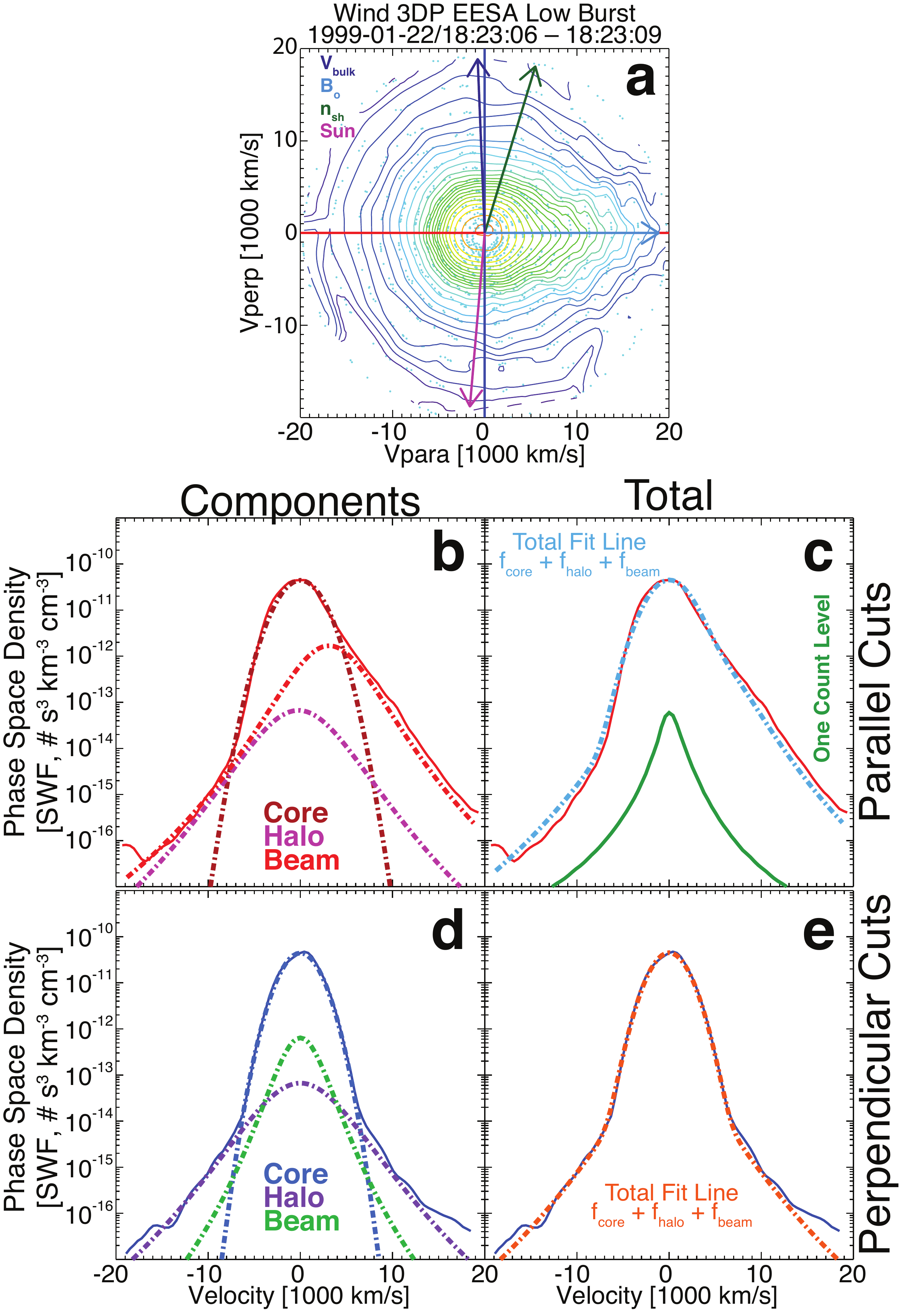}}
    \caption{Another example VDF observed at 18:23:06.116 UTC on 1999-01-22 by the \emph{Wind}/3DP EESA Low detector in burst mode.  The format is the same as Figures \ref{fig:ExampleGoodVDF} and \ref{fig:ExampleBadVDF} where this VDF has $\mathbf{B}{\scriptstyle_{o}}$ $=$ $\left( -0.89, \ -0.32, \ -10.57 \right)$ [nT, GSE], $\mathbf{V}{\scriptstyle_{i}}$ $=$ $\left( -626.59, \ +93.06, \ +76.13 \right)$ [$km \ s^{-1}$, GSE], and $\phi{\scriptstyle_{sc}}$ $=$ 10.67 eV.}
    \label{fig:ExampleIntenseBeamVDF}
\end{figure*}

\indent  Let us use $f^{\left( 0 \right)}$ as the actual data and $f^{\left( m \right)}$ ($=$ $f^{\left( core \right)}$ $+$ $f^{\left( halo \right)}$ $+$ $f^{\left( beam \right)}$) as the total model fit results.  Then one can define the ratio of these two parameters as $\mathcal{R}$ $=$ $f^{\left( 0 \right)}/f^{\left( m \right)}$, which is a two-dimensional array of values.  Then one calculates the median of this array, $\tilde{\mathcal{R}}$, to determine the percent deviation given by:

\begin{equation}
  \label{eq:perc_dev_0}
  \delta \mathcal{R} = \lvert 1 - \tilde{\mathcal{R}} \rvert \ \cdot \ 100\%
\end{equation}

\noindent  where $\delta \mathcal{R}$ is computed for each electron VDF.  The values of $\delta \mathcal{R}$ were then used as uncertainties/error bars for all fit parameters for the associated VDF for all components.  In general, the percent magnitude of the uncertainty in each of the six fit parameters should not be uniform as is used herein (see Appendix \ref{app:NumericalMethodComparisons} for discussion of one-sigma uncertainties).  The uncertainty of any variable calculated using these fit parameters was propagated assuming uncorrelated errors.

\indent  Note that the $\delta \mathcal{R}$ value alone does not always characterize the quality of any given fit.  Therefore, a combination of parameters is chosen to define a set of fit quality flags from best with a value of 10 to worst with a value of 0 (see Appendix \ref{app:DataProduct} for definitions).  In general, fits with flags at least 2 or higher can be used but low fit flags should be treated with caution.  Only $\lesssim$1\% of all core, halo, and beam/strahl fits had flags of 1 while $>$95\% of core, $>$89\% of halo, and $>$61\% of beam/strah flags were at least 2.

\indent  Figure \ref{fig:ExampleGoodVDF} shows an example VDF that had a low $\tilde{\chi}{\scriptstyle_{s}}^{2}$ for each component and a $\delta \mathcal{R}$ $\sim$ 3.0\%, i.e., this is an example of an ideal fit.  The distribution was fit using a symmetric bi-self-similar distribution for the core and a bi-kappa for both the halo and beam/strahl component.  The fit results are as follows:
\begin{itemize}[itemsep=0pt,parsep=0pt,topsep=0pt]
  \item  $n{\scriptstyle_{e\{c,h,b\}}}$ $=$ \{15.43, 2.01, 0.056\} $cm^{-3}$;
  \item  $V{\scriptstyle_{Te\{c,h,b\}, \parallel}}$ $=$ \{1959.6, 2500.0, 3964.7\} $km \ s^{-1}$;
  \item  $V{\scriptstyle_{Te\{c,h,b\}, \perp}}$ $=$ \{1937.9, 2575.5, 4516.2\} $km \ s^{-1}$;
  \item  $V{\scriptstyle_{oe\{c,h,b\}, \parallel}}$ $=$ \{$+$44.58, -0.00, -3898.7\} $km \ s^{-1}$;
  \item  $V{\scriptstyle_{oe\{c,h,b\}, \perp}}$ $=$ \{-0.00, -0.00, -0.00\} $km \ s^{-1}$;
  \item  $\{s{\scriptstyle_{ec}},\kappa{\scriptstyle_{eh}},\kappa{\scriptstyle_{eb}}\}$ $=$ \{2.00, 4.58, 2.57\}, where $s{\scriptstyle_{ec}}$ is the self-similar exponent and $\kappa{\scriptstyle_{es}}$ is the kappa value;
  \item  $\tilde{\chi}{\scriptstyle_{e\{c,h,b\}}}^{2}$ $=$ \{1.07, 1.36, 0.41 \};
  \item  $\tilde{\chi}{\scriptstyle_{tot}}^{2}$ $=$ 6.14; and
  \item  Fit Flag \{c,h,b\} $=$ \{10, 10, 10\}.
\end{itemize}

\indent  In contrast, Figure \ref{fig:ExampleBadVDF} shows an example VDF that had a high $\tilde{\chi}{\scriptstyle_{s}}^{2}$ for two components yet still a small $\delta \mathcal{R}$ $\sim$ 9.4\%, i.e., this is still an example of a good fit despite the bad $\tilde{\chi}{\scriptstyle_{s}}^{2}$ values for the core and beam/strahl fits.  The fit results are as follows:
\begin{itemize}[itemsep=0pt,parsep=0pt,topsep=0pt]
  \item  $n{\scriptstyle_{e\{c,h,b\}}}$ $=$ \{4.41, 0.57, 0.32\} $cm^{-3}$;
  \item  $V{\scriptstyle_{Te\{c,h,b\}, \parallel}}$ $=$ \{3882.6, 2624.5, 4574.5\} $km \ s^{-1}$;
  \item  $V{\scriptstyle_{Te\{c,h,b\}, \perp}}$ $=$ \{2728.2, 2986.3, 2387.6\} $km \ s^{-1}$;
  \item  $V{\scriptstyle_{oe\{c,h,b\}, \parallel}}$ $=$ \{-0.00, -594.9, $+$2000.0\} $km \ s^{-1}$;
  \item  $V{\scriptstyle_{oe\{c,h,b\}, \perp}}$ $=$ \{-0.00, -0.00, -0.00\} $km \ s^{-1}$;
  \item  $\{ p{\scriptstyle_{ec}}, q{\scriptstyle_{ec}}, \kappa{\scriptstyle_{eh}}, \kappa{\scriptstyle_{eb}}\}$ $=$ \{4.00, 2.00, 2.27, 4.61\}, where $p{\scriptstyle_{ec}}$($q{\scriptstyle_{ec}}$) is the parallel(perpendicular) self-similar exponent and $\kappa{\scriptstyle_{es}}$ is the kappa value;
  \item  $\tilde{\chi}{\scriptstyle_{e\{c,h,b\}}}^{2}$ $=$ \{28.5, 0.55, 14.4\};
  \item  $\tilde{\chi}{\scriptstyle_{tot}}^{2}$ $=$ 14.40; and
  \item  Fit Flag \{c,h,b\} $=$ \{4, 6, 5\}.
\end{itemize}

\noindent  Further, the example VDF in Figure \ref{fig:ExampleBadVDF} differs from that in Figure \ref{fig:ExampleGoodVDF} in that an asymmetric self-similar model is used for the former.  The total fit lines also illustrate a weakness of the method used.  Since the components are fit separately, the respective weights change with each fit to prevent the fitting software from giving too much emphasis to, for instance, the core of the distribution when fitting to the halo\footnote{That is, the weights for the halo and beam/strahl fits are modified to force the software to examine only one-side of the velocity distribution at a time.  The weights also remove elements from the core fit to avoid including the core in the fit.}.  Thus, the resultant $f^{\left( m \right)}$ can exceed $f^{\left( 0 \right)}$ in some places.  The software does a post-fit check for instances where either the combined or any component model fit exceeds the data by user-specified factors\footnote{For instance, below $\sim$1000 km/s in Figure \ref{fig:ExampleBadVDF} the magnitude of $f^{\left( m \right)}/f^{\left( 0 \right)}$ stays below $\sim$1.7 and exceed 2.0 on the anti-parallel side above $\sim$10,000 km/s.  The latter was not flagged by the software because it resulted from the beam/strahl fit and that is only fit to the parallel side for this VDF.}.  For most events, the threshold is set between $\sim$2--4 but this varies as some events have known issues.  For instance, the known density from the upper hybrid line is 10 $cm^{-3}$ but no variation of $\phi{\scriptstyle_{sc}}$ yields fit results with $n{\scriptstyle_{e}}$ $\sim$ 10 $cm^{-3}$ without the model exceeding the data at low energies.  The reason is related to known calibration issues (see Appendix \ref{app:DetectorCalibration}).

\indent  Finally, Figure \ref{fig:ExampleIntenseBeamVDF} shows an example VDF that had a high $\tilde{\chi}{\scriptstyle_{s}}^{2}$ for the core component and moderate for beam/strahl but a small $\delta \mathcal{R}$ $\sim$ 2.1\%.  This example VDF was chosen to illustrate a good fit even when $n{\scriptstyle_{eb}} / n{\scriptstyle_{eh}}$ $>$ 1.  As previously discussed, there are post-fit constraints applied to the data based upon statistical and physical constraints.  The constraint relevant to Figure \ref{fig:ExampleIntenseBeamVDF} is that requiring $n{\scriptstyle_{eb}} / n{\scriptstyle_{eh}}$ $<$ 3.  This is why the fit flag value for the beam/strahl is zero and why $\tilde{\chi}{\scriptstyle_{tot}}^{2}$ is larger than a few.  The fit results are as follows:
\begin{itemize}[itemsep=0pt,parsep=0pt,topsep=0pt]
  \item  $n{\scriptstyle_{e\{c,h,b\}}}$ $=$ \{3.37, 0.03, 0.14\} $cm^{-3}$;
  \item  $V{\scriptstyle_{Te\{c,h,b\}, \parallel}}$ $=$ \{2609.8, 5293.2, 4686.9\} $km \ s^{-1}$;
  \item  $V{\scriptstyle_{Te\{c,h,b\}, \perp}}$ $=$ \{2286.9, 5494.9, 2516.2\} $km \ s^{-1}$;
  \item  $V{\scriptstyle_{oe\{c,h,b\}, \parallel}}$ $=$ \{-0.00, -222.8, $+$3273.0\} $km \ s^{-1}$;
  \item  $V{\scriptstyle_{oe\{c,h,b\}, \perp}}$ $=$ \{-0.00, -0.00, -0.00\} $km \ s^{-1}$;
  \item  $\{s{\scriptstyle_{ec}},\kappa{\scriptstyle_{eh}},\kappa{\scriptstyle_{eb}}\}$ $=$ \{2.00, 3.83, 3.53\};
  \item  $\tilde{\chi}{\scriptstyle_{e\{c,h,b\}}}^{2}$ $=$ \{17.84, 0.17, 5.14 \};
  \item  $\tilde{\chi}{\scriptstyle_{tot}}^{2}$ $=$ 13.17; and
  \item  Fit Flag \{c,h,b\} $=$ \{6, 6, 0\}.
\end{itemize}

\noindent  One can see from the figure that the halo component is rather weak compared to the beam/strahl, which could be the result of an enhancement from the electron foreshock of this IP shock or the fast nature of the solar wind upstream of this IP shock.  Regardless, the purpose of this example is to illustrate that stable and good fit solutions can be found that satisfy $n{\scriptstyle_{eb}} / n{\scriptstyle_{eh}}$ $>$ 1 even at 1 AU.

\indent  After examining thousands of fit results, it was determined that the combination of $\delta \mathcal{R}$ with $\tilde{\chi}{\scriptstyle_{s}}^{2}$ and $\tilde{\chi}{\scriptstyle_{tot}}^{2}$ are consistently more reliable quantities used in combination for defining the quality of the fit than using $\tilde{\chi}{\scriptstyle_{s}}^{2}$ alone.  The value is also used as a proxy for the uncertainty of any given fit parameter, e.g., $\delta n{\scriptstyle_{es}}$ $=$ $\pm \ \delta \mathcal{R} \cdot n{\scriptstyle_{es}}/2$ shown as the red error bars in Figure \ref{fig:ExampleIPShock}.  Note that values of 100\% correspond to fill values or bad fit results.  In the following section the one-variable statistics of the $\tilde{\chi}{\scriptstyle_{s}}^{2}$ and $\delta \mathcal{R}$ values are listed for reference to typical/expected values when evaluating the quality of a fit.  In general, the best fits have small values for $\delta \mathcal{R}$ and all $\tilde{\chi}{\scriptstyle_{s}}^{2}$.

\indent  Further tests of consistency were also performed to validate the fit results.  First, the EESA Low detector is known to saturate when the count rate exceeds $\sim$10$^{7}$ counts/second \citep[][]{lin95a}.  Examination of all VDFs found that a total of 10 energy-angle bins (from a total of 20,184,120) or $\sim 5 \times 10^{-5}$\% exceeded the maximum count rate.  Therefore, it is not thought that saturation has a significant impact on the methodology and results of this study.  Second, as illustrated in Figure \ref{fig:ExampleIPShock}, the total electron density satisfies $n{\scriptstyle_{e}}$ $\sim$ $n{\scriptstyle_{p}}$ $+$ 2$n{\scriptstyle_{\alpha}}$ for nearly all intervals.  Statistically, the difference between the fit result for $n{\scriptstyle_{e}}$ $=$ $n{\scriptstyle_{ec}}$ $+$ $n{\scriptstyle_{eh}}$ $+$ $n{\scriptstyle_{eb}}$ and $n{\scriptstyle_{p}}$ $+$ 2$n{\scriptstyle_{\alpha}}$ are within expectations.  The median(lower quartile)[upper quartile] values are 10.3\%(4.9\%)[19.0\%], which is consistent with our $\delta \mathcal{R}$ statistics.

\indent  Finally, the total electron current, $j{\scriptstyle_{e,tot}}$ $=$ $\sum_{s} \ n{\scriptstyle_{es}} \ v{\scriptstyle_{os, \parallel}}$, in the ion rest frame should be zero to maintain a net zero current in the solar wind.  The mean, median, lower quartile, and upper quartile for all data examined are $\sim$22 $km/s \ cm^{-3}$, $\sim$0 $km/s \ cm^{-3}$, $\sim$-214 $km/s \ cm^{-3}$, and $\sim$351 $km/s \ cm^{-3}$, consistent with previously published work on this dataset \citep[e.g.,][]{bale13a, pulupa14a} and consistent with work in progress [\emph{Salem et al.}, in preparation].  Normalizing $j{\scriptstyle_{e,tot}}$ by $n{\scriptstyle_{e}}$ times $V{\scriptstyle_{Tec, tot}}$ yields a mean, median, lower quartile, and upper quartile for all data examined are $\sim$0.17\%, $\sim$10$^{-8}$\%, $\sim$-0.95\%, and $\sim$1.3\%, respectively.  Thus, the values are all small compared to unity.  Quantitatively, $\sim$97.5\% of the $j{\scriptstyle_{e,tot}}$/($n{\scriptstyle_{e}}$ $V{\scriptstyle_{Tec, tot}}$) values satisfy $\lesssim$5.5\%.

\begin{figure*}
  \centering
    {\includegraphics[trim = 0mm 0mm 0mm 0mm, clip, height=130mm]{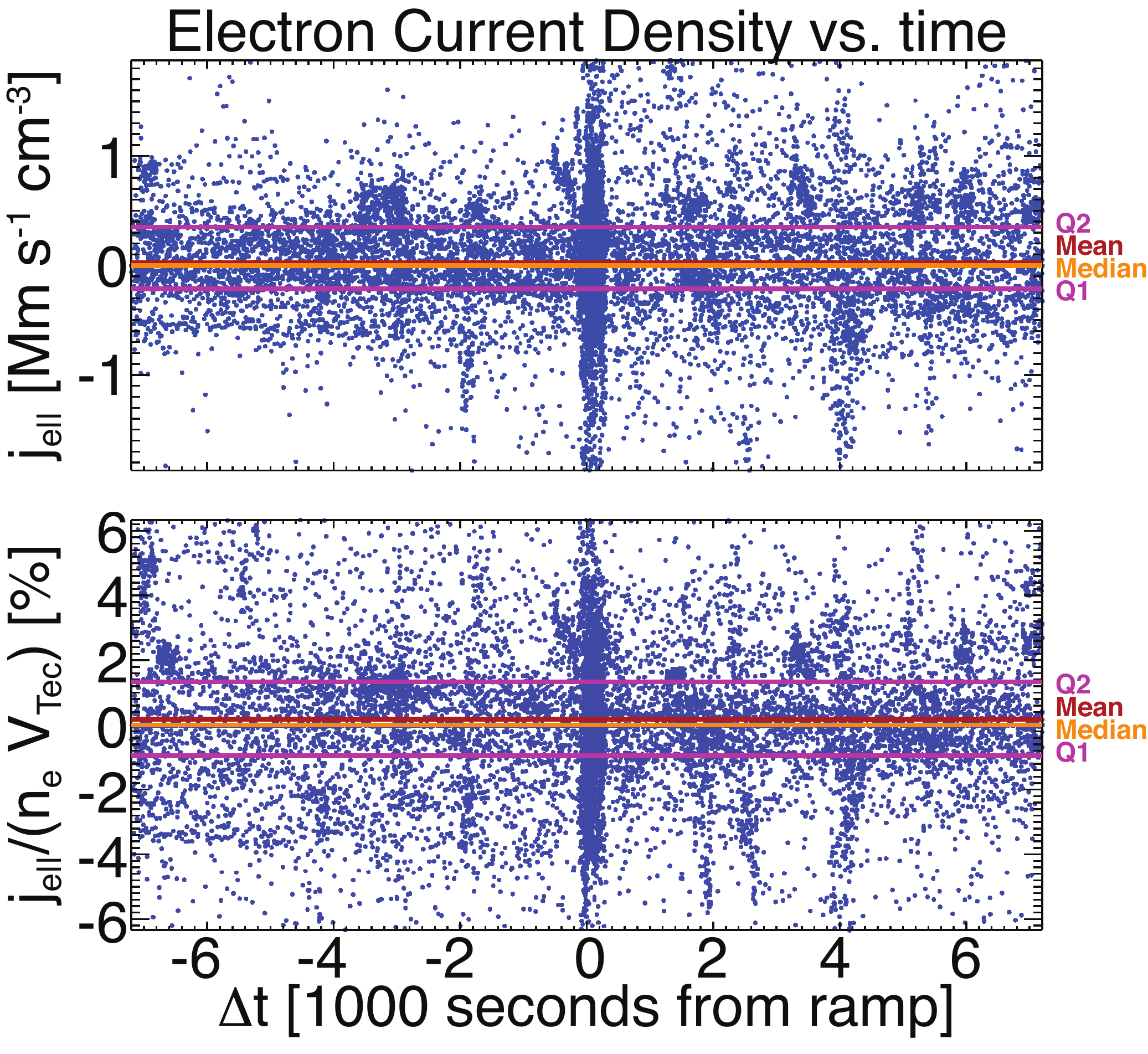}}
    \caption{Two superposed epoch analysis plots of the total electron current density, $j{\scriptstyle_{e,tot}}$ [$Mm \ s^{-1}$] (top panel), and normalized values, $j{\scriptstyle_{e,tot}}$/($n{\scriptstyle_{e}}$ $V{\scriptstyle_{Tec, tot}}$) [\%] (bottom panel), versus seconds from the shock ramp center.  Shown on each plot are the lower (Q1) and upper (Q2) quartiles as magenta lines, the mean as a red line, and the median as an orange line for all data.  That is, the lines are computed for the entire set of data, not at each time stamp.  For reference, the axes ranges were defined as 110\% of the maximum of the absolute value of $X{\scriptstyle_{2.5}}$ and $X{\scriptstyle_{97.5}}$, where $X{\scriptstyle_{2.5}}$ and $X{\scriptstyle_{97.5}}$ are the bottom 2.5$^{th}$ and top 97.5$^{th}$ percentiles.}
    \label{fig:ElectronCurrentDensity}
\end{figure*}

\indent  Figure \ref{fig:ElectronCurrentDensity} shows both $j{\scriptstyle_{e,tot}}$ and $j{\scriptstyle_{e,tot}}$/($n{\scriptstyle_{e}}$ $V{\scriptstyle_{Tec, tot}}$) versus seconds from every shock ramp center time in this study.  One can see that although there are locations with significant deviation from zero (e.g., the shock ramp, which is not tremendously surprising as that is where currents are supposed to exist), the mean (red horizontal line) and median (orange horizontal line) are small for both the raw and normalized current densities.  Note that the data in Figure \ref{fig:ElectronCurrentDensity} includes fit results where there may not be a solution for one or more components (see discussion of first data product ASCII file in Appendix \ref{app:DataProduct}).

\indent  As a final note, there is the question about the validity of using a new model function to describe the thermal core.  Of the 11,874 core VDFs fit with a symmetric bi-self-similar model function there were 9559 or $\sim$80.5\% that satisfied 2.0 $\leq$ $s{\scriptstyle_{ec}}$ $\leq$ 2.05.  That is, the majority of the distributions would be nearly indistinguishable from a bi-Maxwellian on visual inspection.  Therefore, the use of the symmetric bi-self-similar model function is not entirely inconsistent with previous work that modeled the solar wind core with a bi-Maxwellian \citep[e.g.,][]{feldman79a, feldman79b}.  In fact, these results show that most core VDFs are not far from thermal velocity distributions, consistent with results showing evidence for collisional effects on the core \citep[e.g.,][]{bale13a, salem03a}.

\phantomsection   
\subsection{Summary of Fit Results}  \label{subsec:SummaryFitResults}

\indent  For the 52 IP shocks examined there were a total of 15,314 VDFs observed by \emph{Wind}.  Of those 15,314 VDFs, 15,210 progressed to fit analysis and for the core only 534($\sim$4\%) were modeled as bi-kappa VDFs, 12,095($\sim$80\%) were modeled as symmetric bi-self-similar VDFs, and 2581($\sim$17\%) were modeled as asymmetric bi-self-similar VDFs.  All core bi-kappa VDFs were found in the upstream and all downstream core VDFs used either a symmetric or asymmetric bi-self-similar model.  All halo and beam/strahl components were fit to a bi-kappa model.  The justifications for the use of these functions is given in Section \ref{sec:FitMethodology} and Appendix \ref{app:NumericalAnalysis}.  Of those 15,210 that progressed to fit analysis stable solutions were found for 14,847($\sim$98\%) $f^{\left( core \right)}$, 13,871($\sim$91\%) $f^{\left( halo \right)}$, and 9567($\sim$63\%) $f^{\left( beam \right)}$.

\indent  Recall that the fit results presented herein were performed on two-dimensional, (assumed) gyrotropic velocity distributions in the proton bulk flow rest frame.  Most prior work numerically fit to one-dimensional cuts of the VDF or to one-dimensional reduced VDFs.  There are benefits for either method but here it is shown that the method employed is valid by illustrating the consistency with previous work.  The statistical results of the densities are summarized below in the following form \textit{lower quartile}--\textit{upper quartile}(\textit{Mean})[\textit{Median}]
\begin{itemize}[itemsep=0pt,parsep=0pt,topsep=0pt]
  \item[]  \textbf{All}
  \begin{itemize}[itemsep=0pt,parsep=0pt,topsep=0pt]
    \item  $n{\scriptstyle_{ec}}$ $\sim$ 6.44--19.5(13.7)[11.3] $cm^{-3}$;
    \item  $n{\scriptstyle_{eh}}$ $\sim$ 0.21--0.63(0.52)[0.36] $cm^{-3}$;
    \item  $n{\scriptstyle_{eb}}$ $\sim$ 0.09--0.27(0.21)[0.16] $cm^{-3}$;
  \end{itemize}
  \item[]  \textbf{Upstream}
  \begin{itemize}[itemsep=0pt,parsep=0pt,topsep=0pt]
    \item  $n{\scriptstyle_{ec}}$ $\sim$ 4.06--12.5(8.90)[8.09] $cm^{-3}$;
    \item  $n{\scriptstyle_{eh}}$ $\sim$ 0.17--0.49(0.42)[0.27] $cm^{-3}$;
    \item  $n{\scriptstyle_{eb}}$ $\sim$ 0.09--0.26(0.22)[0.16] $cm^{-3}$;
  \end{itemize}
  \item[]  \textbf{Downstream}
  \begin{itemize}[itemsep=0pt,parsep=0pt,topsep=0pt]
    \item  $n{\scriptstyle_{ec}}$ $\sim$ 8.44--24.2(17.3)[16.6] $cm^{-3}$;
    \item  $n{\scriptstyle_{eh}}$ $\sim$ 0.26--0.70(0.59)[0.44] $cm^{-3}$;
    \item  $n{\scriptstyle_{eb}}$ $\sim$ 0.09--0.28(0.21)[0.17] $cm^{-3}$.
  \end{itemize}
\end{itemize}

\noindent  which are consistent with previous results near 1 AU \citep[e.g.,][]{feldman75a, feldman79b, feldman83b, maksimovic97a, nieveschinchilla08a, phillips89a, phillips89b, pierrard16a, salem01a, skoug00a, stverak09a}.  The full statistical results and associated histograms are presented in Paper II.

\indent  The statistical results of the quality analysis are listed below in the following form \textit{lower quartile}--\textit{upper quartile}(\textit{mean})[\textit{median}]
\begin{itemize}[itemsep=0pt,parsep=0pt,topsep=0pt]
  \item[]  \textbf{All}
  \begin{itemize}[itemsep=0pt,parsep=0pt,topsep=0pt]
    \item  $\delta \mathcal{R}$ $\sim$ 5.4\%--15.0\%(11.5\%)[9.1\%];
    \item  $\tilde{\chi}{\scriptstyle_{c}}^{2}$ $\sim$ 0.89--4.28(6.67)[1.95];
    \item  $\tilde{\chi}{\scriptstyle_{h}}^{2}$ $\sim$ 0.41--1.61(2.17)[0.72];
    \item  $\tilde{\chi}{\scriptstyle_{b}}^{2}$ $\sim$ 0.37--1.31(1.56)[0.66];
    \item  $\tilde{\chi}{\scriptstyle_{tot}}^{2}$ $\sim$ 2.82--9.40(694)[4.90];
  \end{itemize}
  \item[]  \textbf{Upstream}
  \begin{itemize}[itemsep=0pt,parsep=0pt,topsep=0pt]
    \item  $\delta \mathcal{R}$ $\sim$ 7.4\%--16.7\%(13.3\%)[11.3\%];
    \item  $\tilde{\chi}{\scriptstyle_{c}}^{2}$ $\sim$ 0.62--2.43(2.05)[1.38];
    \item  $\tilde{\chi}{\scriptstyle_{h}}^{2}$ $\sim$ 0.31--1.01(2.03)[0.50];
    \item  $\tilde{\chi}{\scriptstyle_{b}}^{2}$ $\sim$ 0.32--0.98(1.24)[0.55];
    \item  $\tilde{\chi}{\scriptstyle_{tot}}^{2}$ $\sim$ 2.08--5.51(1557)[3.17];
  \end{itemize}
  \item[]  \textbf{Downstream}
  \begin{itemize}[itemsep=0pt,parsep=0pt,topsep=0pt]
    \item  $\delta \mathcal{R}$ $\sim$ 6.3\%--15.7\%(12.1\%)[10.0\%];
    \item  $\tilde{\chi}{\scriptstyle_{c}}^{2}$ $\sim$ 1.22--8.55(10.1)[2.67];
    \item  $\tilde{\chi}{\scriptstyle_{h}}^{2}$ $\sim$ 0.52--1.95(2.26)[0.91];
    \item  $\tilde{\chi}{\scriptstyle_{b}}^{2}$ $\sim$ 0.41--1.55(1.78)[0.79];
    \item  $\tilde{\chi}{\scriptstyle_{tot}}^{2}$ $\sim$ 3.81--12.8(54.1)[6.98].
  \end{itemize}
\end{itemize}

\indent  The purpose of listing these statistics is to provide a range of typical or expected $\tilde{\chi}{\scriptstyle_{s}}^{2}$ and $\delta \mathcal{R}$ values for reference when determining the quality of any given fit.  Note that the statistics for $\delta \mathcal{R}$ shown above were performed on arrays that excluded the lower and upper boundaries, i.e., 0.1\% and 100\% values.  The statistical results of the model function exponent and drift speed results are presented below and the full data product resulting from this work is described in Appendix \ref{app:DataProduct}.

\phantomsection   
\section{Exponents and Drifts}  \label{sec:ExponentsandDrifts}

\indent  Table \ref{tab:Exponents} shows the one-variable statistics for the exponents from the model fits of the electron VDFs are introduced and discussed, for the core ($s$ $=$ $c$), halo ($s$ $=$ $h$), and beam/strahl ($s$ $=$ $b$).  The VDFs, modeled as bi-kappa ($\kappa{\scriptstyle_{es}}$), symmetric bi-self-similar ($s{\scriptstyle_{es}}$), and asymmetric bi-self-similar velocity distributions ($p{\scriptstyle_{es}}$ for parallel and $q{\scriptstyle_{es}}$ for perpendicular), are summarized for all time periods, upstream only, downstream only, low Mach number only, high Mach number only, quasi-perpendicular only, and quasi-parallel only.  The rows showing N/A (not available) for every entry had no fit results, i.e., the core was only modeled as a bi-kappa in the upstream and an asymmetric bi-self-similar only in the downstream therefore the converse had no results to examine..

\indent  For the VDFs fit to a bi-kappa, the core values typically lie between $\sim$5--10 while the halo and beam/strahl lie between $\sim$3.5--5.4 and $\sim$3.4--5.2, respectively.  Only the core was fit to the bi-self-similar functions and nearly all symmetric exponents are between $\sim$2.00--2.04 while most of the asymmetric parallel and perpendicular exponents lie $\sim$2.2--4.0 and $\sim$2.0--2.5, respectively.

\startlongtable  
\begin{deluxetable}{| l | c | c | c | c | c | c |}
  \tabletypesize{\footnotesize}    
  \tablecaption{Electron Exponent Parameters \label{tab:Exponents}}
  \tablehead{\colhead{Exponent} & \colhead{$X{\scriptstyle_{min}}$}\tablenotemark{a} & \colhead{$X{\scriptstyle_{max}}$}\tablenotemark{b} & \colhead{$\bar{X}$}\tablenotemark{c} & \colhead{$\tilde{X}$}\tablenotemark{d} & \colhead{$X{\scriptstyle_{25\%}}$}\tablenotemark{e} & \colhead{$X{\scriptstyle_{75\%}}$}\tablenotemark{f}}
  \startdata
  \multicolumn{7}{ |c| }{\textbf{All: 15,210 VDFs}} \\
  \hline
  $\kappa{\scriptstyle_{ec}}$  &  2.14 & 100.0 &  9.15 &  7.92 &  5.40 &  10.2  \\
  $s{\scriptstyle_{ec}}$       &  2.00 &  3.00 &  2.03 &  2.00 &  2.00 &  2.04  \\
  $p{\scriptstyle_{ec}}$       &  2.00 &  5.43 &  3.09 &  3.00 &  2.20 &  4.00  \\
  $q{\scriptstyle_{ec}}$       &  2.00 &  3.29 &  2.24 &  2.00 &  2.00 &  2.46  \\
  $\kappa{\scriptstyle_{eh}}$  &  1.51 &  19.7 &  4.62 &  4.38 &  3.58 &  5.34  \\
  $\kappa{\scriptstyle_{eb}}$  &  1.52 &  20.0 &  4.57 &  4.17 &  3.40 &  5.16  \\
  \hline
  \multicolumn{7}{ |c| }{\textbf{Upstream Only: 6546 VDFs}} \\
  \hline
  $\kappa{\scriptstyle_{ec}}$  &  2.14 & 100.0 &  9.15 &  7.92 &  5.40 &  10.2  \\
  $s{\scriptstyle_{ec}}$       &  2.00 &  2.31 &  2.01 &  2.00 &  2.00 &  2.03  \\
  $p{\scriptstyle_{ec}}$       &   N/A &   N/A &   N/A &   N/A &   N/A &   N/A  \\
  $q{\scriptstyle_{ec}}$       &   N/A &   N/A &   N/A &   N/A &   N/A &   N/A  \\
  $\kappa{\scriptstyle_{eh}}$  &  1.52 &  18.4 &  4.16 &  4.10 &  3.25 &  4.83  \\
  $\kappa{\scriptstyle_{eb}}$  &  1.52 &  19.6 &  4.22 &  3.81 &  3.25 &  4.70  \\
  \hline
  \multicolumn{7}{ |c| }{\textbf{Downstream Only: 8664 VDFs}} \\
  \hline
  $\kappa{\scriptstyle_{ec}}$  &   N/A &   N/A &   N/A &   N/A &   N/A &   N/A  \\
  $s{\scriptstyle_{ec}}$       &  2.00 &  3.00 &  2.05 &  2.01 &  2.00 &  2.06  \\
  $p{\scriptstyle_{ec}}$       &  2.00 &  5.43 &  3.09 &  3.00 &  2.20 &  4.00  \\
  $q{\scriptstyle_{ec}}$       &  2.00 &  3.29 &  2.24 &  2.00 &  2.00 &  2.46  \\
  $\kappa{\scriptstyle_{eh}}$  &  1.51 &  19.7 &  4.94 &  4.62 &  3.80 &  5.70  \\
  $\kappa{\scriptstyle_{eb}}$  &  1.53 &  20.0 &  4.82 &  4.45 &  3.61 &  5.44  \\
  \hline
  \multicolumn{7}{ |c| }{\textbf{$\langle M{\scriptstyle_{f}} \rangle{\scriptstyle_{up}}$ $<$ 3 Only: 12,988 VDFs}} \\
  \hline
  $\kappa{\scriptstyle_{ec}}$  &  2.14 & 100.0 &  9.02 &  6.83 &  4.40 &  9.93  \\
  $s{\scriptstyle_{ec}}$       &  2.00 &  3.00 &  2.03 &  2.00 &  2.00 &  2.04  \\
  $p{\scriptstyle_{ec}}$       &  2.00 &  5.43 &  3.10 &  3.00 &  2.18 &  4.00  \\
  $q{\scriptstyle_{ec}}$       &  2.00 &  3.14 &  2.26 &  2.01 &  2.00 &  2.49  \\
  $\kappa{\scriptstyle_{eh}}$  &  1.51 &  19.7 &  4.54 &  4.34 &  3.58 &  5.26  \\
  $\kappa{\scriptstyle_{eb}}$  &  1.52 &  20.0 &  4.62 &  4.20 &  3.46 &  5.19  \\
  \hline
  \multicolumn{7}{ |c| }{\textbf{$\langle M{\scriptstyle_{f}} \rangle{\scriptstyle_{up}}$ $\geq$ 3 Only: 2222 VDFs}} \\
  \hline
  $\kappa{\scriptstyle_{ec}}$  &  4.32 &  27.2 &  9.30 &  8.60 &  6.89 &  10.4  \\
  $s{\scriptstyle_{ec}}$       &  2.00 &  2.30 &  2.03 &  2.00 &  2.00 &  2.08  \\
  $p{\scriptstyle_{ec}}$       &  2.00 &  5.00 &  3.08 &  2.50 &  2.18 &  4.00  \\
  $q{\scriptstyle_{ec}}$       &  2.00 &  3.29 &  2.16 &  2.00 &  2.00 &  2.50  \\
  $\kappa{\scriptstyle_{eh}}$  &  1.60 &  19.2 &  5.06 &  4.68 &  3.62 &  6.05  \\
  $\kappa{\scriptstyle_{eb}}$  &  1.52 &  18.8 &  4.25 &  3.84 &  2.89 &  4.94  \\
  \hline
  \multicolumn{7}{ |c| }{\textbf{$\theta{\scriptstyle_{Bn}}$ $>$ 45$^{\circ}$ Only: 10,940 VDFs}} \\
  \hline
  $\kappa{\scriptstyle_{ec}}$  &  4.05 &  27.2 &  7.77 &  7.18 &  4.84 &  9.11  \\
  $s{\scriptstyle_{ec}}$       &  2.00 &  2.31 &  2.02 &  2.00 &  2.00 &  2.05  \\
  $p{\scriptstyle_{ec}}$       &  2.00 &  5.43 &  3.00 &  2.62 &  2.17 &  4.00  \\
  $q{\scriptstyle_{ec}}$       &  2.00 &  3.29 &  2.28 &  2.04 &  2.00 &  2.56  \\
  $\kappa{\scriptstyle_{eh}}$  &  1.51 &  19.7 &  4.73 &  4.44 &  3.67 &  5.47  \\
  $\kappa{\scriptstyle_{eb}}$  &  1.52 &  20.0 &  4.67 &  4.20 &  3.33 &  5.33  \\
  \hline
  \multicolumn{7}{ |c| }{\textbf{$\theta{\scriptstyle_{Bn}}$ $\leq$ 45$^{\circ}$ Only: 4270 VDFs}} \\
  \hline
  $\kappa{\scriptstyle_{ec}}$  &  2.14 & 100.0 &  16.0 &  11.7 &  10.0 &  14.5  \\
  $s{\scriptstyle_{ec}}$       &  2.00 &  3.00 &  2.06 &  2.00 &  2.00 &  2.04  \\
  $p{\scriptstyle_{ec}}$       &  2.00 &  4.28 &  3.29 &  4.00 &  4.00 &  4.28  \\
  $q{\scriptstyle_{ec}}$       &  2.00 &  3.00 &  2.14 &  2.00 &  2.00 &  2.16  \\
  $\kappa{\scriptstyle_{eh}}$  &  1.55 &  19.4 &  4.32 &  4.18 &  3.37 &  5.09  \\
  $\kappa{\scriptstyle_{eb}}$  &  1.53 &  16.5 &  4.31 &  4.10 &  3.57 &  4.82  \\
  \hline
  \enddata
  \tablenotetext{a}{minimum}
  \tablenotetext{b}{maximum}
  \tablenotetext{c}{mean}
  \tablenotetext{d}{median}
  \tablenotetext{e}{lower quartile}
  \tablenotetext{f}{upper quartile}
  \tablecomments{For symbol definitions, see Appendix \ref{app:Definitions}.}
\end{deluxetable}

\indent  The $\kappa{\scriptstyle_{eh}}$ and $\kappa{\scriptstyle_{eb}}$ values are consistent with previous solar wind observations near 1 AU \citep[e.g.,][]{horaites18a, lazar17a, maksimovic97a, maksimovic05a, pierrard16a, stverak09a, tao16a, tao16b}.  The $\kappa{\scriptstyle_{ec}}$ values are also consistent with previous solar wind observations \citep[e.g.,][]{broiles16a, nieveschinchilla08a}.

\indent  There are several interesting things to note from Table \ref{tab:Exponents}.  The mean, median, and lower/upper quartile values for $\kappa{\scriptstyle_{ec}}$ are slightly higher for high than for low Mach number shocks, though only the median and lower quartile values are significant.  Since a bi-kappa model was only used for upstream core VDFs, this may imply that shock strength is somehow dependent upon the upstream core electron distribution profiles.  One possible physical interpretation would be that the sound speed depends upon the polytropic index for each species, i.e., the equation of state assumed for the system.  A bi-kappa core VDF could effect the estimate of the sound speed, thus altering the fast mode Mach number.  However, the shape of the upstream VDFs will also affect the shock dissipation mechanisms.  For instance, it is known that the existence of power-law tails improves the efficiency of shock acceleration \citep[e.g.,][]{trotta19a}.  Therefore, the larger $\kappa{\scriptstyle_{ec}}$ associated with higher Mach number shocks may imply that lower energy particles have entered the tails thus increasing the exponent\footnote{Recall that $\kappa{\scriptstyle_{ec}}$ values only exist for upstream VDF fits, so the dependence on Mach number is not about thermalization.}.

\indent  In contrast, the asymmetric bi-self-similar exponents, only used in downstream regions, are effectively the same between low and high Mach number shocks.  However, this changes when comparing quasi-parallel and quasi-perpendicular shocks.  The $p{\scriptstyle_{ec}}$ exponent has higher mean, median, and lower/upper quartile values for quasi-parallel than quasi-perpendicular shocks.  The opposite is true for the $q{\scriptstyle_{ec}}$ exponent.

\indent  This is interesting as higher $p{\scriptstyle_{ec}}$ values are predicted to occur in the nonlinear saturation stages of ion-acoustic waves \citep[e.g.,][]{dum74a, dum75a}.  Such waves are driven by relative electron-ion drifts (i.e., currents) and are observed near both quasi-parallel and quasi-perpendicular shocks \citep[e.g.,][]{breneman13a, fuselier84a, wilsoniii07a, wilsoniii10a, wilsoniii12c, wilsoniii14a, wilsoniii14b} but their amplitudes increase with increasing shock strength \citep[e.g.,][]{wilsoniii07a}.  If the largest ion-acoustic waves generate the largest values of $p{\scriptstyle_{ec}}$, then one would expect maximum values downstream of strong quasi-perpendicular shocks, which is not the case here.  This leads to the question of what fraction of energy goes to increasing $p{\scriptstyle_{ec}}$ versus what fraction goes to increasing $T{\scriptstyle_{ec, \parallel}}$.  This would depend upon the effective inelasticity of the wave-particle interactions, where larger inelasticity increases $p{\scriptstyle_{ec}}$ and smaller increases $T{\scriptstyle_{ec, \parallel}}$ \citep[e.g.,][]{dum74a, dum75a, goldman84a, horton76a, horton79a, jain79a}.  The interaction between a wave and a particle can be treated as inelastic if the particle affects the wave amplitude and kinetic energy during the interaction.  Most test-particle treatments do not handle this self-consistently and if the effect is distributed to an entire VDF the net result can be a stochastic heating that increases $p{\scriptstyle_{ec}}$ from 2.0 \citep[e.g.,][]{dum74a, dum75a}.

\indent  Another theory predicts that flattop electron distributions (i.e., $p{\scriptstyle_{ec}}$ $\rightarrow$ $\geq$4 and $q{\scriptstyle_{ec}}$ $\rightarrow$ $\sim$2--3) can result from the combined effects of a quasi-static, cross-shock electric potential and from fluctuation electric fields \citep[e.g.,][]{feldman83a, hull98a} through a process called  maximal filling \citep[e.g.,][]{morse65a}.  However, similar to the predictions for wave-driven flattops this theory should generate stronger flattops (i.e., larger values of $p{\scriptstyle_{ec}}$) for stronger quasi-perpendicular shocks, which we do not observe.  Thus, the evolution of the electron VDFs do not seem consistent with the standard quasi-static, cross-shock electric potential, but rather in agreement with recent high resolution observations at the bow shock \citep[e.g.,][]{chen18a, goodrich18c}.

\indent  Another interesting result is the difference in the $\kappa{\scriptstyle_{eh}}$ values under different conditions.  When the values of $\kappa{\scriptstyle_{eh}}$ are larger(smaller), that implies a less(more) energized halo, i.e., softer(harder) spectra.  One can see that $\kappa{\scriptstyle_{eh}}$ is larger downstream than upstream and near high than low Mach number shocks.  That is, the halo is less energized downstream of IP shocks and near strong IP shocks than the converse, which is somewhat unexpected as strong shocks should more readily energize suprathermal particles \citep[e.g.,][]{caprioli14a, malkov01a, park15a, treumann09a, trotta19a}.  In contrast, $\kappa{\scriptstyle_{eh}}$ is slightly smaller ($\sim$10\%) near quasi-parallel than quasi-perpendicular shocks, which implies more energized halo electrons.  Although quasi-parallel shocks are predicted \citep[e.g.,][]{caprioli14a, malkov01a} and observed \citep[e.g.,][]{wilsoniii16h} to be more efficient particle accelerators, the predictions are usually specific to ions while mildly suprathermal electrons are thought to most efficiently interact with quasi-perpendicular shocks \citep[e.g.,][]{wu84b, park13a, trotta19a}.  Further, very recent simulation results suggest the upstream electron suprathermal tail will become flatter (i.e., smaller kappa values) with increasing Mach number for quasi-perpendicular shocks \citep[][]{trotta19a}.  This may explain why both $\kappa{\scriptstyle_{eh}}$ and $\kappa{\scriptstyle_{eb}}$ are smaller in the upstream than downstream.  The time-evolution of these kappa values will be examined in more detail in Paper III.

\indent  A major caveat of the above discussion is the exchange of particles between the various electron VDF components, i.e., former core electrons can be energized and move to the halo or the converse.  Therefore, one needs to be careful when interpreting the change in a given component-specific parameter.  This will be discussed in more detail in Paper III.

\indent  Finally, the $\kappa{\scriptstyle_{eb}}$ values show a similar behavior between upstream and downstream and shock geometry as $\kappa{\scriptstyle_{eh}}$, but they differ between low and high Mach number shocks.  That is, stronger shocks appear to energize the beam/strahl component more than weaker shocks.  This is likely due to the electron foreshock component observed upstream of strong IP shocks \citep[e.g.,][]{bale99a, pulupa08a, pulupa10a} combined with the usual solar wind beam/strahl component.

\begin{figure*}
  \centering
    {\includegraphics[trim = 0mm 0mm 0mm 0mm, clip, width=170mm]{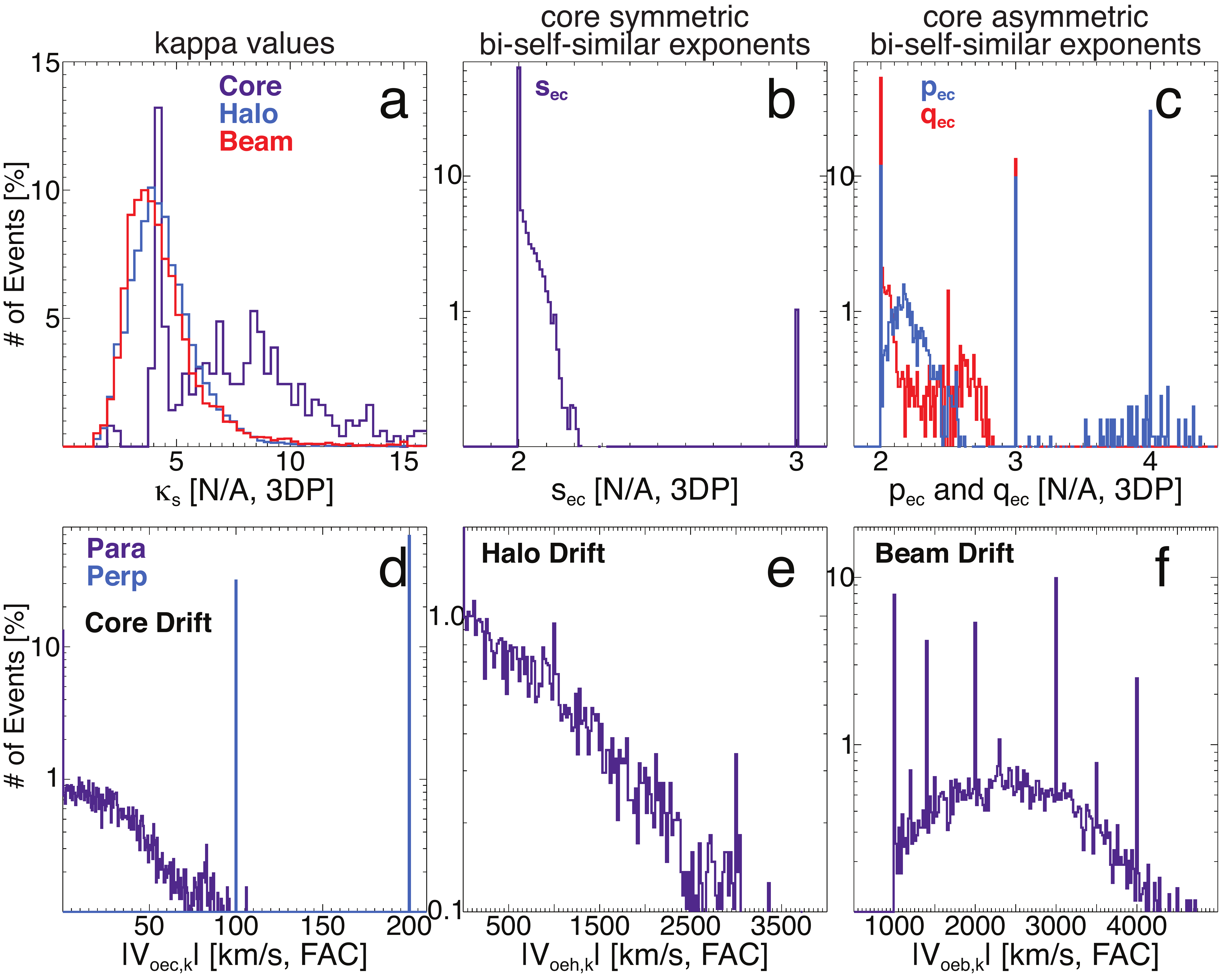}}
    \caption{Histograms of the exponents (top row) and bulk drift velocity magnitudes (bottom row) for the for different electron components for all time periods as percentage of total counts.  Panel a shows the $\kappa{\scriptstyle_{es}}$ values for the core (violet), halo (blue), and beam/strahl (red) components.  Panel b shows the $s{\scriptstyle_{ec}}$ for the core (violet).  Panel c shows the $p{\scriptstyle_{ec}}$ (blue) and $q{\scriptstyle_{ec}}$ (red) values for the core.  Panels d-f show the magnitude of the parallel (violet) and perpendicular (blue) drift velocities for the core, halo, and beam/strahl components, respectively.  The statistics for the exponents are listed in Table \ref{tab:Exponents}.  Note that the tick marks are individually labeled in all panels.}
    \label{fig:Exponents}
\end{figure*}

\indent  Figure \ref{fig:Exponents} shows histograms of $\kappa{\scriptstyle_{es}}$, $s{\scriptstyle_{ec}}$, $p{\scriptstyle_{ec}}$, $q{\scriptstyle_{ec}}$, and the drift speed magnitudes, $V{\scriptstyle_{oes,j}}$ ($s$ for electron components and $j$ for parallel or perpendicular), for the three electron populations.  These histograms show distributions corresponding to the first part of Table \ref{tab:Exponents}, i.e., all VDF solutions.  In many of the panels there are isolated, dominant peaks, nearly all of which result from constraints imposed for specific events, not necessarily an underlying physical reason.  For instance, the peaks for $p{\scriptstyle_{ec}}$ $=$ 3 and $=$ 4 in panel c are for strong shocks exhibiting flattop VDFs in the downstream where the fit routines were not finding stable solutions without imposing constraints on both the exponents and the minimum number density for the core distribution.

\indent  One can see that, as discussed previously, the core parallel drift speeds (violet line, panel d) tend to fall below $\sim$100 km/s, consistent with previous results \citep[e.g.,][]{pulupa14a}. In fact, most of the core and halo drifts are near zero with the number of results satisfying $V{\scriptstyle_{oec, \parallel}}$ $\leq$ 1 km/s and $V{\scriptstyle_{oeh, \parallel}}$ $\leq$ 1 km/s are 8735($\sim$59\%) and 7311($\sim$53\%), respectively.  Note that although there is sometimes a sizable perpendicular core drift (blue line, panel d) for some shock crossings, these were explicitly set after visual inspection of the VDFs during the iterative fitting process.  The non-zero perpendicular drifts almost certainly result from inaccuracies in the calculation of the solar wind rest frame and a dipole correction to $\phi{\scriptstyle_{sc}}$ not included in the present analysis \citep[e.g.,][]{pulupa14a} (see Appendix \ref{app:DetectorCalibration} for more details).

\indent    The magnitudes of $V{\scriptstyle_{oeh, \perp}}$ and $V{\scriptstyle_{oeb, \perp}}$ never deviated from zero\footnote{This was an explicit constraint imposed on all fits but would also have resulted largely from the initial guess that both $V{\scriptstyle_{oeh, \perp}}$ and $V{\scriptstyle_{oeb, \perp}}$ equal zero.  That is, the fit software uses initial guesses to estimate gradient magnitudes for changes between iterations.  So if the initial guess is null, the step size will be null as well.}.  The magnitudes of $V{\scriptstyle_{oeh, \parallel}}$ range from $\sim$0--8860 km/s with a lower to upper quartile range of $\sim$0--850 km/s and a mean(median) of $\sim$580 km/s($\sim$0.1 km/s).  The magnitudes of $V{\scriptstyle_{oeb, \parallel}}$ range from $\sim$1000--9330 km/s with a lower to upper quartile range of $\sim$1750--3090 km/s and a mean(median) of $\sim$2580 km/s($\sim$2480 km/s).  As previously discussed, the lower bound for $V{\scriptstyle_{oeb, \parallel}}$ was imposed on the basis of physical arguments while the magnitude of $V{\scriptstyle_{oeh, \parallel}}$ was allowed to go to zero.  If only magnitudes satisfying $V{\scriptstyle_{oes, \parallel}}$ $>$ 1 km/s are considered, the mean(median) and lower to upper quartile ranges are $\sim$42 km/s($\sim$30 km/s) and $\sim$14--52 km/s for $V{\scriptstyle_{oec, \parallel}}$ and $\sim$1227 km/s($\sim$903 km/s) and $\sim$362--1695 km/s for $V{\scriptstyle_{oeh, \parallel}}$.

\phantomsection   
\section{Discussion}  \label{sec:Discussion}

\indent  A total of 15,314 electron VDFs were observed by the \emph{Wind} spacecraft within $\pm$2 hours of 52 IP shocks of which 15,210 had a stable solution for at least one component.  Stable model function parameters were found for 14,847($\sim$98\%) core fits, 13,871($\sim$91\%) halo fits, and 9567($\sim$63\%) beam/strahl fits.  The fit parameters are consistent with previous studies and will be discussed in detail in the following two parts of this study.  Of the 15,210 VDFs examined herein, the core was modeled as a bi-kappa for 534($\sim$4\%) VDFs, as a symmetric bi-self-similar for 12,095($\sim$80\%) VDFs, and as an asymmetric bi-self-similar for 2581($\sim$17\%) VDFs.  This is the first statistical study to find that the core electron distribution is better fit to a self-similar velocity distribution function than a Maxwellian under all conditions.

\indent  The exponents are summarized below in the following form \textit{lower quartile}--\textit{upper quartile}(\textit{Mean})[\textit{Median}]
\begin{itemize}[itemsep=0pt,parsep=0pt,topsep=0pt]
  \item[]  \textbf{All}
  \begin{itemize}[itemsep=0pt,parsep=0pt,topsep=0pt]
    \item  $s{\scriptstyle_{ec}}$ $\sim$ 2.00--2.04(2.03)[2.00];
    \item  $p{\scriptstyle_{ec}}$ $\sim$ 2.20--4.00(3.09)[3.00];
    \item  $q{\scriptstyle_{ec}}$ $\sim$ 2.00--2.46(2.24)[2.00];
    \item  $\kappa{\scriptstyle_{ec}}$ $\sim$ 5.40--10.2(9.15)[7.92];
    \item  $\kappa{\scriptstyle_{eh}}$ $\sim$ 3.58--5.34(4.62)[4.38];
    \item  $\kappa{\scriptstyle_{eb}}$ $\sim$ 3.40--5.16(4.57)[4.17];
  \end{itemize}
  \item[]  \textbf{Upstream}
  \begin{itemize}[itemsep=0pt,parsep=0pt,topsep=0pt]
    \item  $s{\scriptstyle_{ec}}$ $\sim$ 2.00--2.03(2.01)[2.00];
    \item  $p{\scriptstyle_{ec}}$ $\sim$ N/A;
    \item  $q{\scriptstyle_{ec}}$ $\sim$ N/A;
    \item  $\kappa{\scriptstyle_{ec}}$ $\sim$ 5.40--10.2(9.15)[7.92];
    \item  $\kappa{\scriptstyle_{eh}}$ $\sim$ 3.25--4.83(4.16)[4.10];
    \item  $\kappa{\scriptstyle_{eb}}$ $\sim$ 3.25--4.70(4.22)[3.81];
  \end{itemize}
  \item[]  \textbf{Downstream}
  \begin{itemize}[itemsep=0pt,parsep=0pt,topsep=0pt]
    \item  $s{\scriptstyle_{ec}}$ $\sim$ 2.00--2.06(2.05)[2.01];
    \item  $p{\scriptstyle_{ec}}$ $\sim$ 2.20--4.00(3.09)[3.00];
    \item  $q{\scriptstyle_{ec}}$ $\sim$ 2.00--2.46(2.24)[2.00];
    \item  $\kappa{\scriptstyle_{ec}}$ $\sim$ N/A;
    \item  $\kappa{\scriptstyle_{eh}}$ $\sim$ 3.80--5.70(4.94)[4.62];
    \item  $\kappa{\scriptstyle_{eb}}$ $\sim$ 3.61--5.44(4.82)[4.45];
  \end{itemize}
\end{itemize}
\noindent  Overall the $\kappa{\scriptstyle_{eh}}$ and $\kappa{\scriptstyle_{eb}}$ values are consistent with previous solar wind observations near 1 AU \citep[e.g.,][]{horaites18a, lazar17a, pierrard16a, stverak09a}.  The $\kappa{\scriptstyle_{ec}}$ values are also consistent with previous solar wind observations \citep[e.g.,][]{broiles16a, nieveschinchilla08a}.  The values for $s{\scriptstyle_{ec}}$, $p{\scriptstyle_{ec}}$, and $q{\scriptstyle_{ec}}$ are consistent with previous results as well \citep[e.g.,][]{feldman83a, feldman83b}.

\indent  The interesting aspect of VDFs being well modeled by bi-self-similar functions is that such functions are used to describe the evolution of distributions either for the flow through disordered porous media \citep[e.g.,][]{matyka16a} or the influence of inelastic scattering \citep[e.g.,][]{dum74a, dum75a, goldman84a, horton76a, horton79a, jain79a}.  It is unlikely that the former applies directly but the latter may be interpreted in the following manner.  The typical approach for test particle simulations used to examine wave-particle interactions does not include feedback from the particles on the waves.  In a real plasma, the particles can alter three properties of electromagnetic waves:  their amplitude (potential energy), momentum, and kinetic energy.  Consider a simple scenario whereby a particle reflects off of an electromagnetic wave field along one dimension.  If done self-consistently, the particle can reduce the wave amplitude in addition to affecting the field momentum and kinetic energy.  In the case of a reduced wave amplitude, the resulting scattering problem can be treated as a simple inelastic collision\footnote{That is, the particle kinetic energy may not be preserved through the interaction even if the wave kinetic energy is conserved.}.  Thus, the net result of an ensemble of particles interacting with a wave field can be stochastic  \citep[e.g.,][]{dum74a, dum75a}, which provides one physical justification for the use of the bi-self-similar functions.  These functions are also convenient in that they reduce to bi-Maxwellians in the limit where the exponents go to two, i.e., the deviation from a Maxwellian is a measure of inelasticity in the particles interactions with waves and/or turbulence\footnote{It is also worth noting that a finite time-correlation included in wave-particle interactions, something missing from quasi-linear theory, can yield a similar VDF profile [work in progress by coauthors].}.  Further, as previously discussed, $\sim$80.5\% of the core VDFs modeled with a symmetric bi-self-similar function had exponents satisfying 2.0 $\leq$ $s{\scriptstyle_{ec}}$ $\leq$ 2.05.  Therefore, the majority of the core electron VDFs would be visually indistinguishable from a bi-Maxwellian which supports previous work that used thermal distributions to model the core \citep[e.g.,][]{feldman79a, feldman79b} and work that found evidence for collisional effects in the core distribution \citep[e.g.,][]{bale13a, salem03a}.

\indent  The $\kappa{\scriptstyle_{ec}}$ seem to correlate with $\langle M{\scriptstyle_{f}} \rangle{\scriptstyle_{up}}$, which may suggest a shock strength dependence on the shape of the upstream electron VDFs.  In contrast with expectations from a dependence on quasi-static fields, the values of $p{\scriptstyle_{es}}$ are higher for quasi-parallel shocks while $q{\scriptstyle_{es}}$ are higher for quasi-perpendicular shocks yet neither depends upon $\langle M{\scriptstyle_{f}} \rangle{\scriptstyle_{up}}$.

\indent  Somewhat surprisingly the values of $\kappa{\scriptstyle_{eh}}$ are larger downstream than upstream and they increase with increasing $\langle M{\scriptstyle_{f}} \rangle{\scriptstyle_{up}}$.  That is, the halo spectra are softer downstream and near strong shocks.  Quasi-parallel shocks, however, correlate with smaller $\kappa{\scriptstyle_{eh}}$, i.e., harder halo spectra.  Generally, quasi-parallel shocks are predicted to be more efficient particle accelerators for suprathermal ions and very energetic electrons\footnote{Suprathermal is defined here for ions in the several to 10s of keV energy range while the electrons are many 10s to 100s of keV for typical 1 AU solar wind collisionless shocks.} \citep[e.g.,][]{caprioli14a} but electrons in the halo energy range are predicted to be energized the most efficiently at shocks satisfying $\theta{\scriptstyle_{Bn}}$ $>$ 80$^{\circ}$ \citep[e.g.,][]{park13a}.

\indent  Unlike the halo, $\kappa{\scriptstyle_{eb}}$ are smaller near high Mach number shocks than near low Mach number shocks.  The difference is likely a two-fold consequence of the combined effects from shock-accelerated foreshock electrons and the method used to fit the distributions.  That is, the beam/strahl component is always fit to the anti-sunward, field-aligned side of the VDF while the halo to the opposite.  For nearly all IP shocks at 1 AU, the shock normal is anti-sunward in a direction that would be aligned with the nominal, ambient beam/strahl electron component.  For both the halo and beam/strahl, the ratios of $\langle \kappa{\scriptstyle_{eh}} \rangle{\scriptstyle_{dn}} / \langle \kappa{\scriptstyle_{eh}} \rangle{\scriptstyle_{up}}$ and $\langle \kappa{\scriptstyle_{eb}} \rangle{\scriptstyle_{dn}} / \langle \kappa{\scriptstyle_{eb}} \rangle{\scriptstyle_{up}}$ increase with increasing $\langle M{\scriptstyle_{f}} \rangle{\scriptstyle_{up}}$.  That is, the downstream halo and beam/strahl spectra are softer than the upstream for stronger shocks.  Again, this is likely a consequence of the foreshock electrons that are not observed upstream of weak shocks.  The details of the electron component velocity moments and associated changes will be discussed further in Papers II and III.

\indent  In summary, the first part of this three-part study presented the first statistical study to find that the core electron distribution is better fit to a self-similar velocity distribution function than a bi-Maxwellian under all conditions.  This is an important result for kinetic theory and solar wind evolution.  This work aslo provides the methodology and details necessary to reproduce and qualify the results of the nonlinear least squares fitting performed herein.  In Papers II and III, the statistical and analysis results of the velocity moments will be presented in detail.  These observations are relevant for comparisons with astrophysical plasmas like the intra-galaxy-cluster medium and they provide a statistical baseline of electron parameters near collisionless shocks for the recent \emph{Parker Solar Probe} and upcoming \emph{Solar Orbiter} missions.

\acknowledgments
\noindent  The authors thank A.F.- Vi{\~n}as and D.A. Roberts for useful discussions of basic plasma physics and C. Markwardt for helpful feedback on the usage nuances of his MPFIT software.  The work was supported by the International Space Science Institute's (ISSI) International Teams programme.  L.B.W. was partially supported by \emph{Wind} MO\&DA grants and a Heliophysics Innovation Fund (HIF) grant.  L.-J.C. and S.W. were partially supported by the MMS mission in addition to NASA grants 80NSSC18K1369 and 80NSSC17K0012, NSF grants AGS-1619584 and AGS-1552142, and DOE grant DESC0016278.  D.L.T. was partially supported by NASA grant NNX16AQ50G.  M.L.S. was partially supported by grants NNX14AT26G and NNX13AI75G.  J.C.K. was partially supported by NASA grants NNX14AR78G and 80NSSC18K0986.  D.C. was partially supported by grants NNX17AG30G, GO8-19110A, 80NSSC18K1726, 80NSSC18K1218, and NSF grant 1714658.  S.J.S. was partially supported by the MMS/FIELDS investigation.  C.S.S. was partially supported by NASA grant NNX16AI59G and NSF SHINE grant 1622498.  S.D.B. and C.S.S. were partially supported by NASA grant NNX16AP95G.  M.P.P. and K.A.G. were supported by Parker Solar Probe instrument funds.

\appendix
\section{Definitions and Notation}  \label{app:Definitions}

\indent  In this appendix we define the symbols and notation used throughout.  In the following, all direction-dependent parameters we use the subscript $j$ to represent the direction where $j$ $=$ $tot$ for the entire distribution, $j$ $=$ $\parallel$ for the the parallel direction, and $j$ $=$ $\perp$ for the perpendicular direction.  Note that parallel and perpendicular are with respect to the quasi-static magnetic field vector, $\mathbf{B}{\scriptstyle_{o}}$ [nT].  The use of the generic subscript $s$ to denote the particle species (e.g., electrons, protons, etc.) or the component of a single particle species (e.g., electron core).  For the electron components, the subscript will be $s$ $=$ $ec$ for the core, $s$ $=$ $eh$ for the halo, $s$ $=$ $eb$ for the beam/strahl, and $s$ $=$ $eff$ for the effective, and $s$ $=$ $e$ for the total/entire population.  Below are the symbol/parameters definitions:

\begin{itemize}[itemsep=0pt,parsep=0pt,topsep=0pt]
  \item[]  \textbf{one-variable statistics}
  \begin{itemize}[itemsep=0pt,parsep=0pt,topsep=0pt]
    \item  $X{\scriptstyle_{min}}$ $\equiv$ minimum
    \item  $X{\scriptstyle_{max}}$ $\equiv$ maximum
    \item  $\bar{X}$ $\equiv$ mean
    \item  $\tilde{X}$ $\equiv$ median
    \item  $X{\scriptstyle_{25\%}}$ $\equiv$ lower quartile
    \item  $X{\scriptstyle_{75\%}}$ $\equiv$ upper quartile
  \end{itemize}
  \item[]  \textbf{fundamental parameters}
  \begin{itemize}[itemsep=0pt,parsep=0pt,topsep=0pt]
    \item  $\varepsilon{\scriptstyle_{o}}$ $\equiv$ permittivity of free space
    \item  $\mu{\scriptstyle_{o}}$ $\equiv$ permeability of free space
    \item  $c$ $\equiv$ speed of light in vacuum [$km \ s^{-1}$] $=$ $\left( \varepsilon{\scriptstyle_{o}} \ \mu{\scriptstyle_{o}} \right)^{-1/2}$
    \item  $k{\scriptstyle_{B}}$ $\equiv$ the Boltzmann constant [$J \ K^{-1}$]
    \item  $e$ $\equiv$ the fundamental charge [$C$]
  \end{itemize}
  \item[]  \textbf{plasma parameters}
  \begin{itemize}[itemsep=0pt,parsep=0pt,topsep=0pt]
    \item  $n{\scriptstyle_{s}}$ $\equiv$ the number density [$cm^{-3}$] of species $s$
    \item  $m{\scriptstyle_{s}}$ $\equiv$ the mass [$kg$] of species $s$
    \item  $Z{\scriptstyle_{s}}$ $\equiv$ the charge state of species $s$
    \item  $q{\scriptstyle_{s}}$ $\equiv$ the charge [$C$] of species $s$ $=$ $Z{\scriptstyle_{s}} \ e$
    \item  $T{\scriptstyle_{s, j}}$ $\equiv$ the scalar temperature [$eV$] of the j$^{th}$ component of species $s$
    \item  $\left(T{\scriptstyle_{s'}}/T{\scriptstyle_{s}}\right){\scriptstyle_{j}}$ $\equiv$ the temperature ratio [N/A] of species $s$ and $s'$ of the j$^{th}$ component
    \item  $\left(T{\scriptstyle_{\perp}}/T{\scriptstyle_{\parallel}}\right){\scriptstyle_{s}}$ $\equiv$ the temperature anisotropy [N/A] of species $s$
    \item  $V{\scriptstyle_{Ts, j}}$ $\equiv$ the most probable thermal speed [$km \ s^{-1}$] of a one-dimensional velocity distribution (see Equation \ref{eq:params_2})
    \item  $\mathbf{v}{\scriptstyle_{os}}$ $\equiv$ the drift velocity [$km \ s^{-1}$] of species $s$ in the plasma bulk flow rest frame
    \item  $C{\scriptstyle_{s}}$ $\equiv$ the sound or ion-acoustic sound speed [$km \ s^{-1}$] (see Supplemental Material for definitions)
    \item  $V{\scriptstyle_{A}}$ $\equiv$ the Alfv\'{e}n speed [$km \ s^{-1}$] (see Supplemental Material for definitions)
    \item  $V{\scriptstyle_{f}}$ $\equiv$ the fast mode speed [$km \ s^{-1}$] (see Supplemental Material for definitions)
    \item  $\Omega{\scriptstyle_{cs}}$ $\equiv$ the angular cyclotron frequency [$rad \ s^{-1}$] (see Equation \ref{eq:params_3})
    \item  $\omega{\scriptstyle_{ps}}$ $\equiv$ the angular plasma frequency [$rad \ s^{-1}$] (see Equation \ref{eq:params_4})
    \item  $\lambda{\scriptstyle_{De}}$ $\equiv$ the electron Debye length [$m$] (see Equation \ref{eq:params_5})
    \item  $\rho{\scriptstyle_{cs}}$ $\equiv$ the thermal gyroradius [$km$] (see Equation \ref{eq:params_6})
    \item  $\lambda{\scriptstyle_{s}}$ $\equiv$ the inertial length [$km$] (see Equation \ref{eq:params_7})
    \item  $\beta{\scriptstyle_{s, j}}$ $\equiv$ the plasma beta [N/A] of the j$^{th}$ component of species $s$ (see Equations \ref{eq:params_8} and \ref{eq:params_9})
    \item  $\phi{\scriptstyle_{sc}}$ $\equiv$ the scalar, quasi-static spacecraft potential [eV] \citep[e.g.,][]{pulupa14a, scime94a}
    \item  $E{\scriptstyle_{min}}$ $\equiv$ the minimum energy bin midpoint value [eV] of an electrostatic analyzer \citep[e.g., see Appendices in][]{wilsoniii17c, wilsoniii18b}
  \end{itemize}
\end{itemize}

\noindent  The variables that rely upon multiple parameters are given in the following equations:

\begin{subequations}
  \begin{align}
    T{\scriptstyle_{eff, j}} & = \frac{ \sum_{s} n{\scriptstyle_{s}} \ T{\scriptstyle_{s, j}} }{ \sum_{s} n{\scriptstyle_{s}} } \label{eq:params_0} \\
    T{\scriptstyle_{s, tot}} & = \frac{1}{3} \left( T{\scriptstyle_{s, \parallel}} + 2 \ T{\scriptstyle_{s, \perp}} \right) \label{eq:params_1} \\
    V{\scriptstyle_{Ts, j}} & = \sqrt{ \frac{ 2 \ k{\scriptstyle_{B}} \ T{\scriptstyle_{s, j}} }{ m{\scriptstyle_{s}} } } \label{eq:params_2} \\
    \Omega{\scriptstyle_{cs}} & = \frac{ q{\scriptstyle_{s}} \ B{\scriptstyle_{o}} }{ m{\scriptstyle_{s}} } \label{eq:params_3} \\
    \omega{\scriptstyle_{ps}} & = \sqrt{ \frac{ n{\scriptstyle_{s}} \ q{\scriptstyle_{s}}^{2} }{ \varepsilon{\scriptstyle_{o}} \ m{\scriptstyle_{s}} } } \label{eq:params_4} \\
    \lambda{\scriptstyle_{De}} & = \frac{ V{\scriptstyle_{Te, tot}} }{ \sqrt{ 2 } \ \omega{\scriptstyle_{pe}} } = \sqrt{ \frac{ \varepsilon{\scriptstyle_{o}} \ k{\scriptstyle_{B}} \ T{\scriptstyle_{e, tot}} }{ n{\scriptstyle_{e}} \ e^{2} } } \label{eq:params_5} \\
    \rho{\scriptstyle_{cs}} & = \frac{ V{\scriptstyle_{Ts, tot}} }{ \Omega{\scriptstyle_{cs}} } \label{eq:params_6} \\
    \lambda{\scriptstyle_{s}} & = \frac{ c }{ \omega{\scriptstyle_{ps}} } \label{eq:params_7} \\
    \beta{\scriptstyle_{s, j}} & = \frac{ 2 \mu{\scriptstyle_{o}} n{\scriptstyle_{s}} k{\scriptstyle_{B}} T{\scriptstyle_{s, j}} }{ \lvert \mathbf{B}{\scriptstyle_{o}} \rvert^{2} } \label{eq:params_8} \\
    \beta{\scriptstyle_{eff, j}} & = \frac{ 2 \mu{\scriptstyle_{o}} n{\scriptstyle_{e}} k{\scriptstyle_{B}} T{\scriptstyle_{eff, j}} }{ \lvert \mathbf{B}{\scriptstyle_{o}} \rvert^{2} } \label{eq:params_9} \\
    \intertext{where $n{\scriptstyle_{e}}$ is defined as:}
    n{\scriptstyle_{e}} & = \sum_{s} \ n{\scriptstyle_{es}} \label{eq:params_10}
  \end{align}
\end{subequations}

\indent  For the macroscopic shock parameters, the values are averaged over asymptotic regions away from the shock transition region.

\begin{itemize}[itemsep=0pt,parsep=0pt,topsep=0pt]
  \item[]  \textbf{shock parameters}
  \begin{itemize}[itemsep=0pt,parsep=0pt,topsep=0pt]
    \item  subscripts $up$ and $dn$ $\equiv$ denote the upstream (i.e., before the shock arrives time-wise at the spacecraft for a forward shock) and downstream (i.e., the shocked region)
    \item  $\langle Q \rangle{\scriptstyle_{j}}$ $\equiv$ the average of parameter $Q$ over the $j^{th}$ shock region, where $j$ $=$ $up$ or $dn$
    \item  $\mathbf{n}{\scriptstyle_{sh}}$ $\equiv$ the shock normal unit vector [N/A]
    \item  $\theta{\scriptstyle_{Bn}}$ $\equiv$ the shock normal angle [deg], defined as the acute reference angle between $\langle \mathbf{B}{\scriptstyle_{o}} \rangle{\scriptstyle_{up}}$ and $\mathbf{n}{\scriptstyle_{sh}}$
    \item  $\langle \lvert V{\scriptstyle_{shn}} \rvert \rangle{\scriptstyle_{j}}$ $\equiv$ the $j^{th}$ region average shock normal speed [$km \ s^{-1}$] in the spacecraft frame
    \item  $\langle \lvert U{\scriptstyle_{shn}} \rvert \rangle{\scriptstyle_{j}}$ $\equiv$ the $j^{th}$ region average shock normal speed [$km \ s^{-1}$] in the shock rest frame (i.e., the speed of the flow relative to the shock)
    \item  $\langle M{\scriptstyle_{A}} \rangle{\scriptstyle_{j}}$ $\equiv$ the $j^{th}$ region average Alfv\'{e}nic Mach number [N/A] $=$ $\langle \lvert U{\scriptstyle_{shn}} \rvert \rangle{\scriptstyle_{j}} / \langle V{\scriptstyle_{A}} \rangle{\scriptstyle_{j}}$
    \item  $\langle M{\scriptstyle_{f}} \rangle{\scriptstyle_{j}}$ $\equiv$ the $j^{th}$ region average fast mode Mach number [N/A]
    \item  $M{\scriptstyle_{cr}}$ $\equiv$ the first critical Mach number [N/A]
    \item  $M{\scriptstyle_{ww}}$ $\equiv$ the linear whistler (phase) Mach number [N/A]
    \item  $M{\scriptstyle_{gr}}$ $\equiv$ the linear whistler (group) Mach number [N/A]
    \item  $M{\scriptstyle_{nw}}$ $\equiv$ the nonlinear whistler Mach number
  \end{itemize}
\end{itemize}

\noindent  The critical Mach numbers are phenomenologically defined as follows:  for $\langle M{\scriptstyle_{f}} \rangle{\scriptstyle_{up}} / M{\scriptstyle_{cr}}$ $\ge$ 1 an ion sound wave could not phase stand within the shock ramp \citep[e.g.,][]{edmiston84, kennel85a}; for $\langle M{\scriptstyle_{f}} \rangle{\scriptstyle_{up}} / M{\scriptstyle_{ww}}$ $\ge$ 1 a linear magnetosonic-whistler cannot phase stand upstream of the shock ramp \citep[e.g.,][]{krasnoselskikh02a}; for $\langle M{\scriptstyle_{f}} \rangle{\scriptstyle_{up}} / M{\scriptstyle_{gr}}$ $\ge$ 1 a linear magnetosonic-whistler cannot group stand upstream of the shock ramp; and for $\langle M{\scriptstyle_{f}} \rangle{\scriptstyle_{up}} / M{\scriptstyle_{nw}}$ $\ge$ 1 a nonlinear magnetosonic-whistler is no longer stable/stationary and will result in the shock ramp ``breaking'' and reforming.

\noindent  These definitions are used throughout.

\section{Spacecraft Potential and Detector Calibration}  \label{app:DetectorCalibration}

\indent  The electron electrostatic analyzer data suffer from several sources of uncertainty including differences between the theoretical maximum detector efficiency and actual \citep[e.g.,][]{bordoni71a, goruganthu84a}, unknowns regarding the detector deadtime\footnote{The deadtime is the time period when the detector is unable to measure incident particles due to the channel's discharge recovery time (i.e., time to replenish electrons to wall of conductive material in the microchannel plate), preamp cycle rates, etc.} \citep[e.g.,][]{meeks08a, schecker92a}, and an unknown spacecraft potential \citep[e.g.,][]{lavraud16a, pulupa14a, scime94a, scime94b}.  Significant advances in understanding the response and calibration of electrostatic analyzers have been made in recent years with the development and launch of the Magnetospheric Multiscale (MMS) mission \citep[e.g.,][]{gershman16a, gershman17a, pollock16a}.  However, the improvements resulted from an exhaustive ground calibration campaign that most other missions, including \emph{Wind}, have not had.  Further, the electronic deadtime\footnote{The cycle rate or sample rate of this preamp is listed as 2 MHz but it is not constant.} of the EESA Low preamp (i.e., AMPTEK A111) depends upon the pulse height distribution of the previous pulse [\textit{J.P. McFaddon, Personal Communication}, July 18, 2011].

\indent  Although the corrections for microchannel plate (MCP) degradation etc. have not been updated since very early in the mission, the last calibrations were performed well after the initial and most dramatic scrubbing phase that occurs when the instrument is in space \citep[e.g., see][for further discussions of MCP degradation over time]{mcfadden08a, mcfadden08b}.  The currently used calibrations are those from optical geometric factor corrections, on-ground calibrations, and in-flight calibrations [\emph{D. Larson, Personal Communication}, July 18, 2011].  Although there are expected to be corrections to these calibration values over the course of the time span examined in this work, the same data in the same time range has been presented in numerous refereed publications including but not limited to \citet[][]{bale13a, pulupa14a, pulupa14b, salem01a, salem03a, wilsoniii09a, wilsoniii10a, wilsoniii12c, wilsoniii13a, wilsoniii13b, wilsoniii18b}.  Updating the calibration tables is beyond the scope of this work but is actively being pursued [\emph{Salem et al.}, in preparation].

\indent  Although the \emph{Wind} spacecraft has the capacity to measure electric fields \citep[][]{bougeret95a}, it does not measure the DC-coupled spacecraft potential, $\phi{\scriptstyle_{sc}}$.  It does, however, consistently observe the upper hybrid line (also called the plasma line), which provides an unambiguous measure of the total electron density, $n{\scriptstyle_{e}}$.  For instance, the \emph{Wind}/SWE Faraday Cups (FCs) \citep[][]{ogilvie95} are calibrated to these measurements assuming $n{\scriptstyle_{e}}$ $=$ $n{\scriptstyle_{p}}$ $+$ 2$n{\scriptstyle_{\alpha}}$.  Ions are generally not significantly affected by $\phi{\scriptstyle_{sc}}$ as they typically have $\sim$1 keV of bulk kinetic energy in the solar wind.

\begin{deluxetable}{| l | c | c | c | c | c | c |}
  \tabletypesize{\normalsize}    
  \tablecaption{Spacecraft Potential Statistics \label{tab:SpacecraftPotential}}
  \tablehead{\colhead{$\phi{\scriptstyle_{sc}}$ [eV]} & \colhead{$X{\scriptstyle_{min}}$}\tablenotemark{a} & \colhead{$X{\scriptstyle_{max}}$} & \colhead{$\bar{X}$} & \colhead{$\tilde{X}$} & \colhead{$X{\scriptstyle_{25\%}}$} & \colhead{$X{\scriptstyle_{75\%}}$}}
  \startdata
  \hline
  All: 15,144 Finite Values                                                                   &  1.01 &  26.7 &  7.05 &  6.70 &  5.45 &  7.84  \\
  Upstream Only: 6511 Finite Values                                                          &  1.01 &  26.7 &  7.14 &  6.80 &  5.34 &  7.82  \\
  Downstream Only: 8633 Finite Values                                                        &  1.92 &  24.8 &  6.43 &  6.45 &  4.00 &  7.37  \\
  $\langle M{\scriptstyle_{f}} \rangle{\scriptstyle_{up}}$ $<$ 3 Only: 12,932 Finite Values   &  1.01 &  26.7 &  6.99 &  6.61 &  5.44 &  7.70  \\
  $\langle M{\scriptstyle_{f}} \rangle{\scriptstyle_{up}}$ $\geq$ 3 Only: 2212 Finite Values &  3.58 &  12.0 &  7.35 &  6.90 &  5.50 &  9.63  \\
  $\theta{\scriptstyle_{Bn}}$ $>$ 45$^{\circ}$ Only: 10,894 Finite Values                     &  1.01 &  26.7 &  6.70 &  6.49 &  5.35 &  7.38  \\
  $\theta{\scriptstyle_{Bn}}$ $\leq$ 45$^{\circ}$ Only: 4250 Finite Values                   &  3.53 &  17.6 &  7.94 &  7.14 &  6.10 &  10.2  \\
  \hline
  \enddata
  \tablenotetext{a}{Header symbols match that of Table \ref{tab:Exponents}}
  \tablecomments{For symbol definitions, see Appendix \ref{app:Definitions}.}
\end{deluxetable}

\indent  To estimate $\phi{\scriptstyle_{sc}}$ an initial guess is determined numerically from the ion density.  The value of $\phi{\scriptstyle_{sc}}$ is then adjusted until $n{\scriptstyle_{e}}$ $=$ $n{\scriptstyle_{ec}}$ $+$ $n{\scriptstyle_{eh}}$ $+$ $n{\scriptstyle_{eb}}$ from the fits roughly equals\footnote{Note that the value of $n{\scriptstyle_{e}}$ for a constraint is taken from SWE and the upper hybrid line observed by the WAVES radio receiver \citep[][]{bougeret95a}, when possible.} $n{\scriptstyle_{p}}$ $+$ 2$n{\scriptstyle_{\alpha}}$ and/or when photoelectrons disappear from the VDF plots\footnote{When $\phi{\scriptstyle_{sc}}$ is too low, a discontinuous ``spike'' appears in the cuts of the VDF.  The spike-like feature can also be seen in 1D energy spectra shown in the spacecraft frame with no adjustment for $\phi{\scriptstyle_{sc}}$.}.  Once a reliable estimate of $\phi{\scriptstyle_{sc}}$ determined for each VDF for each IP shock, the software is cycled through all VDFs for that event and the data are saved.  This process is repeated for each IP shock event.  An example time series of $\phi{\scriptstyle_{sc}}$ is shown in Figure \ref{fig:ExampleIPShock}.

\indent  Note that the values of $\phi{\scriptstyle_{sc}}$ determined above should not be treated as the absolute or correct spacecraft potential values.  The reason being that the detector efficiency and gain calibrations suffer from the issues discussed above. \citep[e.g.,][]{bordoni71a, goruganthu84a}.  Therefore, the $\phi{\scriptstyle_{sc}}$ values are proxies for the spacecraft potential that comprise a complicated nonlinear convolution of the real spacecraft potential and the detector deadtime and efficiency.  Despite this uncertainty, the $\phi{\scriptstyle_{sc}}$ values estimated herein are consistent with those in previously published work on the same dataset within the same time span \citep[e.g.,][]{bale13a, pulupa14a}.  Further, the consistency checks discussed in Section \ref{subsec:FitQualityAnalysis} provide further validation of the fit results.

\indent  Table \ref{tab:SpacecraftPotential} provides the one-variable statistics of the $\phi{\scriptstyle_{sc}}$ values for all VDFs, upstream and downstream only, low and high Mach number only, and quasi-parallel and quasi-perpendicular only periods.  There are no dramatic differences other than that the values of $\phi{\scriptstyle_{sc}}$ are slightly smaller downstream than upstream, slightly higher for high than low Mach number shocks, and largest (by mean, median, and quartiles) for quasi-parallel shocks.

\begin{figure*}
  \centering
    {\includegraphics[trim = 0mm 0mm 0mm 0mm, clip, height=180mm]{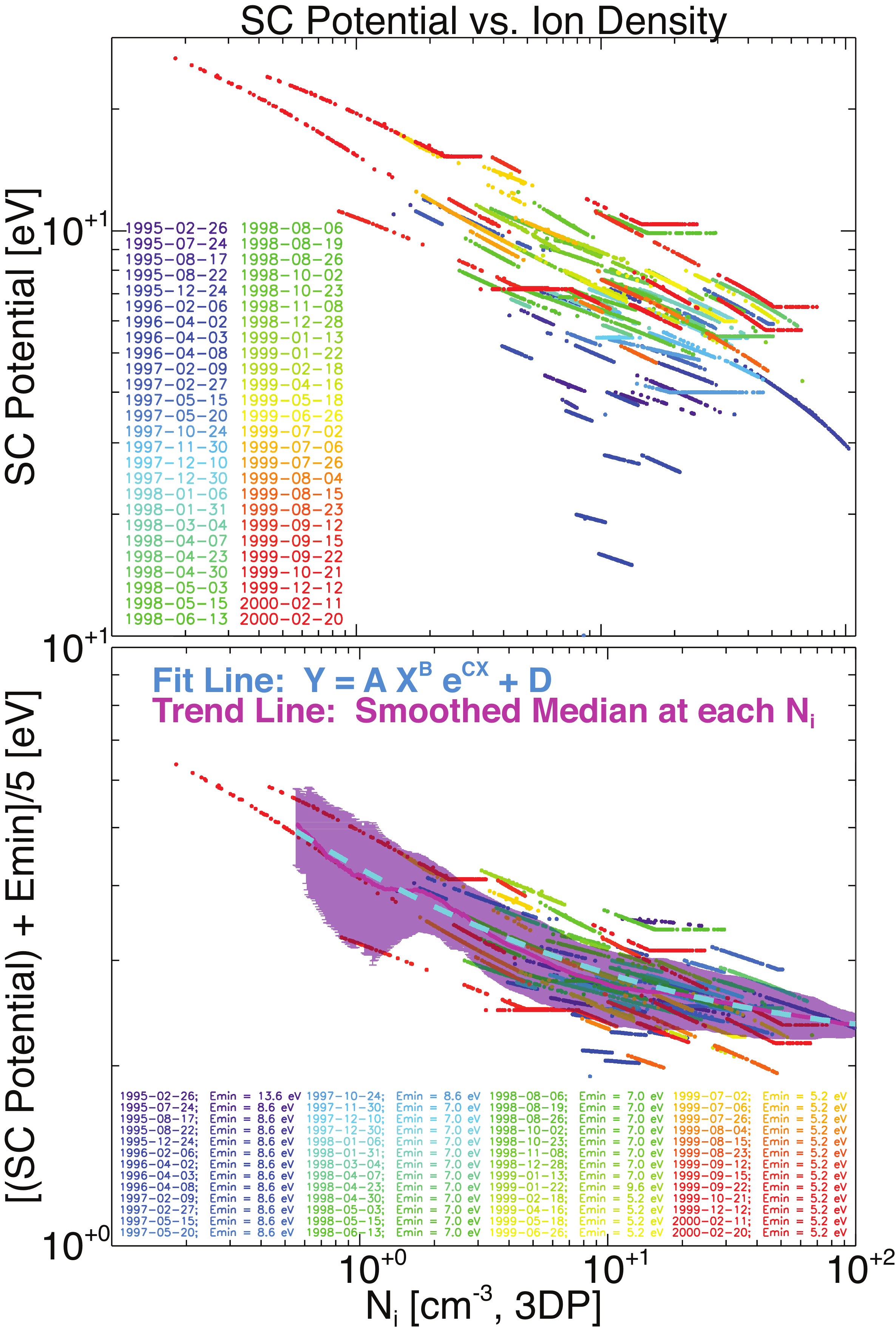}}
    \caption{Spacecraft potential, $\phi{\scriptstyle_{sc}}$, is shown versus the total ion density, $n{\scriptstyle_{i}}$, observed by the \emph{Wind}/3DP ion electrostatic analyzer (PESA Low).  The top panel shows the value of $\phi{\scriptstyle_{sc}}$ [eV] determined iteratively, as described in this appendix, versus $n{\scriptstyle_{i}}$ [cm$^{-3}$] where the color-code is defined by the IP shock data given in the lower left-hand corner.  The bottom panel shows the same data but now $\phi{\scriptstyle_{sc}}$ is offset by the detector minimum energy, $E{\scriptstyle_{min}}$, and divided by the constant 5.0 to keep the magnitudes near unity.  The $E{\scriptstyle_{min}}$ are color-coded and date-specific, as in the top panel.  The solid magenta line is a smoothed median trend line and the magenta shaded region indicates the standard deviation of the values at each $n{\scriptstyle_{i}}$.  The cyan dashed line indicates a fit line to the data using the model function defined near the top-center of this panel.}
    \label{fig:SCPotentialvsDensity}
\end{figure*}

\indent  Figure \ref{fig:SCPotentialvsDensity} shows $\phi{\scriptstyle_{sc}}$ versus $n{\scriptstyle_{i}}$ as both the raw values and a renormalized version where the EESA Low detector $E{\scriptstyle_{min}}$ is used as an offset.  The data were fit to a power-law-exponential, $Y$ $=$ $X^{B} \ e^{C \ X} + D$, where $Y$ $=$ $\left( \phi{\scriptstyle_{sc}} + E{\scriptstyle_{min}} \right)/5$ [eV] and $X$ $=$ $n{\scriptstyle_{i}}$ [$cm^{-3}$].  The fit parameters producing the cyan dashed line are $A$ $=$ 2.272$\pm$0.013 [$cm^{+3 \ B}$], $B$ $=$ -0.431$\pm$0.019 [N/A], $C$ $=$ 0.00115$\pm$0.00155 [$cm^{+3}$], and $D$ $=$ 2.0$\pm$0.0 [eV], with a reduced chi-squared value of $\tilde{\chi}^{2}$ $\sim$ 0.144.

\indent  The choice of the form of the fit line is empirical and matches the observations in trend.  The typical approach is to measure the spacecraft potential and number density then fit to a function of the spacecraft potential for the number density, i.e., $n{\scriptstyle_{i}}$ $=$ $n{\scriptstyle_{i}} \left( \phi{\scriptstyle_{sc}} \right)$ \citep[e.g.,][]{scudder00a}.  As previously stated, \emph{Wind} cannot actively measure $\phi{\scriptstyle_{sc}}$ and the values shown in Figure \ref{fig:SCPotentialvsDensity} are really a proxy due to the uncertain values for the deadtime and efficiency for each detector anode.  The purpose of the above approach is to find a semi-analytical expression for $\phi{\scriptstyle_{sc}}$ that only depends upon $n{\scriptstyle_{i}}$ (or $n{\scriptstyle_{e}}$) as an initial estimate.  The unexpected result here is that the trend depends upon $E{\scriptstyle_{min}}$ as an offset, which is likely only reflecting a one-sided measurement boundary preventing the detector from observing the entire VDF.

\indent  Note that similar analysis on the same dataset has also found a small dipolar correction to the typical monopolar approximation used herein \citep[e.g.,][]{pulupa14a}.  The dipole term is typically less than 1 eV, however, and only $\sim$1.5\% of all the VDFs examined in our study satisfied $\phi{\scriptstyle_{sc}}$ $<$ 1.5 eV.  Further, the dipole correction will only affect the odd velocity moments, i.e., the drift velocity and heat flux.  We did not calculate the heat flux but we did observe perpendicular core velocity drifts previously shown to be affected by the dipole correction \citep[e.g.,][]{pulupa14a}.

\phantomsection   
\section{Numerical Analysis Procedure}  \label{app:NumericalAnalysis}

\indent  The data are fit to a user defined model function using the nonlinear least squares fit algorithm called the Levenberg-Marquardt Algorithm (LMA) \citep[][]{more78a}.  The generalized LMA software, called MPFIT \citep[][]{markwardt09a}, requires at minimum the following inputs when fitting to a two-dimensional array of data:
\begin{description}[itemsep=0pt,parsep=0pt,topsep=0pt]
  \item[\textbf{FUNC}]  A scalar [string] defining the model function routine file name;
  \item[\textbf{X}(\textbf{Y})]  N(M)-element [numeric] array defining the first(second) dimension coordinate abscissa values;
  \item[\textbf{Z}]  NxM-element [numeric] array defining the dependent data associated with \textbf{X} and \textbf{Y} abscissa values;
  \item[\textbf{ERR}]  NxM-element [numeric] array defining the error associated with each element of \textbf{Z}; and
  \item[\textbf{PARAM}]  K-element [numeric] array defining the initial guesses for the fit parameters supplied to the model function routine \textbf{FUNC}.
\end{description}
\noindent  The error array will be ignored if the user supplies an array of weights, $\mathcal{W}$.  The details of the use of the software and documentation are provided by the author at \\
\url{https://www.physics.wisc.edu/\~craigm/idl/fitting.html} \\
and in the publication \citet[][]{markwardt09a}.

\indent  For the purposes of finding numerical fits to electron VDFs in the solar wind, a substantial set of wrapping routines were written for use with the MPFIT libraries and can be found at \\
\url{https://github.com/lynnbwilsoniii/wind\_3dp\_pros}. \\
\noindent  The wrapping software also provides detailed documentation with extensive manual pages and numerous comments throughout.

\noindent  The approach used for each electron VDF is as follows:
\begin{itemize}[itemsep=0pt,parsep=0pt,topsep=0pt]
  \item  The raw VDF data, $f^{\left( 0r \right)}$, is retrieved as an IDL structure with the data in units of counts.  A copy is created and the data structure tag is replaced with the square root of the number of counts, $f^{\left( 0cr \right)}$, i.e., Poisson statistics are assumed.
  \item  A unit conversion is applied to change to units of phase space density [i.e., cm$^{-3}$ km$^{-3}$ s$^{+3}$] then the energies are adjusted to account for the spacecraft potential \citep[e.g.,][]{salem01a, wilsoniii14a, wilsoniii16h} (details are discussed in Appendix \ref{app:DetectorCalibration}) giving $f^{\left( 0sc \right)}$ and $f^{\left( 0csc \right)}$.
  \item  Then $f^{\left( 0sc \right)}$ and $f^{\left( 0csc \right)}$ are transformed into the ion bulk flow rest frame \citep[e.g.,][]{compton35a, ipavich74a} following the methods described in \citet[][]{wilsoniii16h} (see also the associated Supplemental Material) using a relativistically correct Lorentz transformation.  The data are then interpolated onto a regular grid using Delaunay triangulation in the plane defined by the quasi-static magnetic field, $\mathbf{B}{\scriptstyle_{o}}$, along the horizontal and the transverse component of the ion bulk flow velocity, $\mathbf{V}{\scriptstyle_{i}}$, i.e., $\left( \mathbf{B}{\scriptstyle_{o}} \times \mathbf{V}{\scriptstyle_{i}} \right) \times \mathbf{B}{\scriptstyle_{o}}$.  The result is a two-dimensional gyrotropic VDF, $f^{\left( 0 \right)}$, and the associated Poisson errors/uncertainties, $f^{\left( 0c \right)}$, both as functions of the parallel, $V{\scriptstyle_{\parallel}}$, and perpendicular, $V{\scriptstyle_{\perp}}$, velocity with respect to $\mathbf{B}{\scriptstyle_{o}}$.
  \item  Numerous weighting schemes were tried and the best results (for \emph{Wind}/3DP) were achieved by defining $\mathcal{W}$ $=$ $\left(f^{\left( 0c \right)}\right)^{-2}$ for the weights\footnote{Several approaches were tried for the $\mathcal{W}$ values but the most reliable and robust was to use Gaussian weights on Poisson errors.  Reliable and robust here mean that the fitting software required the fewest number of constraints and user-imposed limits to find fit parameters that well represent the observations.}.
  \item  Every $f^{\left( 0 \right)}$ is fit to the sum of three model functions in two-dimensions\footnote{That is, the data are not fit to two one-dimensional cuts of a two-dimensional VDF separately but rather both dimensions are fit simultaneously.} for the core, halo, and beam/strahl components.  Again, the components can be fit separately because the solar wind is a non-equilibrium, weakly collisional, kinetic gas\footnote{It should also be noted that initial approaches tried to fit all electron components simultaneously, but failed.  Later approaches tried to fit the combination of only the core and halo simultaneously, but again the analysis was too unstable.  Thus, the final approach fit to each component sequentially from core-to-beam/strahl.}.  The allowed model functions (defined in Section \ref{subsec:VelocityDistributionFunctions}) and are bi-Maxwellian \citep[e.g.,][]{kasper06a}, bi-kappa  \citep[e.g.,][]{livadiotis15a, mace10a, vasyliunas68a}, symmetric bi-self-similar \citep[e.g.,][]{dum74a, dum75a}, and asymmetric bi-self-similar (defined in Section \ref{subsec:VelocityDistributionFunctions}).
  \begin{itemize}[itemsep=0pt,parsep=0pt,topsep=0pt]
    \item  It is important to note that the fit is not done for all components simultaneously.  This was the initial approach but proved to require stringent constraints for nearly all fit parameters and the software exited before all fit parameters were varied due to numerical instabilities\footnote{There is also an issue of threshold tests for convergence.  The software allows the user to define the thresholds for various gradients in the Jacobian.  If the gradient magnitudes fall below these thresholds, the software exits with a specific fit status parameter associated with the specific threshold.  For numerous reasons, the initial approach of fitting to all three components simultaneously prevented accurate fit results due to these thresholds being satisfied too early in the iteration process.} \citep[e.g.,][]{liavas99a}, discussed in Appendix \ref{app:NumericalInstability}.
    \item  Thus, the core fit, $f^{\left( core \right)}$, is performed first and then the model result subtracted from the data to yield the first residual, $f^{\left( 1 \right)}$.
    \item  The halo fit, $f^{\left( halo \right)}$, is next but only to the side of $f^{\left( 1 \right)}$ opposite to that expected for the strahl/beam, where the latter is defined as the anti-sunward direction along $\mathbf{B}{\scriptstyle_{o}}$.  The entire two-dimensional halo fit is then subtracted from $f^{\left( 1 \right)}$ to yield the second residual, $f^{\left( 2 \right)}$, i.e., both sides are subtracted but only one side is used for the fit.
    \item  The beam/strahl fit, $f^{\left( beam \right)}$, is last and fit to only the side of $f^{\left( 2 \right)}$ that is in the anti-sunward direction along $\mathbf{B}{\scriptstyle_{o}}$.
  \end{itemize}
  \item  Not all VDFs will have fit results for all three components.  In fact, $f^{\left( beam \right)}$ is often not found either because $f^{\left( halo \right)}$ left too few finite elements in $f^{\left( 2 \right)}$ or numerical instability reasons (discussed in Appendix \ref{app:NumericalInstability}).
\end{itemize}

\indent  All model functions are defined with six input parameters to be varied by the LMA software in the following order:  PARAM[0] is the number density, $n{\scriptstyle_{s}}$ [$cm^{-3}$];  PARAM[1] and PARAM[2] are the parallel and perpendicular thermal speeds, $V{\scriptstyle_{Ts, j}}$ [$km \ s^{-1}$];  PARAM[3] and PARAM[4] are the parallel and perpendicular drift speeds, $V{\scriptstyle_{os, j}}$ [$km \ s^{-1}$]; and PARAM[5] is the function exponent.  The exponent input is ignored for the bi-Maxwellian routine as it is always 2.0 but can vary in the other routines.  For the asymmetric bi-self-similar routine PARAM[4] is the parallel exponent and PARAM[5] is the perpendicular exponent (see Section \ref{subsec:VelocityDistributionFunctions} for functional form).

\indent  Initial guesses are defined for all elements of PARAM that are specific to each shock event determined through an iterative trial-and-error approach.  For each event, a zeroth order guess is used on a subset of all VDFs and the PARAM arrays for each component are adjusted accordingly to maximize the number of stable fit results for all components.  Note that the PARAM arrays for each component differ depending on whether the VDF is located upstream or downstream of the shock ramp.  In stronger shocks, the function used also varies (i.e., use symmetric bi-self-similar upstream and asymmetric bi-self-similar downstream).

\section{Numerical Instability}  \label{app:NumericalInstability}

\indent  The Levenberg-Marquardt Algorithm (LMA) software works by minimizing the the chi-squared value given by:

\begin{equation}
  \label{eq:chisqrd_0}
  \chi{\scriptstyle_{s}}^{2} = \sum_{i=0}^{N - 1} \ \sum_{j=0}^{M - 1} \ \left( f{\scriptstyle_{ij,s}}^{\left( 0 \right)} - f{\scriptstyle_{ij,s}}^{\left( mod \right)} \right)^{2} \ \lvert \mathcal{W}{\scriptstyle_{ij,s}} \rvert
\end{equation}

\noindent  where $f{\scriptstyle_{s}}^{\left( mod \right)}$ is the model fit function of component $s$ returned by the model function routine \textbf{FUNC} (see Section \ref{app:NumericalAnalysis}), $\chi{\scriptstyle_{s}}^{2}$ is the chi-squared value of the fit of component $s$, and the $i$ and $j$ subscripts correspond to the indices of the parallel and perpendicular velocity space coordinates, respectively.

\indent  A total reduced chi-squared, $\tilde{\chi}{\scriptstyle_{tot}}^{2}$, value was also calculated for all VDFs analyzed herein.  The difference in calculation is that the weights were not offset and the model function and distribution function are for the entire VDF, not the components.  Further, unlike the components, the $\tilde{\chi}{\scriptstyle_{tot}}^{2}$ values used all data points in $f^{\left( 0 \right)}$ and $\mathcal{W}$ even if they were excluded during the fit process\footnote{Specific energy-angle bins were excluded for various physical reasons in some VDFs including, for instance, energy and/or pitch-angle range constraints to avoid ``contamination'' by other components as is done to examine the halo-only and beam/strahl-only parts of the VDF.}.  However, the $\tilde{\chi}{\scriptstyle_{tot}}^{2}$ calculation excluded data below the nine-count level to avoid non-Gaussian weights in low-count values removed ``spiky'' solutions in the beam or halo fits defined by small $T{\scriptstyle_{es, j}}$ and $\kappa{\scriptstyle_{es}}$.  That is, ``spiky'' solutions are defined as those satisfying $\left( \kappa{\scriptstyle_{es}} \leq 3 \right)$ $\wedge$ $\left( \left( T{\scriptstyle_{es, \parallel}} \leq 11.8 \right) \vee \left( T{\scriptstyle_{es, \perp}} \leq 11.8 \right) \right)$ for model fit parameters.  As evidenced by Figures \ref{fig:ExampleGoodVDF}--\ref{fig:ExampleIntenseBeamVDF}, the $\tilde{\chi}{\scriptstyle_{tot}}^{2}$ parameter alone is not necessarily an accurate measure of the quality of the fit.

\indent  An unexpected nuance arose during the development and testing of the software.  The typical phase space density of any given element of $f^{\left( 0 \right)}$ for electrons near 1 AU varies from $\sim$ 10$^{-18}$ to 10$^{-8}$ cm$^{-3}$ km$^{-3}$ s$^{+3}$.  The LMA software uses a combination of gradients by constructing a Jacobian matrix of the input model fit function\footnote{i.e., the partial derivatives are with respect to the fit parameters, not the velocity coordinates}.  This is problematic when the magnitude of the input data and output model function are much much less than unity as it results in numerical instabilities \citep[e.g.,][]{liavas99a}.  That is, the partial derivative of a number on the order of 10$^{-18}$ with respect to a number slightly greater than unity can produce exceedingly small gradients.

\indent  While the limits of double-precision are not, in general, challenged by such computations, the LMA software \citep[][]{markwardt09a} was designed such that all the inputs be near unity.  The solution was to multiply $\mathcal{W}$ by a constant offset to increase the contrast in the Jacobian components that are used to minimize $\chi^{2}$.  A consequence of this approach is that the output $\chi^{2}$, $f^{\left( m \right)}$, and one-sigma error estimates of the fit parameters must be re-normalized by this offset factor.  The more standard approach is to perform the fit in logarithmic space, which reduces the dynamic range of the data.  However, as discussed in Appendix \ref{app:NumericalMethodComparisons}, this does not necessarily produce better fit results.

\indent  The above approach worked well except for cases with so called flattop distributions \citep[e.g.,][]{feldman83a, thomsen87b}, modeled using the self-similar distributions \citep[e.g.,][]{dum74a, dum75a, goldman84a, horton76a, horton79a, jain79a} given by either Equation \ref{eq:app4_8} or \ref{eq:genbissvdf_3b}.  In cases where the phase space densities were independent of energy for the core, the use of the weights above was not sufficient to constrain the fits.  In these cases, shock-specific constraints/limits were imposed on the least number of fit parameters necessary to reliably and robustly produce good results (see Supplemental Material ASCII files described in Appendix \ref{app:DataProduct} for list of constraints by shock).

\section{Numerical Method Comparisons}  \label{app:NumericalMethodComparisons}

\indent  As stated in Appendix \ref{app:NumericalInstability}, the standard approach to avoiding numerical instabilities due to the small magnitude of $f^{\left( 0 \right)}$ usually involves fitting to the logarithm of $f^{\left( 0 \right)}$ \citep[e.g.,][]{stverak09a}.  To illustrate the validity of the method used herein, an example VDF was chosen from a different study [\textit{Farrugia et al.}, in preparation] that examines a single shock-magnetic-cloud system.

\begin{figure*}
  \centering
    {\includegraphics[trim = 0mm 0mm 0mm 0mm, clip, height=150mm]{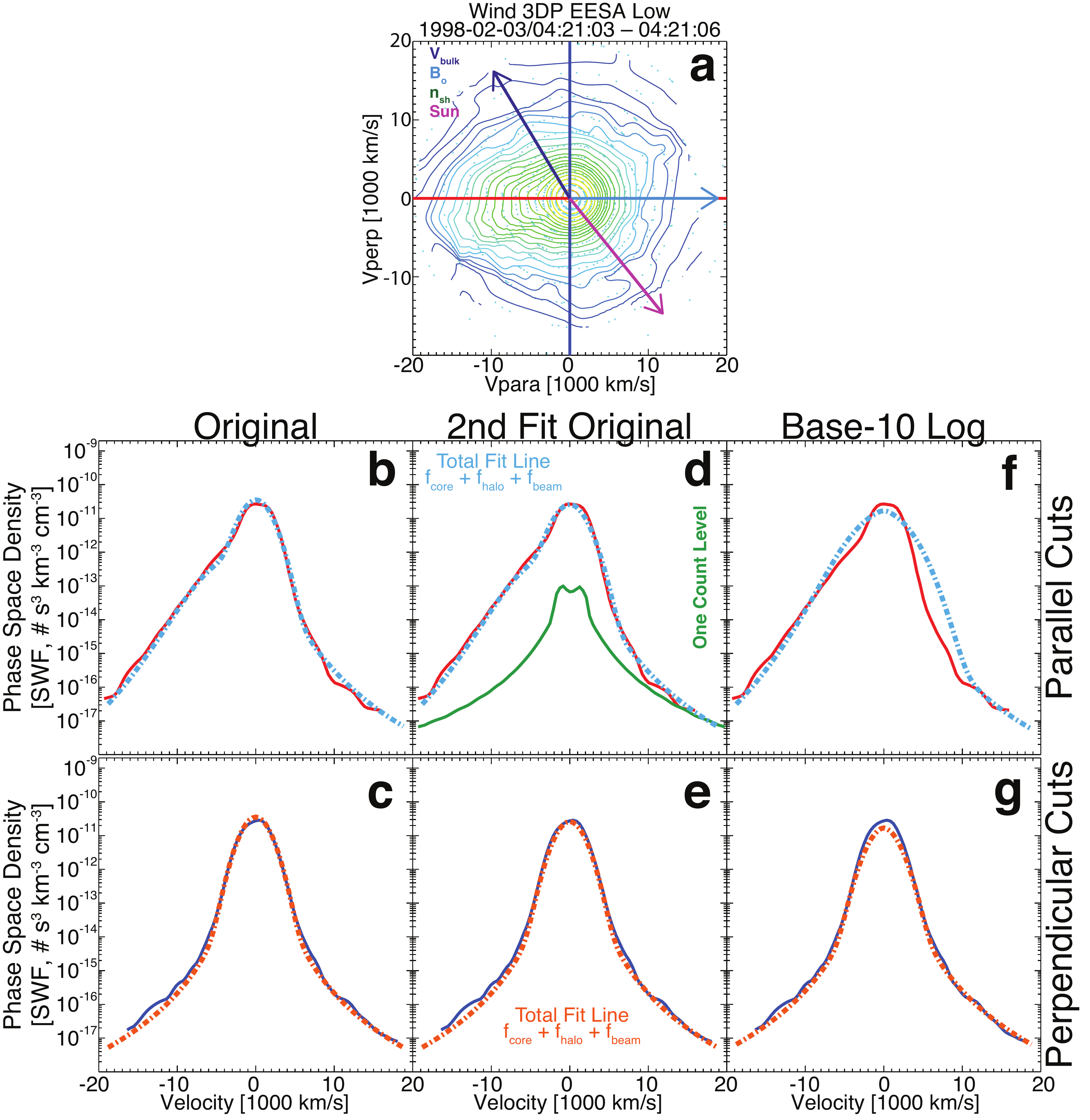}}
    \caption{An example VDF observed at 04:21:03.646 UTC on 1998-02-03 by the \emph{Wind}/3DP EESA Low detector.  The format is similar to Figures \ref{fig:ExampleGoodVDF}--\ref{fig:ExampleIntenseBeamVDF} except that each of the one-dimensional cut panel columns show a different fit result and only the total model fits are shown.  The values of the relevant parameters for this VDF are $\mathbf{B}{\scriptstyle_{o}}$ $=$ $\left( +4.59, \ -5.43, \ +1.79 \right)$ [nT, GSE], $\mathbf{V}{\scriptstyle_{i}}$ $=$ $\left( -323.98, \ -36.55, \ +30.66 \right)$ [$km \ s^{-1}$, GSE], and $\phi{\scriptstyle_{sc}}$ $=$ 12.04 eV.  Panels b, d, and f show the 1D parallel cuts along the horizontal (solid red line is data in both panels) and panels c, e, and g show the 1D perpendicular cuts along the vertical (solid blue line is data in both panels).  Panel d shows the one-count level for reference.}
    \label{fig:ExampleCompareFitVDF}
\end{figure*}

\indent  Figure \ref{fig:ExampleCompareFitVDF} shows a comparison of three different fit results to illustrate the validity of the method used herein.  Given the hindsight and statistics of the results from the present analysis, more refined constraints and better initial guesses were available. The fit shown in Panels b and c, referred to as the \textit{test fit} from hereon, was found following the automated method used for the 52 events examined in this study, i.e., the software is given initial guesses for parameters and constraints defined by knowns like $n{\scriptstyle_{e}}$ then allowed to find the best fit.  The test fit results shown in Panels b and c were then used as initial guesses (first perturbed, of course) on the same VDF to compare the method used herein (referred to as \textit{linear method}) to the base-10 logarithm approach (referred to as \textit{log method}).  A larger range of constraints were used to provide a more open parameter space.  Thus, in the following a comparison between the linear and log methods is presented as an illustrative test.

\indent  Panels d and e show the fit results using the linear method with the new initial guesses and parameter constraints while Panels f and g show the log method fit results.  Unexpectedly, the log method did much worse in the core fit than the linear method but did well for the halo and beam/strahl fits.  The numerical fit results are as follows:
\begin{itemize}[itemsep=0pt,parsep=0pt,topsep=0pt]
  \item[]  \textbf{Test Fit (Panels b and c)}
  \begin{itemize}[itemsep=0pt,parsep=0pt,topsep=0pt]
    \item  $n{\scriptstyle_{ec(h)[b]}}$ $\sim$ 1.407(0.054)[0.060] $cm^{-3}$;
    \item  $V{\scriptstyle_{Tec(h)[b], \parallel}}$ $\sim$ 2028.0(3621.7)[4183.7] $km \ s^{-1}$;
    \item  $V{\scriptstyle_{Tec(h)[b], \perp}}$ $\sim$ 1927.2(3486.6)[2833.1] $km \ s^{-1}$;
    \item  $V{\scriptstyle_{oec(h)[b], \parallel}}$ $\sim$ $+$50.4(0.0)[-3752.6] $km \ s^{-1}$;
    \item  $V{\scriptstyle_{oec(h)[b], \perp}}$ $\sim$ 0.0(0.0)[0.0] $km \ s^{-1}$;
    \item  $s{\scriptstyle_{ec}}$ $\sim$ 2.002;
    \item  $\kappa{\scriptstyle_{eh}}$ $\sim$ 1.908;
    \item  $\kappa{\scriptstyle_{eb}}$ $\sim$ 5.151;
    \item  $\delta \mathcal{R}$ $\sim$ 14.1\%;
    \item  $\tilde{\chi}{\scriptstyle_{c(h)[b]}}^{2}$ $\sim$ 4.52(1.82)[2.98];
    \item  $\tilde{\chi}{\scriptstyle_{tot}}^{2}$ $\sim$ 1.30;
    \item  Fit Flag \{c,h,b\} $=$ \{8, 8, 8\}.
  \end{itemize}
  \item[]  \textbf{Linear Method Fit (Panels d and e)}
  \begin{itemize}[itemsep=0pt,parsep=0pt,topsep=0pt]
    \item  $n{\scriptstyle_{ec(h)[b]}}$ $\sim$ 1.122(0.051)[0.055] $cm^{-3}$;
    \item  $V{\scriptstyle_{Tec(h)[b], \parallel}}$ $\sim$ 2183.7(3694.7)[4154.0] $km \ s^{-1}$;
    \item  $V{\scriptstyle_{Tec(h)[b], \perp}}$ $\sim$ 1947.4(3557.0)[2863.1] $km \ s^{-1}$;
    \item  $V{\scriptstyle_{oec(h)[b], \parallel}}$ $\sim$ 0.0(0.0)[-3960.1] $km \ s^{-1}$;
    \item  $V{\scriptstyle_{oec(h)[b], \perp}}$ $\sim$ 0.0(0.0)[0.0] $km \ s^{-1}$;
    \item  $s{\scriptstyle_{ec}}$ $\sim$ 2.000;
    \item  $\kappa{\scriptstyle_{eh}}$ $\sim$ 1.901;
    \item  $\kappa{\scriptstyle_{eb}}$ $\sim$ 5.073; and
    \item  $\delta \mathcal{R}$ $\sim$ 12.5\%;
    \item  $\tilde{\chi}{\scriptstyle_{c(h)[b]}}^{2}$ $\sim$ 3.73(1.82)[2.98];
    \item  $\tilde{\chi}{\scriptstyle_{tot}}^{2}$ $\sim$ 1.01;
    \item  Fit Flag \{c,h,b\} $=$ \{8, 8, 8\}.
  \end{itemize}
  \item[]  \textbf{Log Method Fit (Panels f and g)}
  \begin{itemize}[itemsep=0pt,parsep=0pt,topsep=0pt]
    \item  $n{\scriptstyle_{ec(h)[b]}}$ $\sim$ 1.086(0.089)[0.062] $cm^{-3}$;
    \item  $V{\scriptstyle_{Tec(h)[b], \parallel}}$ $\sim$ 3248.5(2938.9)[3762.2] $km \ s^{-1}$;
    \item  $V{\scriptstyle_{Tec(h)[b], \perp}}$ $\sim$ 2086.3(3043.6)[2652.4] $km \ s^{-1}$;
    \item  $V{\scriptstyle_{oec(h)[b], \parallel}}$ $\sim$ 0.0(0.0)[-4206.0] $km \ s^{-1}$;
    \item  $V{\scriptstyle_{oec(h)[b], \perp}}$ $\sim$ 0.0(0.0)[0.0] $km \ s^{-1}$;
    \item  $s{\scriptstyle_{ec}}$ $\sim$ 2.000;
    \item  $\kappa{\scriptstyle_{eh}}$ $\sim$ 1.852;
    \item  $\kappa{\scriptstyle_{eb}}$ $\sim$ 4.016;
    \item  $\delta \mathcal{R}$ $\sim$ 20.0\%;
    \item  $\tilde{\chi}{\scriptstyle_{c(h)[b]}}^{2}$ $\sim$ 82.7(2.51)[2.15];
    \item  $\tilde{\chi}{\scriptstyle_{tot}}^{2}$ $\sim$ 0.71;
    \item  Fit Flag \{c,h,b\} $=$ \{2, 6, 6\}.
  \end{itemize}
\end{itemize}
\noindent  Thus, one can see that the log method did not produce a better fit for this specific example, which was not the expected outcome.  This is almost certainly a consequence of the large constraint ranges and a better fit would be found for a tighter range.  That is, this example is not meant to argue that the linear method is better than the log method.  Rather the example is meant to illustrate that the linear method is a viable approach.

\indent  A point should also be made about the initiation stability of the LMA software.  During the course of fitting all the VDFs in the present study, it was found that the choice of initial guess parameters was critical.  For instance, in the example shown in Figure \ref{fig:ExampleCompareFitVDF}, the initial guess values used for the core fit were $n{\scriptstyle_{ec}}$ $\sim$ 2.0 $cm^{-3}$, $V{\scriptstyle_{Tec, \parallel [\perp]}}$ $\sim$ 2297[2297] km/s (i.e., 15 eV temperatures), $V{\scriptstyle_{oec, \parallel [\perp]}}$ $\sim$ $+$10.0[0.0] km/s, and $s{\scriptstyle_{ec}}$ $\sim$ 2.0.  If any of the parameters were perturbed by $\sim$20--30\% away from these initial guesses, the log method would not initiate fit iterations due to diverging deviates and/or diverging model results, i.e., the software could not establish an initial Jacobian\footnote{This is associated with a fit status code of -16 as reported in the fit results ASCII file discussed in Appendix \ref{app:DataProduct}.}.  Unexpectedly, the linear method was more tolerant of perturbed initial guess parameters.  There are still several checks for each component fit to address this possible non-initiation error but even so this sometimes did not fix the issue, which is one reason why not all VDFs had stable solutions.

\indent  Finally a note about the one-sigma uncertainties of every fit parameter.  These values are not reported because it was found they do not accurately or consistently reflect the quality of fit.  For instance, the one-sigma uncertainties of $n{\scriptstyle_{eh}}$ and $V{\scriptstyle_{Teh, \parallel}}$ for the log method in the example VDF shown in Figure \ref{fig:ExampleCompareFitVDF} (Panels f and g) are $\sim$19,988 km/s (i.e., $\sim$617\% error) and $\sim$3.53 $cm^{-3}$ (i.e., $\sim$5163\% error), respectively, even though $\tilde{\chi}{\scriptstyle_{h}}^{2}$ $\sim$ 2.51.  The one-sigma uncertainties for the same parameters but for the fit in Panels d and e are $\sim$110.1 km/s (i.e., $\sim$3.1\% error) and $\sim$0.0047 $cm^{-3}$ (i.e., $\sim$8.5\% error) and $\tilde{\chi}{\scriptstyle_{h}}^{2}$ $\sim$ 1.82.  That is, the reduced chi-squared values differ by only $\sim$39\% but the one-sigma uncertainties by 100s to 1000s percent.  The one-sigma uncertainties determined by the LMA software that are assigned to the output fit parameters are not representative of the actual uncertainties.  The reason is related to the orthogonal basis constructed during the qr-factorization (ultimately used to minimize $\tilde{\chi}{\scriptstyle_{s}}^{2}$) is not the same basis as that for the fit parameters.  The output uncertainties thus contain nonlinear convolution of one-sigma uncertainties from potentially multiple fit parameters.  The effect is analogous to electric field measurements from two antenna with differing noise levels.  If the electric field data are rotated to a new coordinate basis from the original instrument basis, the resulting field components will have a nonlinear convolution of noise from the original components.  Thus, the one-sigma uncertainties were not used as errors for each parameter.

\indent  The one-sigma errors are also forced to zero in the software when the fit value reaches a user-defined boundary/constraint/limit.  This is reported in the fit constraints ASCII file described in Appendix \ref{app:DataProduct} (i.e., under the heading ``Peg'' in the ASCII file).  As previously stated, the $\delta \mathcal{R}$ value alone does not always characterize the quality of any given fit.  Therefore, a combination of parameters were used to define fit quality flags (see Appendix \ref{app:DataProduct} for definitions), which should be used for determining the reliability of any given fit.

\phantomsection   
\section{Data Product}  \label{app:DataProduct}

\indent  One of the primary purposes of this first part of this three-part study is to describe the methodology and nuances of the fit procedure to provide context and documentation for the resulting data product.  This will serve as the reference document for use of the data product by the heliospheric and astrophysical communities.  The nuances and details of the procedure are critical for reproducibility and quality control in the use of the data product described in this section.  While Papers II and III discuss the statistics and analysis results in detail, this first part is critical for any statistical or physical interpretation of the data and it includes analysis of the exponents and drifts.

\indent  The fit results are provided in two ASCII files.  The first contains all fit parameters for the three electron components in addition to several other relevant parameters.  The non-electron data products are linearly interpolated to the midpoint time stamp of each electron VDF.  The ASCII file contains a detailed header with descriptions and explanations of the parameters with associated units.  The data included are as follows:  UTC time of electron VDF midpoint time stamp; $n{\scriptstyle_{p}}$ and $n{\scriptstyle_{\alpha}}$ measured by SWE [$cm^{-3}$]; $n{\scriptstyle_{i}}$ measured by 3DP [$cm^{-3}$]; $T{\scriptstyle_{p, j}}$ and $T{\scriptstyle_{\alpha, j}}$ measured by SWE [$eV$]; $T{\scriptstyle_{i, j}}$ measured by 3DP [$eV$]; $B{\scriptstyle_{o,j}}$ measured by MFI [$nT$]; $V{\scriptstyle_{p, j}}$ and $V{\scriptstyle_{\alpha, j}}$ measured by SWE [$km \ s^{-1}$]; $V{\scriptstyle_{i, j}}$ measured by 3DP [$km \ s^{-1}$]; $\phi{\scriptstyle_{sc}}$ determined from fit process [$eV$]; $\delta \mathcal{R}$ calculated from fit process [\%]; $n{\scriptstyle_{es}}$ from 3DP fits [$cm^{-3}$]; $T{\scriptstyle_{es, j}}$ from 3DP fits [$eV$]; $V{\scriptstyle_{oes, j}}$ from 3DP fits [$km \ s^{-1}$]; $\kappa{\scriptstyle_{es}}$, $p{\scriptstyle_{es}}$, and $q{\scriptstyle_{es}}$ from 3DP fits [N/A]; $\tilde{\chi}{\scriptstyle_{s}}^{2}$ from 3DP fits [N/A]; and the numeric fit status value for each electron component [N/A].  The total reduced chi-squared values for all fits are also included in the ASCII file.  The fit flags for each component fit are also included.  Let $\Xi$ $\equiv$ $\sum_{s} \ \tilde{\chi}{\scriptstyle_{s}}^{2}$, then the list is as follows:
\begin{itemize}[itemsep=0pt,parsep=0pt,topsep=0pt]
  \item  Fit Flag $\{c,h,b\}$ $=$ 0  :  $\left( 100\% \leq \delta \mathcal{R} \right)$ $\vee$ non-finite for any of the following:  $\Xi$, $\tilde{\chi}{\scriptstyle_{s}}^{2}$, $\delta \mathcal{R}$
  \item  Fit Flag $\{c,h,b\}$ $=$ 1  :  $\left( 100 \leq \tilde{\chi}{\scriptstyle_{tot}}^{2} < 10^{30} \right)$ $\wedge$ $\left( \left( \Xi < 200 \right) \vee \left( \tilde{\chi}{\scriptstyle_{s}}^{2} \leq 200 \right) \right)$ $\wedge$ $\left( \delta \mathcal{R} < 95\% \right)$
  \item  Fit Flag $\{c,h,b\}$ $=$ 2  :  $\left( 0 \leq \tilde{\chi}{\scriptstyle_{tot}}^{2} < 100 \right)$ $\wedge$ $\left( \left( \Xi < 100 \right) \vee \left( \tilde{\chi}{\scriptstyle_{s}}^{2} \leq 100 \right) \right)$ $\wedge$ $\left( \delta \mathcal{R} < 75\% \right)$
  \item  Fit Flag $\{c,h,b\}$ $=$ 3  :  $\left( 0 \leq \tilde{\chi}{\scriptstyle_{tot}}^{2} < 100 \right)$ $\wedge$ $\left( \left( \Xi < 50 \right) \vee \left( \tilde{\chi}{\scriptstyle_{s}}^{2} \leq 40 \right) \right)$ $\wedge$ $\left( \delta \mathcal{R} < 55\% \right)$
  \item  Fit Flag $\{c,h,b\}$ $=$ 4  :  $\left( 0 \leq \tilde{\chi}{\scriptstyle_{tot}}^{2} < 100 \right)$ $\wedge$ $\left( \left( \Xi < 40 \right) \vee \left( \tilde{\chi}{\scriptstyle_{s}}^{2} \leq 30 \right) \right)$ $\wedge$ $\left( \delta \mathcal{R} < 50\% \right)$
  \item  Fit Flag $\{c,h,b\}$ $=$ 5  :  $\left( 0 \leq \tilde{\chi}{\scriptstyle_{tot}}^{2} < 100 \right)$ $\wedge$ $\left( \left( \Xi < 30 \right) \vee \left( \tilde{\chi}{\scriptstyle_{s}}^{2} \leq 20 \right) \right)$ $\wedge$ $\left( \delta \mathcal{R} < 45\% \right)$
  \item  Fit Flag $\{c,h,b\}$ $=$ 6  :  $\left( 0 \leq \tilde{\chi}{\scriptstyle_{tot}}^{2} < 100 \right)$ $\wedge$ $\left( \left( \Xi < 20 \right) \vee \left( \tilde{\chi}{\scriptstyle_{s}}^{2} \leq 10 \right) \right)$ $\wedge$ $\left( \delta \mathcal{R} < 40\% \right)$
  \item  Fit Flag $\{c,h,b\}$ $=$ 7  :  $\left( 0 \leq \tilde{\chi}{\scriptstyle_{tot}}^{2} < 30 \right)$ $\wedge$ $\left( \left( \Xi < 15 \right) \vee \left( \tilde{\chi}{\scriptstyle_{s}}^{2} \leq 9 \right) \right)$ $\wedge$ $\left( \delta \mathcal{R} < 30\% \right)$
  \item  Fit Flag $\{c,h,b\}$ $=$ 8  :  $\left( 0 \leq \tilde{\chi}{\scriptstyle_{tot}}^{2} < 30 \right)$ $\wedge$ $\left( \left( \Xi < 10 \right) \vee \left( \tilde{\chi}{\scriptstyle_{s}}^{2} \leq 7 \right) \right)$ $\wedge$ $\left( \delta \mathcal{R} < 20\% \right)$
  \item  Fit Flag $\{c,h,b\}$ $=$ 9  :  $\left( 0 \leq \tilde{\chi}{\scriptstyle_{tot}}^{2} < 15 \right)$ $\wedge$ $\left( \left( \Xi < 7 \right) \vee \left( \tilde{\chi}{\scriptstyle_{s}}^{2} \leq 5 \right) \right)$ $\wedge$ $\left( \delta \mathcal{R} < 15\% \right)$
  \item  Fit Flag $\{c,h,b\}$ $=$ 10  :  $\left( 0 \leq \tilde{\chi}{\scriptstyle_{tot}}^{2} < 7 \right)$ $\wedge$ $\left( \left( \Xi < 5 \right) \vee \left( \tilde{\chi}{\scriptstyle_{s}}^{2} \leq 3 \right) \right)$ $\wedge$ $\left( \delta \mathcal{R} \leq 10\% \right)$
\end{itemize}

\indent  The second ASCII file contains the fit constraints, initial guesses, whether the fit parameters reached a fit constraint boundary, the number of iterations required to reach a stable fit, the chi-squared of the fit, the degrees of freedom of the inputs, and a two-letter code for the model function used.

\indent  Both ASCII files contain fit results even if they are not high quality or reliable results, which can be determined from the combination of $\tilde{\chi}{\scriptstyle_{s}}^{2}$, $\tilde{\chi}{\scriptstyle_{tot}}^{2}$, and $\delta \mathcal{R}$ used to define the fit flags in the first ASCII file, as discussed previously.  The entries with fill values (listed in the header) resulted because a stable fit was not found or the fit was determined to be ``bad,'' as defined in Section \ref{subsec:FitQualityAnalysis} and Appendix \ref{app:NumericalAnalysis}.  When there is a significant discrepancy between $n{\scriptstyle_{p}}$ and $n{\scriptstyle_{i}}$ (e.g., differ by a factor exceeding $\sim$40\%), the more reliable/accurate of the two is $n{\scriptstyle_{p}}$.  Under these circumstances, $T{\scriptstyle_{i, j}}$ and $V{\scriptstyle_{i, j}}$ should be subject to scrutiny as well.  The model function used for the core is given in the second ASCII file.

\indent  Note that the second ASCII file will contain non-fill, fit values for the same parameters that are all fill values in the first ASCII file.  Although many constraints were set as far from the expected values as possible to avoid a parameter from being limited during the fit, some were imposed after all the fits were found for a given shock crossing.  These were imposed for physical reasons (e.g., see Section \ref{subsec:FitPhysicalConstraints}) and to avoid issues during regridding and/or interpolation for comparison with other datasets (e.g., magnetic fields).  These post-fit constraints are 1.5 $<$ $\kappa{\scriptstyle_{eh}}$ $\leq$ 20, 1.5 $<$ $\kappa{\scriptstyle_{eb}}$ $\leq$ 20, 0 $\leq$ $n{\scriptstyle_{eh}}/n{\scriptstyle_{ec}}$ $\leq$ 0.75, 0 $\leq$ $n{\scriptstyle_{eb}}/n{\scriptstyle_{ec}}$ $\leq$ 0.50, 0.0 $\leq$ $n{\scriptstyle_{eb}}/n{\scriptstyle_{eh}}$ $\leq$ 3.0, 11.4 eV $\leq$ $T{\scriptstyle_{eh, j}}$ $\leq$ 285 eV, and 11.4 eV $\leq$ $T{\scriptstyle_{eb, j}}$ $\leq$ 285 eV.  All statistics and fit results presented herein are with respect to the first ASCII file values but we include all the fit results in the second ASCII file for reference.  This is because some of our post-fit constraints eliminated good fits like that shown in Figure \ref{fig:ExampleIntenseBeamVDF}, which failed the $n{\scriptstyle_{eb}} / n{\scriptstyle_{eh}}$ $<$ 3 test.  Most of the fits that failed this specific test were clearly bad fits but not all.

\indent  The purpose of providing the detailed inputs for the fit results is for reproducibility and for quality control/sanity-checks for users interested in future use by the heliospheric and astrophysical communities.  The data product will benefit current and future missions like Parker Solar Probe in addition to providing a statistical comparison with astrophysical shocks, which was currently not available.

\clearpage

\end{document}